\documentclass[12pt]{article}
\usepackage[toc,page]{appendix}
\usepackage{epsf}
\usepackage{amsmath,amssymb}
\usepackage{pdfpages}
\usepackage{graphicx}
\usepackage{physics}
\usepackage{mathtools}
\usepackage{slashed}
\usepackage[hidelinks]{hyperref}
\usepackage{tikz}
\usetikzlibrary{positioning}
\usepackage[compat=1.0.0]{tikz-feynman}
\usepackage{circuitikz}
\usepackage{pgfplots}
\usepackage{subfig}
\usepackage{multirow}
\usepackage{cite}
\usepackage{mhchem}
\usepackage{siunitx}

\usepackage{MnSymbol}

\begin{document}

\newcommand{\rem}[1]{{$\spadesuit$\bf #1$\spadesuit$}}
\newcommand{\SC}[1]{{\textcolor{cyan}{$\spadesuit$\bf [SC] #1$\spadesuit$}}}

\renewcommand{\thefootnote}{\fnsymbol{footnote}}
\setcounter{footnote}{0}

\begin{titlepage}

\def\thefootnote{\fnsymbol{footnote}}

\begin{center}

\hfill July 2023\\

\vskip .5in

{\Large \bf
  Dark matter detection using nuclear magnetization in magnet with hyperfine interaction
  \\

}

\vskip .3in

{\large
  So Chigusa$^{(a,b,e)}$, Takeo Moroi$^{(c,e)}$, Kazunori Nakayama$^{(d,e)}$\\[1mm]
  and Thanaporn Sichanugrist$^{(c)}$
  \\
}

\vskip 0.3in

$^{(a)}$
{\em
Theoretical Physics Group,
Lawrence Berkeley National Laboratory,\\
Berkeley, CA 94720, USA
}

\vskip 0.1in

$^{(b)}$
{\em
Berkeley Center for Theoretical Physics, Department of Physics,\\
University of California, Berkeley, CA 94720, USA
}

\vskip 0.1in

$^{(c)}$
{\em
Department of Physics, The University of Tokyo, Tokyo 113-0033, Japan
}

\vskip 0.1in

$^{(d)}$
{\em
Department of Physics, Tohoku University, Sendai 980-8578, Japan
}

\vskip 0.1in

$^{(e)}$
{\em
International Center for Quantum-field Measurement Systems for Studies of the Universe and Particles (QUP), KEK, 1-1 Oho, Tsukuba, Ibaraki 305-0801, Japan
}

\end{center}
\vskip .5in

\begin{abstract}

We consider the possibility to detect cosmic light dark matter (DM), i.e., axions and dark
photons, of mass $\sim \SI{e-6}{eV}$ and  $\sim \SI{e-4}{eV}$, by magnetic excitation in a magnet with strong hyperfine interaction. In particular, we consider a canted antiferromagnet, \ce{MnCO3}, as a concrete candidate material. With spin transfer between nuclear and electron spins allowed by the hyperfine interaction, nuclear spins become naturally highly polarized due to an effective (electron-spin-induced) magnetic field, and have long-range interactions with each other. The collective precession of nuclear spins, i.e., a nuclear magnon, can be generated by the DM field through the nucleon-DM interaction, while they are also sensitive to the electron-DM interaction through the electron-nuclear spin mixing. Compared with conventional nuclear-spin precession experiments, this system as a DM sensor is sensitive to higher frequency needing only a small static magnetic field applied. The system also has collective precession of electron spins, mixed with nuclear spins, as the additional channels that can be used for DM probes. We estimate the sensitivity under appropriate readout setups such as an inductive pick-up loop associated with an \textit{LC} resonant circuit, or a photon cavity with a photon counting device. We show that this method covers an unexplored parameter region of light bosonic DM.

\end{abstract}

\end{titlepage}

\renewcommand{\thepage}{\arabic{page}}
\setcounter{page}{1}
\renewcommand{\thefootnote}{\#\arabic{footnote}}
\setcounter{footnote}{0}
\renewcommand{\theequation}{\thesection.\arabic{equation}}

\tableofcontents
\newpage
\section{Introduction}
\label{sec:intro}
\setcounter{equation}{0}

Dark matter (DM) has been the unanswered mystery of particle physics for decades. Its existence, covering $80\%$ of matter in the Universe, has been vividly suggested and supported by both astrophysical and cosmological evidence (for a review, see, e.g., Ref.~\cite{Workman:2022ynf}). However, DM cannot be reasonably accounted for by particles already known in the standard model (SM), and its particle-physics properties are still unknown. Undoubtedly, DM would provide us priceless hints for understanding physics beyond the standard model, even if we discover just a few secrets of its anatomy directly.

Light bosonic particles with mass $\SI{e-22}{eV}$--$\si{keV}$ are possible candidates of DM. These include axions, pseudo Nambu-Goldstone bosons emerging from the Peccei--Quinn (PQ) symmetry breaking \cite{Peccei:1977hh, Peccei:1977ur, Weinberg:1977ma, Wilczek:1977pj}, including axion-like-particles (ALPs) motivated from string theory \cite{Svrcek:2006yi, Arvanitaki:2009fg, Cicoli:2012sz}, and dark photons \cite{Jaeckel:2010ni, Jaeckel:2012mjv, Arias:2012az, Fabbrichesi:2020wbt}, spin-1 vector bosons kinetically mixing with ordinary photons. Generally, they have small mass with large number density which causes them to effectively behave like a coherent classical field.  This type of DM is out of reach from the nuclear (electron) recoil experiments, e.g., CDMS \cite{CDMS-II:2009ktb, CDMS:2009fba, SuperCDMS:2018mne}, XENON \cite{XENON:2007uwm, XENON:2020kmp}, and PANDA \cite{PandaX-II:2017hlx, PandaX-II:2018xpz}; these experiments have excellent sensitivity for $\sim \ \si{GeV}$ ($\si{MeV}$) DM, however, for the DM of mass below $\si{GeV}$ ($\si{MeV}$) the sensitivities become worse rapidly because the recoil energy is severely suppressed due to the smallness of the momentum of DM. So far, constraints on the sub-MeV and light DM are weak and require dedicated methods for their detection.

Interestingly, physical excitations in condensed matter show their unique and bizarrely various range of excitation energy which is sensitive to the energy deposition of the sub-eV scale. They also provide rich types of interaction between them and DM. The scattering and absorption of light DM with/via the excitation in a solid have been studied and experimented on by, e.g., SENSEI \cite{SENSEI:2020dpa}, QUAX \cite{Barbieri:2016vwg}, and CASPEr \cite{Budker:2013hfa,JacksonKimball:2017elr,PhysRevLett.126.141802}. They aim to utilize electronic states, collective electron spins, and nuclear spins in condensed matter, respectively. Various types of excitation are showing their potential: for example, electrons~\cite{Hochberg:2015pha,Hochberg:2015fth,Hochberg:2016sqx,Bloch:2016sjj,Hochberg:2017wce,Coskuner:2019odd,Trickle:2019nya,Griffin:2019mvc,Geilhufe:2019ndy,Trickle:2020oki,Hochberg:2021pkt,Knapen:2021run,Knapen:2021bwg,Mitridate:2021ctr,Chen:2022pyd,Mitridate:2022tnv,Trickle:2022fwt}, phonons~\cite{Knapen:2017ekk,Griffin:2018bjn,Cox:2019cod,Mitridate:2020kly,Marsh:2022fmo}, electron spins or magnons~\cite{Trickle:2019ovy,Chigusa:2020gfs,Mitridate:2020kly,Esposito:2022bnu,Chigusa:2023hms}, electron states in topological material~\cite{Marsh:2018dlj,Schutte-Engel:2021bqm,Chigusa:2021mci}, and qubits~\cite{Dixit:2020ymh,Chen:2022quj}.

In this paper, we consider the nuclear magnetization system with a strong hyperfine interaction \cite{PhysRev.129.1105, borovik1984spin, Shiomi:2019aa, Kikkawa:2021aa,doi:10.1063/5.0107157}, which allows the mixing between electron spin and nuclear spin. 
At low temperature, nuclear spins are highly polarized due to the electron-spin-induced effective magnetic field of $ O(10) \, \si{T}$. Besides, nuclear spins have a long-range interaction with each other via the so-called Suhl-Nakamura exchange interaction \cite{PhysRev.109.606,10.1143/PTP.20.542}, which ensures the existence of nuclear spin waves (or nuclear magnons in the quantized version \cite{doi:10.1063/5.0107157}). The presence of DM can excite the mixed state between electron and nuclear spin precessions, resulting in macroscopically observable magnetization, which is enhanced by a factor of $O(10)-O(10^3)$ compared with the typical nuclear magnetization signal.  
According to the eigenfrequency profile of material with a strong hyperfine interaction such as \ce{MnCO3} \cite{doi:10.1143/JPSJ.15.2251,doi:10.1063/1.1713505, Shiomi:2019aa,Kikkawa:2021aa}, \ce{CsMnF3} \cite{PhysRev.132.144, PhysRev.143.361, PhysRev.156.370}, 
\ce{CoCO3} \cite{borovik1984spin}, and \ce{FeBO3} \cite{borovik1984spin}, they may probe $\sim \si{\micro eV}$ -- $\si{m eV}$ DM with good sensitivity to various interaction parameters through either electron-DM or nucleon-DM coupling.\footnote{To achieve sensitivity in such a range of frequency, technically, a nuclear magnetic system without a strong hyperfine interaction needs an applied magnetic field of magnitude larger than 30 T for the experimental setup.} 
This kind of system also has electron-spin precession modes, mixed with nuclear spins, as an additional channel that can be used for DM probes. 
In order to show the possibility of using magnetically ordered material with strong hyperfine interaction for DM detection, we focus on the canted antiferromagnetic material \ce{MnCO3}, which has strong hyperfine interaction with $100\%$ magnetic isotope, as a concrete example. We explore in detail its resonance profile and response to the DM field composed of axions or dark photons under the sensible readout setup such as an inductive pick-up loop associated with an \textit{LC} resonant circuit, or a photon cavity with a photon counting device.
We show that a sizable signal from a DM axion or dark photons can be expected.

We summarize the advantage of a magnetic system with hyperfine interaction as follows:
\begin{enumerate}
\item  Nuclear spins are naturally highly polarized, which leads to large resonance signal.
\item Sensitivity at high DM mass is achievable with a small applied static field, compared with a nuclear magnetic system without hyperfine interaction.
\item There are sensitivities to both electron-DM or nucleon-DM couplings.
\item Readout is enhanced or becomes possible by the presence of mixing of electron spins to nuclear spins.
\end{enumerate}

The construction of this paper is as follows. We start with a review of DM focusing on the two promising candidates, axions and dark photons in Sec.~\ref{sec: DM target}. We elaborate on the properties of the material in Sec.~\ref{sec: nuclear MD}: the magnetic material with strong hyperfine interaction and excitation of hybridized precession state of electron and nuclear spins induced by DM. In Sec. \ref{sec: sensitivity}, we show the sensitivity of light DM detection with several concrete proposals of experimental setup. We conclude in Sec.~\ref{sec: discuss and conclusion}.

%%%%%%%%%%%%%%%%%%%%%%%%%%%%%%%%%%%%%%%%%%%%%%%%%%%
\section{Dark matter target}
\label{sec: DM target}
\setcounter{equation}{0}
%%%%%%%%%%%%%%%%%%%%%%%%%%%%%%%%%%%%%%%%%%%%%%%%%%%

We consider two candidates of light bosonic DM, axions and dark photons, whose masses correspond to the energy which can cause the magnetic resonance in a magnet with hyperfine interaction. Because their masses are small, their number densities are large such that they behave coherently as a classical field and may act as an oscillating magnetic field in the view of spins of SM particles. If the coupling is strong enough, spins inside magnetic material can be perturbed from the ground state and macroscopically produce an observable signal as oscillating magnetization. We review models of axions and dark photons, and describe the oscillating effective magnetic field induced by them.
We illustrate in Sec.~\ref{sec: nuclear MD} the response of magnet excited by DM, based on \ce{MnCO3} as a concrete example material with strong hyperfine interaction. 

In this section, natural unit is adopted.

\subsection{DM axion}

QCD axions and ALPs are candidates for the DM particle. QCD axions are pseudo Nambu-Goldstone bosons arising from the PQ symmetry breaking \cite{Peccei:1977hh, Peccei:1977ur, Weinberg:1977ma, Wilczek:1977pj}. They are proposed to solve the strong \textit{CP} problem.  QCD axions interact with gluons, photons, and SM fermions with interaction strength determined by the axion mass. On the other hand, ALPs are the generalization of QCD axions; they are particles that interact with SM particles with a similar form of interactions but do not have roles to solve the strong \textit{CP} problem and do not have a specific relation between the mass and the coupling with SM particles. We collectively call QCD axions and ALPs as simply ``axions.''

\subsubsection{Axion model and effective Lagrangian}

Typically, the QCD axion is introduced as the Nambu-Goldstone boson from spontaneous symmetry breaking of a new U(1) symmetry, the so-called U(1) PQ symmetry, with chiral anomaly associated with the color SU(3) symmetry and also electromagnetic symmetry in some models.
With this mechanism called the PQ mechanism, axion effective Lagrangian reads as
\begin{equation}
\mathcal{L}_a= \frac{1}{2} (\partial_\mu a)^2 +\frac{1}{2} m_a^2 a^2+\frac{1}{4} g_{a \gamma \gamma } a F_{\mu \nu} \tilde{F}^{\mu \nu} + g_{aNN} \frac{\partial_\mu a}{2 m_N} \bar{N} \gamma^\mu \gamma_5 N
+ g_{aee} \frac{\partial_\mu a}{2 m_e} \bar{e} \gamma^\mu \gamma_5 e, \label{axioneffgluon0}
\end{equation}
where $a$, $N (\equiv p,n)$, and $e$ represent axion, nucleon (proton, neutron), and electron fields, respectively; $m_a$ is the axion mass; and $F_{\mu\nu}$ is the field strength tensor of photons.
Coupling constants $g_{a\gamma \gamma}$, $g_{aNN}$, and $g_{aee}$ are model dependent and can be written in the form
\begin{gather}
g_{app}=m_p \frac{c_{ap}}{f_a}, \quad g_{ann}=m_n \frac{c_{an}}{f_a}, \quad  g_{aee}=m_e \frac{c_{ae}}{f_a},
\label{gaffQCD}
\end{gather}
showing the dependence on the axion decay constant $f_a$ and model-dependent constants $c_{af}$ with $f=p,n,e$. The QCD axion mass $m_a$ is related to scale $f_a$ by the relation \cite{Gorghetto:2018ocs}:
\begin{equation}
m_a \simeq 5.7 \times \left( \frac{\SI{e12}{GeV}}{f_a} \right) \, \si{\micro eV},
\end{equation}
which implies a model-dependent relation between the coupling constant $g_{aff}$ and the axion mass $m_a$.

As explicit examples of the QCD axion model, there are several famous ones adopted as building blocks of others. The main difference lies in how the anomaly of the $\mathrm{U(1)_{PQ}}$ symmetry is introduced. The model of Kim-Shifman-Vainshtein-Zakharov (KSVZ) type introduces heavy quarks $\mathcal{Q}$s which transform under PQ charge chirally~\cite{PhysRevLett.43.103, SHIFMAN1980493}. The other one is the Dine-Fischler-Srednicki-Zhitnitsky (DFSZ) type model which contains two Higgs doublets responsible for the electroweak symmetry breaking and assigns the PQ charges to SM particles and to Higgs bosons~\cite{DINE1981199, Zhitnitsky:1980tq}. For the KSVZ model of axions, the model-dependent parameters are
\begin{gather}
c^\mathrm{KSVZ}_{ap}=-0.47(3), \quad c^\mathrm{KSVZ}_{an}=-0.02(3),\\
c^\mathrm{KSVZ}_{ae}=0.
\end{gather}
For the DFSZ model of axions, 
\begin{gather}
c^\mathrm{DFSZ}_{ap}= -0.182 -0.435 \sin^2 \beta \pm 0.025, \quad c^\mathrm{DFSZ}_{an}=0.160 +0.414 \sin^2\beta \pm 0.025,\\
c^\mathrm{DFSZ}_{ae}=\frac{1}{3} \sin^2 \beta,
\end{gather}
where $\beta$ is the ratio of the vacuum expectation values of two Higgs doublets.  It should be noted that the value of parameter $\beta$ is constrained by the perturbativity of the Yukawa coupling as $0.28 <\tan \beta < 140$ \cite{Chen:2013kt}. 
Recently flavorful axion models are also considered \cite{Ema:2016ops,Calibbi:2016hwq}, which also predict sizable axion-fermion couplings.

On the other hand, ALPs in general are expected to possess the same Lagrangian as shown in Eq.~\eqref{axioneffgluon0} while there is no relationship between its mass $m_a$ and the coupling $g_{aff}$.

\subsubsection{Axion-induced magnetic field} \label{subsubsec: axionfield}

We now estimate the magnitude of the axion-induced effective magnetic field acting on spins of SM particles assuming that all of the DM is composed of axions.
Recall that the axion-nucleon interaction is given by the fourth term of Eq.~\eqref{axioneffgluon0}.
In the nonrelativistic limit, the interaction reduces to the form
\begin{equation}
\mathcal{L}_{aNN}=\frac{g_{aNN}}{m_N} ( \vec{\nabla} a) \cdot  \vec{S}_N, \label{spincouplingaNN}
\end{equation}
with $\vec{S}_N$ representing the nucleon spin.
This shows us that the axion field acts as an effective magnetic field interacting with nuclear spins. Below, we estimate the amplitude of the effective magnetic field from the axion properties.

We consider the axion DM in the mass range $\SI{e-6}{eV}\lesssim m_a\lesssim \SI{e-4}{eV}$. 
We adopt the standard cold DM velocity profiles that the DM velocity $v_\mathrm{DM}$ is expected to be $\sim 10^{-3}$ with spreading $\Delta v_\mathrm{DM}$ of the same order. 
The de Broglie wavelength of axion DM,
\begin{equation}
\lambda_\mathrm{DM}=\frac{2\pi}{m_a v_\mathrm{DM}} \sim O(0.01)-O(1) \, \si{km},
\end{equation}
is much longer than the size of the magnetic material used for the experiment.
The occupation number of the axion DM is also large due to the small mass, so we treat the axion DM as a classical field that interacts coherently with nuclear spins inside the magnetic material within coherence time
 \begin{equation}
\tau_\mathrm{DM} = \frac{2 \pi }{m_a v_\mathrm{DM}^2}\sim O(0.01)-O(1) \, \si{ms}. \label{taucohDM}
\end{equation}
We parametrize the axion DM classical field as
\begin{equation}
a(\vec{x},t)=a_0 \sin (m_a t+ m v^2_\mathrm{DM}t/2- m_a \vec{v}_\mathrm{DM}\cdot \vec{x} + \delta),
\end{equation}
where $a_0$ is the oscillation amplitude with which the energy density is given by $\rho_a=\frac{1}{2}m^2_a a_0^2$, while $\delta$ is a random phase. We assume $\rho_a=\rho_\mathrm{DM}$, where $\rho_\mathrm{DM}$ is the local DM density around the Earth.  In our numerical calculation, we take 
\begin{equation}
  \rho_\mathrm{DM} = \SI[per-mode=symbol]{0.4}{\GeV\per\cm\tothe{3}}.
\end{equation}
We assume that the oscillation persists in its coherent phase for time interval $\tau_\mathrm{DM}$, each interval connected with a discrete jump of $\vec{v}_{\rm DM}$ and $\delta$. Owing to the velocity distribution, the bandwidth of the field is given by
\begin{equation}
\Delta \omega_\mathrm{DM}=m_av_\mathrm{DM}^2=\frac{2\pi}{\tau_\mathrm{DM}}, \label{DMBW}
\end{equation}
which translates to quality factor 
\begin{equation}
Q_\mathrm{DM}=\frac{m_a}{\Delta \omega_\mathrm{DM}}\sim10^6.
\end{equation}

To consider the interaction between nuclear spin $\vec{I}$ and the axion, we need to take into account the nucleon spin contribution to the nuclear spin. This is shown in detail in Appendix \ref{exotic interaction}.
Matching the nucleon spin-axion interaction to the interaction between the nuclear magnetic moment $\vec{\mu}_I$ and an arbitrary magnetic field $\vec{h}$ of the form $\mathcal{L}_\mathrm{int}=\vec{\mu}_I \cdot\vec{h}$, we can define the effective magnetic field induced by the axion, which is felt by the nuclear spin as
\begin{equation}
\vec{h}_n^\mathrm{axion}= \frac{1}{ \gamma_n}\frac{\tilde{g}_{aI}}{m_N} \sqrt{2\rho_\mathrm{DM}} \vec{v}_\mathrm{DM}   \sin \left(m_a t + \delta \right), \label{axionfieldI}
\end{equation}
where $\tilde{g}_{aI} \equiv g_{app} \sigma_p + g_{ann} \sigma_n$ with $\sigma_{p,n}$ denoting the spin contribution of proton and neutron to nuclear spin, and $\gamma_n \equiv g_I \mu_N$ is the nuclear gyromagnetic ratio defined by the nuclear $g$ factor $g_I$ and the nuclear magneton $\mu_\mathrm{N}\equiv e/2m_p$. Note that the value of $\gamma_n$ depends on the magnetic isotope, while the typical order of magnitude is $O(10^7) \, \si[inter-unit-product=\cdot]{\radian\per\second\per\tesla}$.

On the other hand, the axion also interacts with an electron through the last term of Eq.~\eqref{axioneffgluon0}.
Similarly to the case of the nuclear spin, we introduce the axion-induced effective magnetic field for the electron spin as
\begin{equation}
\vec{h}_e^\mathrm{axion}=\frac{1}{\gamma_e} \frac{g_{aee}}{m_e} \sqrt{2 \rho_\mathrm{DM}} \vec{v}_\mathrm{DM}  \sin (m_a t + \delta), \label{axionfielde}
\end{equation}
where $\gamma_e=\SI[inter-unit-product=\cdot]{1.760e11}{\radian\per\second\per\tesla}$ is the electron gyromagnetic ratio.

As a result, we obtain the amplitude of the effective magnetic field as:
\begin{align}
h^\mathrm{axion}_n= \SI{4.0e-18}{\tesla}& 
\times
\left(\frac{g_{aNN}}{10^{-10}} \right)
\left(\frac{\rho_\mathrm{DM}}{\SI{0.4}{GeV / cm^3}} \right)^{1/2}
\left( \frac{v_\mathrm{DM}/c}{10^{-3}}  \right) \nonumber\\
\times
&
\left( \frac{\SI[inter-unit-product=\cdot]{5e7}{\radian\per\second\per\tesla}}{\gamma_n }  \right)
\left( \frac{\sigma_p+\sigma_n}{0.5}  \right),\\
h^\mathrm{axion}_e=\SI{4.2e-18}{\tesla} &  
\times
\left(\frac{g_{aee}}{10^{-10}} \right)
\left(\frac{\rho_\mathrm{DM}}{\SI{0.4}{GeV / cm^3}} \right)^{1/2}
\left( \frac{v_\mathrm{DM}/c}{10^{-3}}  \right),
\end{align}
for $g_{app}=g_{ann}=g_{aNN}$.

\subsection {DM dark photon}

The dark photon is introduced as a new gauge boson of a new dark U(1) gauge symmetry in addition to the SM symmetry. Even when SM particles do not have dark U(1) charges, kinetic mixing between the dark photon and the SM photon becomes a portal for the SM particles to interact with dark photons.
The massive dark photon that is light enough and weakly interacts with SM particles is a viable candidate of DM from the viewpoint of the up-to-present observable constraints. See Refs.~\cite{Graham:2015rva,Ema:2019yrd,Agrawal:2018vin,Co:2018lka,Bastero-Gil:2018uel,Kitajima:2023pby,Dror:2018pdh,Nakayama:2021avl,Long:2019lwl,Kitajima:2022lre,Kitajima:2023fun} for production mechanisms of dark photon DM. 
The general review of the dark photon is given in Ref.~\cite{Fabbrichesi:2020wbt}, while the careful treatment of dark photon polarization and its effect on actual experiments are discussed in Ref.~\cite{PhysRevD.104.095029}. 

\subsubsection{Dark photon model}

We introduce the dark photon, a massive vector field (denoted as $A_\mu'$) which couples to the SM fields only through the kinetic mixing with the ordinary photon.  The Lagrangian includes the following terms:
\begin{align}
  \mathcal{L} \ni \frac{1}{2} m_{\gamma'} A'_\mu A'^{\mu} - \frac{1}{4}(F'_{\mu \nu} )^2-\frac{1}{4}( \mathcal{F}_{\mu \nu} )^2 + \frac{\epsilon}{2} F_{\mu \nu}' \mathcal{F}^{\mu \nu}+e J_\mu^\mathrm{EM}\mathcal{A}^\mu, \label{startdphoton}
\end{align}
where $\mathcal{A}_\mu,A'_\mu$ are the gauge fields associated with the electromagnetic U(1) symmetry and dark U(1) symmetry, respectively, $\mathcal{F}_{\mu \nu}\equiv\partial_\mu \mathcal{A}_\nu-\partial_\nu \mathcal{A}_\mu$ and $F'_{\mu \nu}\equiv\partial_\mu A'_\nu-\partial_\nu A'_\mu$ are the corresponding field strength tensors, while $J^\mathrm{EM}_\mu$ is the electromagnetic current of the SM particles.  In addition, $\epsilon$ is the kinetic mixing parameter that should be much smaller than unity.  In this basis, there is no direct coupling between the dark photon and the SM fermions.  

We can also work in the basis in which the kinetic mixing of dark and ordinary photons vanishes.  There is a mass eigenstate with zero mass, which corresponds to the ordinary photon as
\begin{align}
  A_\mu \equiv \mathcal{A}_\mu - \epsilon A'_\mu.
\end{align}
The Lagrangian can be expressed as
\begin{equation}
  \mathcal{L} \ni \frac{1}{2} m^2_{\gamma'} {A'}_\mu {A'}^\mu  - \frac{1}{4} ( F'_{\mu\nu} )^2 -\frac{1}{4} ( F_{\mu\nu} )^2 +e J^\mathrm{EM}_\mu (A^\mu + \epsilon {A'}^\mu). \label{masseigenDP}
\end{equation}
Here and hereafter, we neglect terms of $O(\epsilon^2)$ because $\epsilon\ll 1$. We can see that, in this basis, the dark photon $A_\mu'$ obeys the Proca Lagrangian with $\epsilon  e J^\mathrm{EM}_\mu{A'}^\mu$ as a source term. By taking the Lorenz gauge for the SM photon, the equations of motion are given by
\begin{gather}
\square A_\mu =J^\mathrm{EM}_\mu, \quad \partial_\mu A^\mu=0,\\
(\square  +m^2_{\gamma '}) {A'}_\mu=\epsilon J_\mu^\mathrm{EM}, \quad \partial_\mu {A'}^\mu=0, \label{MaxwellDP}
\end{gather}
with $\square=\partial_t^2-\vec{\nabla}^2$. 

\subsubsection{Dark-photon-induced magnetic field} \label{subsubsec: DPfield}

Let us focus on the interaction between the dark photon and the SM particles in the basis of mass eigenstates:
\begin{equation}
\mathcal{L}_{\mathrm{int}} = e\epsilon J^\mathrm{EM}_{\mu} A'^{\mu}.
\end{equation}
We assume the following form of the vector potential of dark photon:
\begin{equation}
  \vec{A}' (t)= \vec{A}_0' \sin (m_{\gamma'} t+m_{\gamma'} v_\mathrm{DM}^2t/2 - m_{\gamma '}\vec{v}_\mathrm{DM}\cdot \vec{x} + \delta).
\end{equation}
The spread of amplitude in frequency space is given also by Eq.~\eqref{DMBW} with mass replaced by $m_{\gamma'}$.
Then, assuming that the whole DM abundance is explained by the dark photon, we obtain $\rho_\mathrm{DM}=(1/2) m^2_{\gamma'} \vec{A}_0'^2$. Similarly to the ordinary relation between the vector potential and the magnetic field, the effective magnetic field induced by the dark photon is expressed as
\begin{equation}
\vec{h}^{\gamma '} (t)=\epsilon \vec{\nabla} \times \vec{A} '(t)
= \epsilon v_\mathrm{DM}\sqrt{2 \rho_\mathrm{DM} } \sin ( m_{\gamma '} t + \delta)  \hat{v}_\mathrm{DM} \times \hat{A}',
\label{Beff_DP}
\end{equation}
where the hat symbol represents the unit vector pointing in the direction of the original vector. Numerically, we obtain the amplitude of the effective magnetic field as
\begin{equation}
h^{\gamma'}=\SI{9e-19}{\tesla} \times
\left(\frac{\epsilon}{10^{-10}} \right)
\left(\frac{\rho_\mathrm{DM}}{\SI{0.4}{GeV / cm^3}} \right)^{1/2}
\left( \frac{v_\mathrm{DM}/c}{10^{-3}}  \right)\left( \frac{\sin \varphi}{\sqrt{0.5}} \right),
\end{equation}
where $\sin \varphi \equiv \hat{v}_\mathrm{DM} \times \hat{A}'$ is the angle between the DM velocity and the vector potential $\vec{A}'$. 

Note that the effective magnetic/electric field induced by the dark photon is significantly affected by the conductor shield around the experimental apparatus~\cite{Chaudhuri:2014dla}.
As a rough estimate, for the typical length scale of the shield $L_{\rm shield}$, the effective magnetic field (\ref{Beff_DP}) should be multiplied by an additional factor of $(m_{\gamma'} L_{\rm shield} v_{\rm DM})^{-1}$ for $1< m_{\gamma'} L_{\rm shield} < v^{-1}_{\rm DM}$ and by $m_{\gamma'} L_{\rm shield}/v_{\rm DM}$ for $m_{\gamma'} L_{\rm shield} < 1$, which can either enhance or suppress the signal depending on $m_{\gamma'} L_{\rm shield}$. Note that $v_\mathrm{DM}\sim 10^{-3}. $ Numerically, the signal is suppressed when
\begin{align}
    L_{\rm shield} \lesssim \SI{20}{\centi\meter} \left(
        \frac{\SI{e-6}{eV}}{m_{\gamma'}}
    \right),
\end{align}
which could be avoided by using a reasonably large magnetic shield. On the other hand, thanks to the absence of a suppression factor $v_\mathrm{DM}$ in the shielding effect, the effective magnetic field (\ref{Beff_DP}) can be enhanced up to a factor of $10^3$.
In this paper, we do not discuss a real experimental apparatus including the shield and simply use Eq.~\eqref{Beff_DP} as a conservative estimate.

%%%%%%%%%%%%%%%%%%%%%%%%%%%%%%%%%%%%%%%%%%%%%%%%%%%%%%%%%%%%%%%%%%%%%%%%%%%%%%%%
\section{Magnetization dynamics in \ce{MnCO3}} \label{sec: nuclear MD}
\setcounter{equation}{0}
%%%%%%%%%%%%%%%%%%%%%%%%%%%%%%%%%%%%%%%%%%%%%%%%%%%%%%%%%%%%%%%%%%%%%%%%%%%%%%%%

The DM-induced magnetic field discussed in the previous section can cause magnetic resonance in the material and give an observable oscillating magnetization signal.
We propose to use materials with strong hyperfine interaction that allows spin transfer between nuclear and electron spins for DM detection.

The hyperfine interaction between electron and nuclear spins originates from the magnetic dipole interaction between them. The following properties are realized due to this interaction. Nuclear spins are highly polarized by the large effective magnetic field due to electron spin even in the low external static magnetic field. It provides large magnetic signals in the high-frequency region not easily accessible by other approaches using ordinary nuclear magnetization. Through the exchange interaction with electron spins, nuclear spins effectively obtain an exchange interaction among themselves called the Suhl-Nakamura interaction~\cite{PhysRev.109.606,10.1143/PTP.20.542}, to realize the nuclear spin wave modes (or the nuclear magnons in the quantum picture). In this mode dominated by the precession of nuclear spins, the electron spins with a larger gyromagnetic ratio also contribute to the total observable magnetization enhancing the overall magnetization signal compared with nuclear magnetization alone. On the other hand, there is also an electron-spin-dominated mode mixed with nuclear spins.
They are sensitive to both the DM-electron and the DM-nucleon interactions. As a concrete example, we focus on a canted antiferromagnetic material \ce{MnCO3}.

We introduce the property of \ce{MnCO3} in Sec.~\ref{subsec: introMn}. Then, we illustrate the magnetic dynamics of \ce{MnCO3} in Sec.~\ref{subsec: MDMnCO3} and its response to the DM coupled only to either the nuclear spin or the electron spin in Sec.~\ref{subsec: response}. 

The calculation in the following sections is done in the SI unit. However, we simply omit the vacuum permeability factor $\mu_0$ in the formulas for convenience.\footnote{
  The formulas are then the same as that in the centimetre-gram-second unit except for the definition of susceptibility.
}
The factor $\mu_0$ is restored in Sec.~\ref{sec: sensitivity}.

\subsection{Introduction to \ce{MnCO3}} \label{subsec: introMn}

\ce{MnCO3} is a canted antiferromagnetic material. It has large nuclear spins (with $I=5/2$) associated with $100 \%$ magnetic isotope of $^{55}\mathrm{Mn}$, which couple to localized electron spins (with $S=5/2$) through strong hyperfine interactions.
The electron ground state configuration of $\ce{^55Mn}$ is $\mathrm{[Ar] \ 4s^2 \ 3d^5}$ with multiplet $\mathrm{^6S_{5/2}}$.

\begin{figure}[t]
  \centering
  \includegraphics[width=0.3 \linewidth]{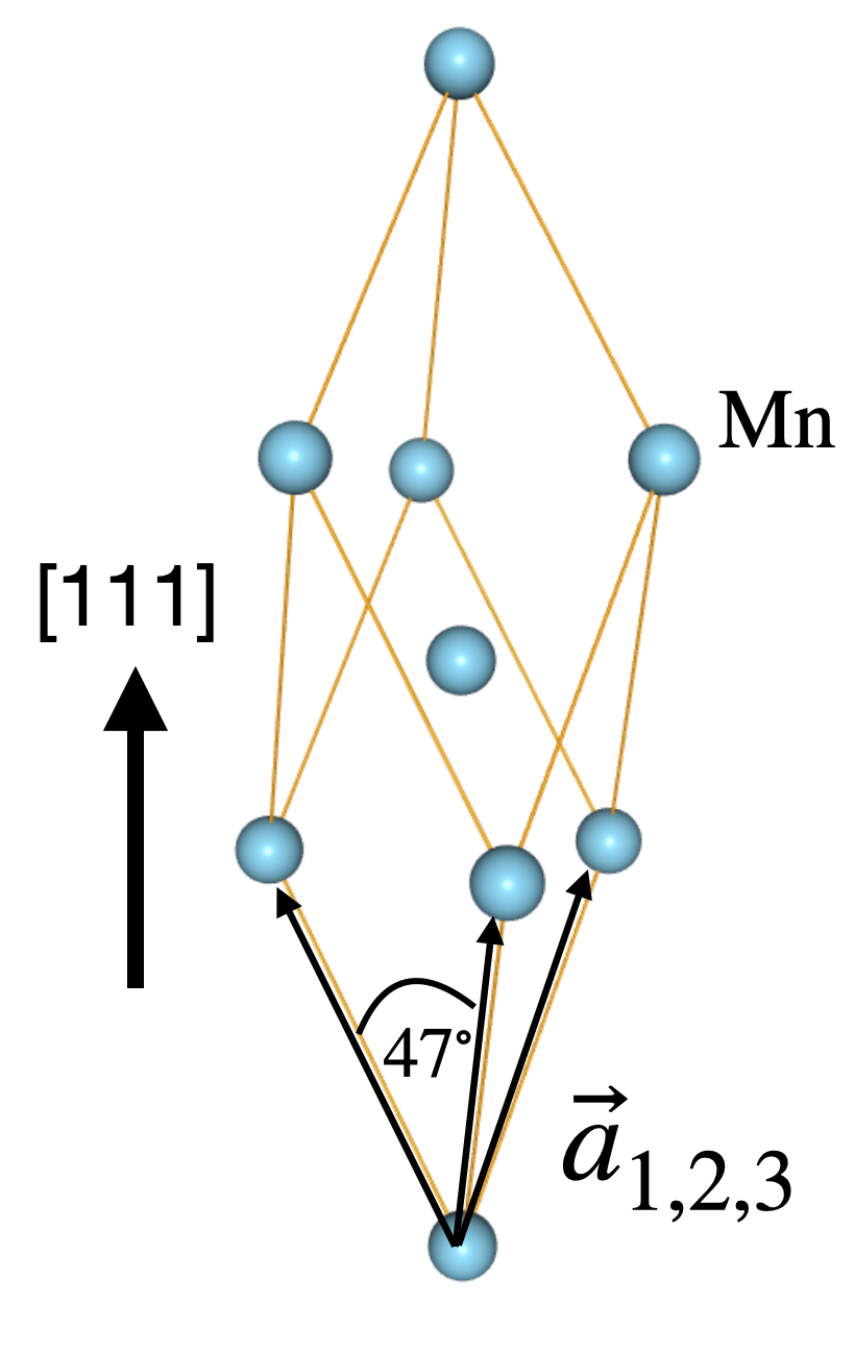}
  \caption{Rhombohedral unit cell of \ce{MnCO3}. We show here for simplicity only Mn ions. Translation vector $\vec{a}_{1,2,3}$  with length $a_\mathrm{rh}=\SI{5.84}{\angstrom}$ connects Mn of the same sublattice. Note that the Mn ion in the middle and those connected to it by $\vec{a}_{1,2,3}$ belong to different sublattices. In the antiferromagnetic phase, spins localized at \ce{Mn}s of one sublattice align in the same direction in the (111) plane perpendicular to the [111] direction, and in the opposite direction to spins of the other sublattice. See also Refs.~\cite{Shiomi:2019aa,PhysRevB.86.224407,Liang:2020aa} for crystallography of \ce{MnCO3}.}
  \label{fig: crysyalMnCO}
\end{figure}

The \ce{MnCO3} lattice structure can be represented by the rhombohedral unit cell. The parameters for this representation are interval length $a_\mathrm{rh}=\SI{5.84}{\angstrom}$ with angle $\alpha=\SI{47}{\degree}$. The unit cell including only \ce{Mn} ions is shown in Fig.~\ref{fig: crysyalMnCO}. See also Refs.~\cite{Shiomi:2019aa,PhysRevB.86.224407,Liang:2020aa} for crystallography of \ce{MnCO3}. In the following, lattice directions such as [111] or $[10\bar{1}]$ are to be understood as of the rhombohedral representation; the [$nml$] direction means the direction of $n\vec{a}_1+m\vec{a}_2+l\vec{a}_3$ where $\vec{a}_{1,2,3}$ is the translation vector connecting lattice sites of the same sublattice and bars on numbers (e.g., that in $[10\bar{1}]$) indicate minus signs.

Electron spins, which are localized at each Mn ion, lie in the basal plane ((111) plane) perpendicular to the [111] direction showing hard-axis anisotropy. Importantly, owing to an exchange interaction between electron spins, they align antiparallel to the one in the nearest site, forming two sublattices in an antiferromagnetic order below the N\`{e}el temperature ($T^*\approx \SI{35}{\K}$ \cite{Shiomi:2019aa}). Besides, the Dzyaloshinskii–Moriya interaction \cite{DZYALOSHINSKY1958241,PhysRev.120.91,PhysRev.117.635} slightly bends spins of two sublattices producing the weak-ferromagnetic properties of \ce{MnCO3} in the basal plane.

In general, the interaction between nuclear spin $\vec{I}$ and electron spin $\vec{s}$ originates from the dipolar interaction between them. It is given as follows \cite{jiseikei}: 
\begin{equation}
H_\mathrm{hy}
=\gamma_e \gamma_n \hbar^2
\left[ \frac{(\vec{l}-\vec{s}) \cdot \vec{I}}{r^3} +3 \frac{1}{r^5}(\vec{s}\cdot \vec{r}) (\vec{I} \cdot \vec{r})+\frac{8}{3} (\vec{s}\cdot \vec{I})\delta (\vec{r})
\right],
\label{hyperdef}
\end{equation}
where $\gamma_e$ and $\gamma_n$ are the gyromagnetic ratios of the electron and nucleus, respectively; $\vec{l}$ is the orbital angular momentum of the electron and $\vec{r}$ is the position vector of the electron with the origin $\vec{r}=0$ being the nucleus position.
In the \ce{MnCO3} case, the unpaired 3d electrons have Coulomb interaction with the 2s electrons of the opposite spin more efficiently than the 2s electrons of the same spin.
Thus the spin-polarization occurs at the core and is proportional to the magnitude of 3d electron spins.
The last term of Eq.~\eqref{hyperdef} shows the interaction between an electron-spin polarized core and nuclear spins, which indicates an effective but strong interaction between 3d electron spins and nuclear spins of the form $H_\mathrm{hy}\propto \vec{I}\cdot\vec{S}$ where $\vec{S}$ is the total electron spin associated with each Mn atom. 

In \ce{MnCO3}, the nuclear spin is sensitive to a hyperfine field and becomes highly polarized pointing to the direction correlated to the electron spin. Besides, nuclear magnetic resonance occurs at a very high frequency $\sim \SI{500}{\MHz}$ compared with that of typical nuclear spin precession. Under the presence of the strong hyperfine interaction and the exchange interaction of electron spins, there exists an effective exchange interaction between nuclear spins ensuring the existence of nuclear spin wave. On the other hand, the system can also be viewed as a hybrid system of nuclear and electron spins. 
Detailed dynamics of such a system and its response to the DM will be discussed in the following subsections.

Materials with $\ce{^55Mn}$ ions are frequently used in nuclear-spin-wave experiments. The reasons are the following:
\begin{itemize}
\item \ce{^55Mn} magnetic isotopes have large localized nuclear and electron spins ($I=5/2,S=5/2$).

\item The electron spin wave modes have low eigenfrequency close to that of nuclear modes, leading to a large mixing between nuclear and electron spins.\footnote{Note that the 3d electrons of \ce{^55Mn} are all unpaired and fill each 3d shell, and hence the total orbital angular momentum of the electron spin associated with each atom is zero. This only leads to a weak magnetocrystalline anisotropy, which causes a small gap of the electron-spin system.}
\end{itemize}

Other example materials with $\ce{^55Mn}$ ions besides the canted antiferromagnet \ce{MnCO3}\footnote{
  \ce{MnCO3} has a weak-ferromagnetic property due to the Dzyaloshinkii--Moriya interaction, which forces the ground state spins to be canted even in the absence of an applied static magnetic field. Therefore, there is no need to worry about the phase transition from the antiferromagnetic phase to the spin-flop phase as in usual antiferromagnetic materials.
} \cite{doi:10.1063/1.1713505, PhysRevLett.114.226402, PhysRevB.97.024425, Shiomi:2019aa,Kikkawa:2021aa}, include the hexagonal antiferromagnet \ce{CsMnF3} \cite{PhysRev.132.144, PhysRev.143.361, PhysRev.156.370} with biaxial anisotropy, the antiferromagnet \ce{RbMnF3} \cite{PhysRev.184.574, PhysRevB.4.1572} and \ce{KMnF3} \cite{doi:10.1063/1.1729366} with cubic crystalline anisotropy, the antiferromagnet \ce{MnF2} \cite{PhysRevLett.37.533} with uniaxial anisotropy, and the ferrimagnet \ce{MnFe2O4} \cite{PhysRev.135.A661}. Other materials are also studied in the context of nuclear spin waves: \ce{CoCO3} \cite{borovik1984spin} since the \ce{^59Co} isotope is $100\%$ abundance, and \ce{FeBO3} \cite{borovik1984spin} since its electron magnetic precession has low eigenfrequencies.

Recently, Shiomi et al. \cite{Shiomi:2019aa} and Kikkawa et al. \cite{Kikkawa:2021aa} reported experiments combining nuclear-spin waves with spintronics in \ce{MnCO3}, showing the nuclear-spin pumping effect and nuclear-spin Seebeck effect of the system and establishing a new area of spin technology, nuclear spintronics.

The spin transfer between electron and nuclear spins allows us to deal with nuclear spins more easily through more-accessible electron spins. Here, for DM detection, it provides a unique probe for nucleon-DM and electron-DM interactions of the favorable frequency range, with a ``nature" tool (electron spins and their magnetization) supporting signal readout. At the same time, the precession of electron spins, mixed with nuclear spins, is also sensitive to DM (specifically to electron-DM coupling) and we then include it in the discussion.
Next, we move to the details of the magnetic system of \ce{MnCO3}.

\subsection{Magnetic system of \ce{MnCO3}} \label{subsec: MDMnCO3}

We discuss the (macroscopic) magnetization dynamics of \ce{MnCO3} within a classical theory, with some details shown in Appendix \ref{appendix: classic}.  We also show that the results are consistent with those derived in the quantum magnon picture in Appendix \ref{appendix: magnonpic}. 

We define $\vec{M}_{1}$ and $\vec{M}_2$ as electron magnetization vectors and $\vec{m}_{1}$ and $\vec{m}_2$ as nuclear magnetization vectors, where the subscripts indicate the sublattices they belong to.
Magnetization is defined as the magnetic dipole moment per unit volume, and thus\footnote{The gyromagnetic ratio of the electron is negative. However, we define $\gamma_e$ and $\gamma_n$ to be positive and hence the additional minus sign appears for electron magnetization.}
\begin{gather}
\vec{M}_1=-\gamma_e \hbar \frac{\sum^{\mathrm{lattice1}}_i \vec{S}_i}{V}, \quad 
\vec{M}_2=-\gamma_e \hbar \frac{\sum^{\mathrm{lattice2}}_j \vec{S}_j}{V}, \label{MSrelation} \\
\vec{m}_1=\gamma_n \hbar \frac{\sum^{\mathrm{lattice1}}_i \vec{I}_i}{V}, \quad 
\vec{m}_2=\gamma_n \hbar \frac{\sum^{\mathrm{lattice2}}_j \vec{I}_j}{V}, \label{mIrelation}
\end{gather}
where $\vec{S}_{i}$ and  $\vec{I}_i$ are the electron-spin operator and nuclear-spin operator at the spin site $i$, respectively, and $V$ is the volume of the sample. The summations of spins run over sublattice 1 and 2 for the corresponding magnetization vector.
At the ground state, the magnitude of magnetization of each sublattice is assumed to be $M_0$ and $m_0$ for electron magnetization and nuclear magnetization, respectively. They can be expressed by 
\begin{equation}
M_0=\gamma_e \hbar S\frac{\rho_s}{2}, \quad m_0 =\gamma_n \hbar \langle I \rangle \frac{\rho_s}{2}, \label{M0m0def}
\end{equation}
where $\rho_s\equiv N_\mathrm{total}/V$ is the number density of \ce{Mn} ions with $N_\mathrm{total}$ denoting the total number of \ce{Mn} (which is equal to the number of spin sites); $S=5/2$ is the total value of electron spins localized at each Mn ion, and $\langle I \rangle$ is the thermal average of nuclear spins localized at each Mn ion, which generally takes a smaller value than $I=5/2$ as we will argue later.

In the antiferromagnetic phase, the potential per unit volume for magnetizations in \ce{MnCO3} is given by (see, e.g., Refs \cite{doi:10.1143/JPSJ.15.2251, PhysRev.136.A218,turov1973nuclear, PhysRevB.97.024425})
\begin{align}
U=&\frac{H_E}{M_0} \vec{M}_1 \cdot \vec{M}_2 + \frac{H_D}{M_0} \{ M_1^x M_2^z- M_1^z M_2^x \} \nonumber\\
&+ \frac{H_{K}}{2M_0} \{ [ M_1^y ]^2+ [ M_2^y ]^2  \} -\frac{H_{K'}}{2M_0} \{ [ M_1^z ]^2+ [ M_2^z ]^2  \} \nonumber\\
& -(\vec{M}_1+\vec{M}_2) \cdot (\vec{H}+\vec{h}_e(t))  - (\vec{m}_1 +\vec{m}_2) \cdot( \vec{H}+\vec{h}_n(t)) \nonumber\\
& -A_\mathrm{hy} \vec{M}_1 \cdot \vec{m}_1  -A_\mathrm{hy} \vec{M}_2 \cdot \vec{m}_2.  \label{potentialpervolume}
\end{align}
The subscripts $x,y,z$ attached to the magnetization parameters refer to the component of the vector which points in those directions. The potential contains the (1) antiferromagnetic exchange interaction, (2) Dzyaloshinskii-Moriya interaction, (3) hard-axis anisotropy, (4) in-plane uniaxial anisotropy, (5) Zeeman effect, and (6) hyperfine interaction.
The coefficients $H_E, H_D, H_K$, $H_{K'}$ and $A_\mathrm{hy}$ correspond to constants associated with the exchange interaction, Dzyaloshinskii-Moriya interaction, hard-axis anisotropic effect, in-plane anisotropic effect,  and hyperfine interaction, respectively.
The coordinate setup is assumed such that the $y$ direction is the hard axis corresponding to the [111] lattice direction.
External magnetic fields are assumed to include an applied static field $\vec{H}$ pointing in the $[10\bar{1}]$ lattice direction (which we call $x$):
\begin{equation}
\vec{H}=H_0\vec{e}_x,
\end{equation}
with $\vec{e}_i$ being the unit vector pointing in the $i$ direction,
an oscillating field $\vec{h}_e$ coupling to electron spins, and an oscillating field $\vec{h}_n$ coupling to nuclear spins. The latter two account for the exotic fields originating from DM.
For convenience of later discussion, we define effective fields
\begin{equation}
H_a\equiv A_\mathrm{hy} m_0, \quad H_n\equiv A_\mathrm{hy}M_0,
\end{equation} 
which come from hyperfine interactions and are felt by electron and nuclear spins in their ground state, respectively.

\begin{figure}[t]
  \centering
  \includegraphics[width=0.7 \linewidth]{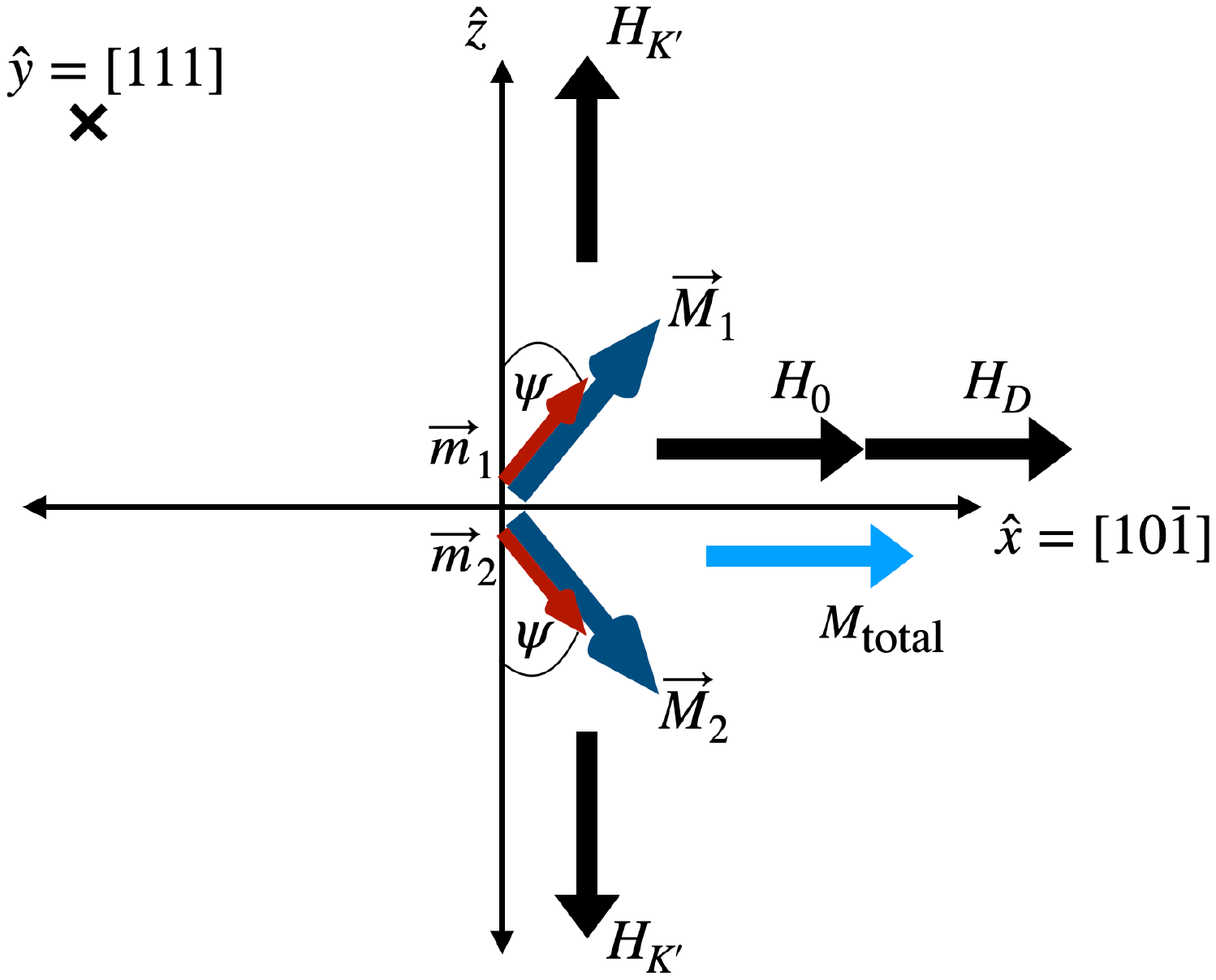}
  \caption{The ground state of magnetization in \ce{MnCO3} under the Hamiltonian given by Eq.~\eqref{potentialpervolume}. The red and blue arrows represent the nuclear and electron magnetizations of each sublattice. The light blue arrow shows the total magnetization. The black (thick) arrows represent some of the effective fields present in the system. The tilted angle $\psi$ is given by Eq.~\eqref{sinpsi}.}
  \label{fig: groundstateM}
\end{figure}

\begin{figure}[t]
  \centering
  \includegraphics[width=0.5 \linewidth]{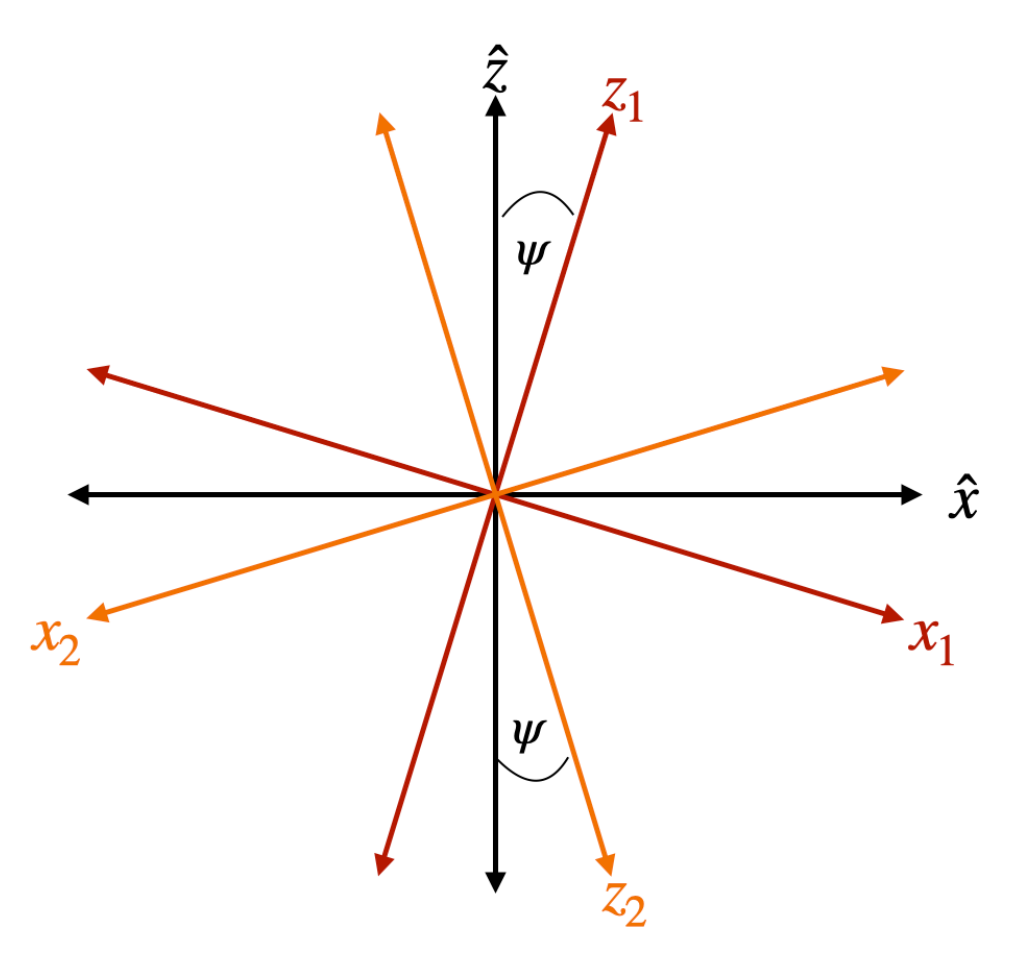}
  \caption{Coordinate transformation according to the tilt angle of the spin configuration of the ground state. The red and orange coordinate systems are those of sublattices 1 and 2, respectively. See also Eqs.~\eqref{coordinate trans} and \eqref{coordinate trans nuclear} for the explicit representation of the transformation applied to electron and nuclear magnetizations.}
  \label{fig: coortransf}
\end{figure}

\begin{figure}[t]
  \centering
  \includegraphics[width=0.6 \linewidth]{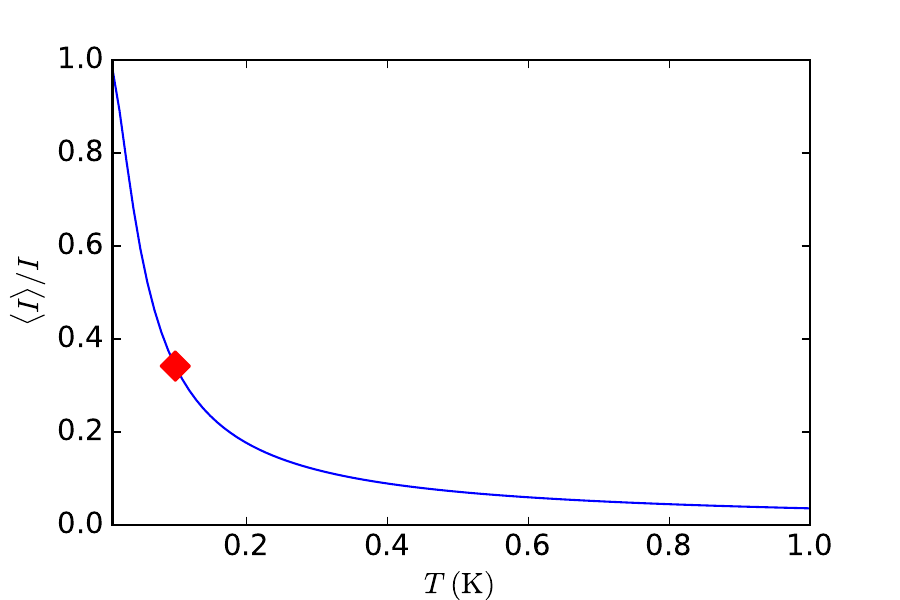}
  \caption{The nuclear spin polarization ratio $\langle I \rangle/I$ as a function of temperature.  For $T=0.1 \ \mathrm{K}$, the alignment of nuclear spins is $\langle I \rangle/I \simeq 0.4$. Note that, in the temperature range shown in the plot, electron spins align in two sublattices in the antiferromagnetic phase with $\langle S \rangle \approx S=5/2$. }
  \label{fig: BF}
\end{figure}

In the ground state of this system, there are two sublattices of spins pointing in almost anti-parallel directions along the $z$ axis and in the basal plane ($xz$ plane) due to the in-plane and hard-axis anisotropic effects, respectively.
The Dyaloshinskii-Moriya interaction and an applied magnetic field in the $x$ direction tilt the spins of two sublattices; the tilt angle is denoted as $\psi$.
Then, the magnetic system shows effective ferromagnetism in the basal plane. The schematic picture of the ground state and effective fields are shown in Fig.~\ref{fig: groundstateM}.
For convenience, we use magnetization parameters based on the coordinate systems tilted by angles $\psi$, $\pi-\psi$. For electron magnetization, we apply
\begin{subequations} \label{coordinate trans}
\begin{gather} 
\begin{pmatrix}
M_1^x \\
M_1^y\\
M_1^z
\end{pmatrix}
=
\begin{pmatrix}
\cos \psi & 0 & \sin \psi \\
0 & 1 & 0 \\
-\sin \psi & 0 & \cos \psi
\end{pmatrix}
\begin{pmatrix}
M_1^{x_1} \\
M_1^{y_1}\\
M_1^{z_1}
\end{pmatrix}
, \\
\begin{pmatrix}
M_2^x \\
M_2^y\\
M_2^z
\end{pmatrix}
=
\begin{pmatrix}
-\cos \psi & 0 & \sin \psi \\
0 & 1 & 0 \\
-\sin \psi & 0 & -\cos \psi
\end{pmatrix}
\begin{pmatrix}
M_2^{x_2} \\
M_2^{y_2}\\
M_2^{z_2}
\end{pmatrix},
\end{gather}
\end{subequations}
and similarly, for nuclear magnetization, we apply
\begin{subequations}  \label{coordinate trans nuclear}
\begin{gather}
\begin{pmatrix}
m_1^x \\
m_1^y\\
m_1^z
\end{pmatrix}
=
\begin{pmatrix}
\cos \psi & 0 & \sin \psi \\
0 & 1 & 0 \\
-\sin \psi & 0 & \cos \psi
\end{pmatrix}
\begin{pmatrix}
m_1^{x_1} \\
m_1^{y_1}\\
m_1^{z_1}
\end{pmatrix}
, \\
\begin{pmatrix}
m_2^x \\
m_2^y\\
m_2^z
\end{pmatrix}
=
\begin{pmatrix}
-\cos \psi & 0 & \sin \psi \\
0 & 1 & 0 \\
-\sin \psi & 0 & -\cos \psi
\end{pmatrix}
\begin{pmatrix}
m_2^{x_2} \\
m_2^{y_2}\\
m_2^{z_2}
\end{pmatrix}.
\end{gather}
\end{subequations}
In these tilted frames with
\begin{equation}
\sin\psi=\frac{H_0+H_D}{2H_E+H_{K'}},  \label{sinpsi}
\end{equation}
the ground state expectation values are given by $M_{1}^{x_1,y_1}=0, M_{2}^{x_2,y_2}=0$ with $M_1^{z_1}=M_2^{z_2}=M_0$ and $m_{1}^{x_1,y_1}=0,m_2^{x_2,y_2}=0$ with $m_1^{z_1}=m_2^{z_2}=m_0$.
The coordinate system adopted in the transformation is shown in Fig.~\ref{fig: coortransf}.
The polarization factor $\langle I \rangle/I$
is expressed by the thermal average in the presence of magnetic field $H_n$ exerted on the nuclear spins:
\begin{equation}
  \frac{\langle I\rangle}{I}= B_{5/2}\left( \frac{5}{2} \frac{\gamma_n \hbar H_n}{k T} \right), \label{Iavg}
\end{equation}
with
\begin{equation}
B_J (x)\equiv \frac{2J+1}{2J} \coth \left( \frac{ 2J+1 }{2J} x\right)- \frac{1}{2 J} \coth \frac{x}{2J}.
\end{equation}
We plot it in Fig.~\ref{fig: BF}, which clearly shows a paramagnetic property of nuclear spins.
However, owing to the large magnetic field $H_n \sim \SI{60}{\tesla}$ through the hyperfine interaction, the polarization is naturally large without any other external field.
This is one benefit of the strong hyperfine interaction.
For concrete estimation, we adopt the temperature of the sample and of nuclear spins to be $T=\SI{0.1}{\K}$.

With this coordinate, the hyperfine interaction (the last line of (\ref{potentialpervolume})) can be written in the form
\begin{equation}
U_\mathrm{hy}= U_{\parallel} +U_{\mathrm{mix}},
\end{equation}
where
\begin{gather}
U_{\parallel} =-A_\mathrm{hy }\left(  M^{z_1}_1 m_1^{z_1}  +  M_2^{z_2} m_2^{z_2} \right),\\
U_{\mathrm{mix}}=-A_\mathrm{hy} \left(  M_1^{x_1} m_1^{x_1} + M_1^{y_1} m_1^{y_1} +  M_2^{x_2} m_2^{x_2} + M_2^{y_2} m_2^{y_2}\right).
\end{gather}
The term $U_{\parallel}$ represents the hyperfine interaction in the direction of spin alignment in the ground state, which makes the Larmor frequency of both nuclear and electron spin precessions higher.
On the other hand, the term $U_{\mathrm{mix}}$ causes the mixing of the nuclear- and electron-spin precessions. 

In the presence of an effective oscillating magnetic field induced by DM, the magnetic resonance of this system may occur. We derive the equation of motion for magnetization similarly as done in Refs.~\cite{turov1973nuclear, PhysRevB.97.024425}.  Under the potential $U(\vec{M}_{1,2},\vec{m}_{1,2})$, the magnetizations $\vec{M}_{1,2}$ and $\vec{m}_{1,2}$ feel the effective magnetic fields determined by
\begin{equation}
\vec{H}^{M_{1,2}}_\mathrm{eff} \equiv -\frac{\partial U}{\partial \vec{M}_{1,2}}, \quad \vec{H}^{m_{1,2}}_\mathrm{eff} \equiv -\frac{\partial U}{\partial \vec{m}_{1,2}},
\end{equation}
from which they receive torque and precess according to the equations of motion given by
\begin{equation}
\frac{d\vec{M}_{1,2}}{dt}= - \gamma_e \vec{M}_{1,2} \times \vec{H}^{M_{1,2}}_\mathrm{eff}, \quad \frac{d\vec{m}_{1,2}}{dt}= \gamma_n \vec{m}_{1,2} \times \vec{H}^{m_{1,2}}_\mathrm{eff}. \label{EOMgeneral}
\end{equation}
We can linearize these equations by focusing on the small perturbations around the ground state.
We consider the precession of magnetization with a small deflect angle such that $M_1^{x_1,y_1} \ll M_{1}^{z_1}$ and   $M_2^{x_2,y_2} \ll M_{2}^{z_2}$, and approximate $M^{z_1}_1, M^{z_2}_2$ to be $M_0$.
Similarly, we consider a situation where the precession angle of the nuclear magnetization vector from the ground state is small such that $m_1^{x_1,y_1} \ll m_{1}^{z_1}$ and   $m_2^{x_2,y_2} \ll m_{2}^{z_2}$, and approximate $m^{z_1}_1, m^{z_2}_2$ to be $m_0$.

Defining ``plus'' and ``minus'' modes as
\begin{equation}
M^\alpha_{\pm} \equiv  M_1^{\alpha_1}\pm M_2^{\alpha_2}, \quad m^\alpha_{\pm} \equiv m_1^{\alpha_1}\pm m_2^{\alpha_2}, \label{defMnpm}
\end{equation}
with $\alpha=x,y,z$, it turns out that the equation of motions for the magnetization vector are decoupled for $+$ and $-$ combinations, which are usually called the in-phase and out-phase modes, respectively. 
The total magnetization vector can be written in terms of in-phase and out-phase modes as:
\begin{subequations} \label{totalpm}
\begin{gather}
\vec{M}_1+\vec{M}_2 = 
(M_-^x \cos\psi + 2 M_0 \sin \psi ) \vec{e}_x
+ M_+^y \vec{e}_y
- M_+^x \sin \psi \vec{e}_z,\\
\vec{m}_1+\vec{m}_2 = 
(m_-^x \cos\psi + 2 m_0 \sin \psi ) \vec{e}_x
+ m_+^y \vec{e}_y
- m_+^x \sin \psi \vec{e}_z. 
\end{gather}
\end{subequations}
For electron magnetization, we obtain the equation of motion for the in-phase mode as
\begin{subequations} \label{M-inphase-EOM}
\begin{align}
-\frac{1}{\gamma_e} \frac{d M^x_+}{dt} =& \left\{ 2H_E +H_K+H_{K'}+ H_a +\frac{H_D(H_0+H_D)}{2H_E} \right\}M_+^y \nonumber  \\
&-H_n m_+^y -2M_0 h_e^y\;, \\
-\frac{1}{\gamma_e} \frac{d M^y_{+}}{dt} =& - \left\{ \frac{2H_E (H_{K'}+ H_a) + H_0 (H_0 + H_D)}{2H_E } \right\} M_+^x \nonumber\\
&- 2M_0 h_e^z \sin\psi+H_n m_+^x\;,
\end{align}
\end{subequations}
and for the out-phase mode as
\begin{subequations}  \label{M-outphase-EOM}
\begin{align}
-\frac{1}{\gamma_e} \frac{d M^x_{-}}{dt} =& \left\{ H_K+H_{K'}+H_a+\frac{H_D(H_0+H_D)}{2H_E} \right\} M_-^y-H_n m_-^y\;, \\
-\frac{1}{\gamma_e} \frac{d M^y_{-}}{dt} =& -M^x_- \left\{2H_E+H_{K'}+H_a+\frac{(2H_D-H_0)(H_D+H_0)}{2H_E}\right\}  +H_n m^x_-\nonumber\\
&+\left(2M_0 -\frac{M_0 (H_0+H_D)^2}{4H^2_E} \right)h_e^x\;,
\end{align}
\end{subequations}
where the magnetic field $h^{x,y,z}_e$ is an exotic field interacting only with electron spins and oscillates in the $x$, $y$, or $z$ direction, respectively. 
For nuclear magnetization, the equations of motion for the in-phase mode are
\begin{subequations}  \label{m-inphase-EOM}
\begin{align}
\frac{1}{\gamma_n} \frac{d m^x_+}{dt} =& H_n m^y_{+} -2 m_0 h_n^y - H_a M_+^y,  \label{Eommp}\\
\frac{1}{\gamma_n} \frac{d m^y_+}{dt} =& H_a M^x_{+}  - H_n m^x_{+}  - 2m_{0} h_n^z \sin \psi, 
\end{align}
\end{subequations}
and those for the out-phase mode are
\begin{subequations}  \label{m-outphase-EOM}
\begin{align}
\frac{1}{\gamma_n} \frac{d m^x_{-}}{dt} =& H_n m^y_{-}- H_a M_-^y ,  \label{Eommn}\\
\frac{1}{\gamma_n} \frac{d m^y_{-}}{dt} =& H_a M^x_{-}  - H_n m^x_{+}  + 2 m_{0} h_n^x , 
\end{align}
\end{subequations}
which include an exotic oscillating field $h^{x,y,z}_n$ that interacts only with nuclear spins and oscillates in the $x$, $y$ or $z$ direction, respectively.
Note that the in-phase modes can only be excited by oscillating fields in $y$ and $z$ directions. On the other hand, the out-phase modes can be excited only via the $x$ direction. 

The eigensystem of the in-phase mode can be solved with the ansatz:
\begin{gather}
M^x_{+}(t)=M^x_{+} e^{i\omega t}, \quad  M^y_{+}(t)=M^y_{+} e^{i\omega t}, \\
m^x_{+}(t)=m^x_{+} e^{i\omega t}, \quad  m^y_{+}(t)=m^y_{+} e^{i\omega t},
\end{gather}
in the absence of exotic fields ($h_{e,n}=0$).
The problem is reduced to an ordinary eigenproblem. The detailed calculation is shown in Appendix \ref{appendix: classic}. The same can be done for the out-phase mode. For convenience, we give first the eigenfrequency
of the in-phase/out-phase precession mode of the electron and nuclear magnetization without mixing between them (i.e., neglecting $U_\mathrm{mix}$). The eigenfrequencies of the electron-magnetization system are 
\begin{gather}
\omega_{e,-}=\gamma_e\sqrt{2H_E (H_{K}+H_{K'}+H_a) +H_D(H_0+H_D)}, \label{e-freq}\\
\omega_{e,+}=\gamma_e \sqrt{2H_E(H_{K'}+H_a)+H_0(H_0+H_D)},\label{e+freq},
\end{gather}
and the eigenfrequency of the nuclear-magnetization system is 
\begin{equation}
\omega_{n}=\gamma_n H_n.
\end{equation}
Note that nuclear magnetization precessions have two degenerate modes (with angular frequency $\omega_n$) in the absence of the mixing term $U_\mathrm{mix}$.

\begin{figure}[t]
  \centering
  \includegraphics[width=0.75 \linewidth]{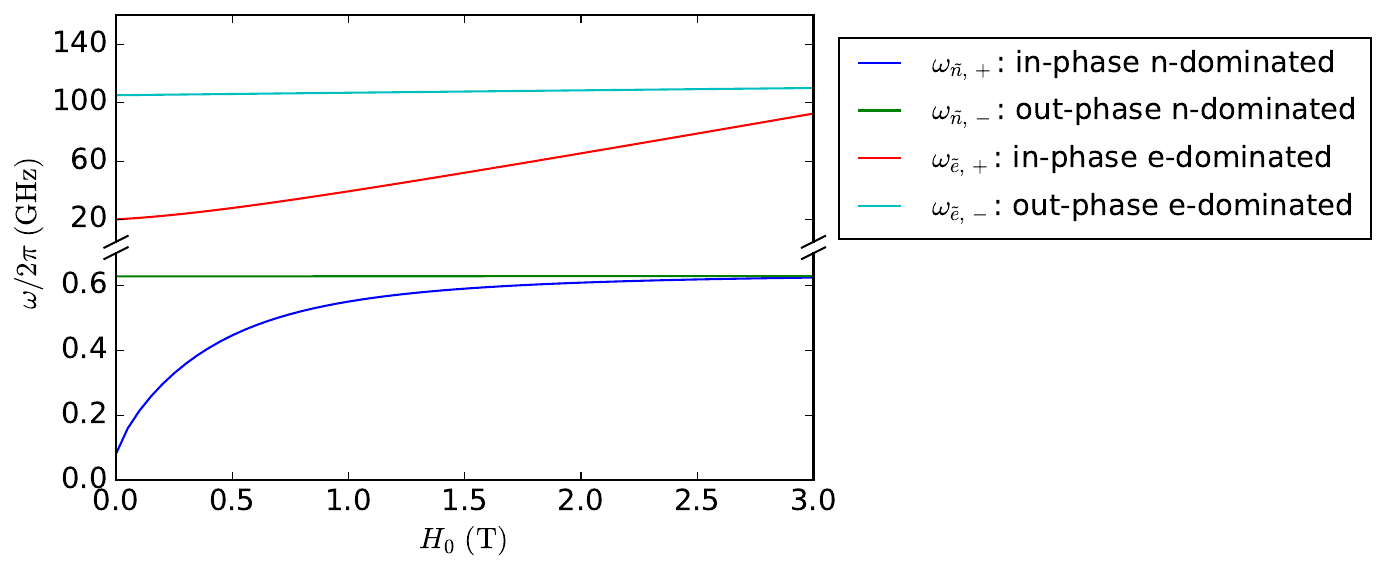}
  \caption{Field-dispersion relation $\omega(H_0)$ of diagonalized modes for both nuclear-dominated and electron-dominated hybridized modes at $T=\SI{0.1}{\K}$.
  }
  \label{fig: omega0}
\end{figure}

Once we take into account the mixing $U_\mathrm{mix}$, we obtain the following.
For the out-phase mode, the eigenfrequencies are
\begin{gather}
\omega_{\tilde{e},-} \approx \omega_{e,-}, \label{eoutMnCO3}\\
\omega_{\tilde{n},-} \approx \omega_{n} \left[ 1-  \frac{\langle I \rangle }{S}  \frac{\omega_{n} \omega_E}{\omega^{2}_{e,-}} \right] \ ; \quad \omega_E=\gamma_e H_E, \label{noutMnCO3}
\end{gather}
which correspond to the eigenmodes dominated by electron and nuclear spins, respectively.
For the in-phase mode, we can obtain a similar expression:
\begin{gather}
\omega_{\tilde{e},+} \approx \omega_{e,+}, \label{einMnCO3}\\
\omega_{\tilde{n},+} \approx \omega_{n} \left[ 1- 2 \frac{\langle I \rangle }{S}  \frac{\omega_{n} \omega_E}{\omega^2_{e,+}} \right]^{1/2} \ , \label{ninMnCO3}
\end{gather}
corresponding to the electron- and nuclear-dominated modes, respectively. The eigenfrequencies of the magnetic precession in \ce{MnCO3} are plotted in Fig.~\ref{fig: omega0}.
The gyromagnetic ratio $\gamma_n$ of $^{55} \mathrm{Mn}$ is taken from Refs.~\cite{PhysRev.82.651,STONE200575}, while the other parameters in the Hamiltonian are taken from Refs.~\cite{PhysRev.136.A218,doi:10.1063/1.1713505,PhysRev.142.300}, which we show in Table \ref{table: parameter}.
There is a relation among magnitudes of effective fields
\begin{equation}
H_{K'}, H_a \ll H_0, H_D, H_K \ll H_E, H_n \quad \mathrm{while} \quad H_E H_{K'}, H_E H_a \sim H_0^2,H_D^2, H_0 H_D.
\end{equation}

\begin{table}[t]
\begin{center}
\begin{tabular}{|| c | c  ||} 
 \hline
 Parameters & Magnitude\\
 \hline \hline
$\gamma_e$  & $\SI{1.76e11}{\per\second\per\tesla}$  \\
 \hline
 $\gamma_n$ &  $\SI{6.63e7}{\per\second\per\tesla}$  \\
 \hline
 $H_E$ & $ \SI{33.4}{\tesla}$\\
 \hline
  $H_D$& $ \SI{0.46}{\tesla}$\\
 \hline
   $ H_{K}$ & $ \SI{0.2}{\tesla}$\\
    \hline
  $ H_{K'}$ & $\SI{1.0e-4}{\tesla}$ \\
 \hline
    $ H_{n}\equiv A_\mathrm{hy}M_0$ & $ \SI{60.6}{\tesla}$\\
 \hline
     $ H_{a}\equiv A_\mathrm{hy}m_0$ & $ 0.02 \times \langle I \rangle/I \,\si{\tesla}$\\
 \hline
\end{tabular} 
\caption{\label{table: parameter} Parameters in the Hamiltonian of \ce{MnCO3}.  The gyromagnetic ratio $\gamma_n$ of $^{55} \mathrm{Mn}$ is taken from Refs.~\cite{PhysRev.82.651,STONE200575}, while the other parameters in the Hamiltonian are taken from Refs.~\cite{PhysRev.136.A218,doi:10.1063/1.1713505,PhysRev.142.300}. }
\end{center}
\end{table} 

Because of the relatively large difference between the eigenfrequencies of the out-phase modes of nuclear and electron magnetizations, the mixing angle between their precessions is expected to be small.\footnote{The shift $\Delta \omega_{p,\pm} \equiv \omega_n-\omega_{\tilde{n},\pm}$ of nuclear eigenfrequency due to the hyperfine mixing is inversely proportional to $\omega_{e,\pm}^2$ when $\omega_{e,\pm}\gg \omega_n$ as can be seen from Eqs.~\eqref{noutMnCO3} and \eqref{ninMnCO3}. 
The mixing angle between nuclear and electron magnetizations is expected to scale as $\sim \Delta \omega_{p,\pm} \propto 1/\omega_{e,\pm}^2$.}
On the other hand, the in-phase modes of nuclear and electron magnetizations, which have similar eigenfrequencies, significantly mix with each other showing the bending shape of the spectrum of the nuclear-dominated mode (the so-called pulling effect).
Therefore, from now on, we focus on and discuss only the in-phase modes and their response to the external oscillating field, while neglecting the out-phase modes. 

%%%%inphaseoutphase
%%%%

We want to emphasize that the magnetic resonance of the system has a relatively high frequency compared with the nuclear magnetic system without hyperfine interaction. For instance, \ce{^3He} needs a magnetic field of $\SI{30}{\tesla}$ to realize a magnetic resonance of the same frequency.

\subsection{Response to DM field} \label{subsec: response}

As discussed in Sec.~\ref{subsubsec: axionfield}, when the entire DM is composed of axions, a magnetized sample is affected by oscillating magnetic fields $\vec{h}_n=\vec{h}_n^\mathrm{axion}$ and $\vec{h}_e=\vec{h}_e^\mathrm{axion}$ defined in Eqs.~\eqref{axionfieldI} and \eqref{axionfielde}, respectively.
It should be noted that, based on the nuclear shell model of nuclei with odd atomic number, the nuclear spin of \ce{MnCO3} mainly comes from the proton spin \cite{RevModPhys.24.63,Ye_2023} and hence is sensitive to the axion-proton coupling $g_{app}$, not to the axion-neutron coupling $g_{ann}$. For simplicity, we take $\sigma_p=0.1,\sigma_n=0.0$ as a spin contribution from protons and neutrons for the numerical estimation of axion-induced magnetic field $\vec{h}^\mathrm{axion}_n$. See Appendix \ref{exotic interaction} for more detail. For the case of dark photon DM, we have instead $\vec{h}_n=\vec{h}_e=\vec{h}^{\gamma '}$ defined in \eqref{Beff_DP}.

Here we show the response of the system by evaluating the magnetization and power absorbed by the system, in Secs.~\ref{subsubsec: magnetization sig} and \ref{subsubsec: power absorbed}, respectively, when the DM mass is equal to the excitation energy of the magnetic material.

\subsubsection{Magnetization signal} \label{subsubsec: magnetization sig}

Now suppose the existence of a nonzero oscillating field from axion or dark-photon DM, $\vec{h}_{n,e} \propto \sin \omega_\mathrm{DM} t$ with $\hbar \omega_\mathrm{DM}=m_a,m_{\gamma'}$.
We can write the response in terms of susceptibility $\chi$ defined by the relation of total magnetization $\vec{M}_\mathrm{total}$ and external oscillating field $\vec{h}_n$ and $\vec{h}_e$ as
\begin{equation}
M_\mathrm{total}^{\alpha}\equiv M_1^\alpha+M_2^\alpha+m_1^\alpha+m_2^\alpha= \sum_\beta \left(\chi_n^{\alpha \beta}h_n^{\beta}+\chi_e^{\alpha \beta}h_e^{\beta}\right),
\end{equation}
with $\alpha,\beta=x,y,z$ denoting the direction of vectors $\vec{M}_\mathrm{total},\vec{h}_n,\vec{h}_e$ or the corresponding component of the tensor $\chi$. 
The magnetization susceptibility $\chi$ can be found by solving the equation of motion for magnetization given by Eqs.~\eqref{M-inphase-EOM} and \eqref{m-inphase-EOM}.
The relaxation time of magnetization should also be taken into account. 
The detailed calculation for signal susceptibility is shown in Appendix \ref{appendix: classic}. 
We assume that the \ce{MnCO3} sample is magnetized along the $x$ direction, as in Fig.~\ref{fig: groundstateM}, and consider only in-phase modes. The relevant direction of the oscillating field induced by DM is the $y$ or $z$ direction, corresponding to the precession of total magnetization in the $yz$ plane (see Eq.~\eqref{totalpm}).
We define a magnetization signal $M_\mathrm{signal}$ as the $z$ component of the total magnetization vector: \begin{equation}
M_\mathrm{signal} \equiv M^z_\mathrm{total} =\sum_{\alpha=y,z} \left(\chi^{z\alpha}_n h_n^{\alpha}+\chi^{z\alpha}_e h_e^{\alpha}\right). \label{MsigMztotal}
\end{equation}
Note that because the gyromagnetic ratio of the electron spin is much larger than that of the nuclear spin, a large part of magnetization of the sample is induced by electrons.
One can then estimate $M^z_\mathrm{total}$ from $M^z_\mathrm{total} \simeq M_1^z+M_2^z=M^x_{+} \sin \psi$.
 
 %%%%chitable
 \begin{table}[t]
\begin{center}
\begin{tabular}{||c| c c  ||} 
 \hline
 $\omega_\mathrm{DM}$& $\chi_n^{zy}(\omega_\mathrm{DM})$ &  $\chi_n^{zz}(\omega_\mathrm{DM})$  \\[0.5ex] 
 \hline\hline
 $\omega_\mathrm{DM} \sim \omega_{\tilde{n},+}$ &  $\largestar \ \eta m_0   T_{2n}  \gamma_n$ &  $\left(   \frac{\omega_{n}}{\omega_{\tilde{n},+}} \sin \psi  \right)\eta m_0   T_{2n}  \gamma_n$   \\ 
 \hline
 $\omega_\mathrm{DM} \sim \omega_{\tilde{e},+}$ &  $\largestar \ \eta m_0   T_{2e}  \gamma_n$ &  $\left(   \frac{\omega_{n}}{\omega_{\tilde{e},+}} \sin \psi  \right)\eta m_0   T_{2e}  \gamma_n  \left( 1-\frac{\omega_{e,+}^2}{2H_E \gamma_e \omega_n} \right)$  \\[0.5ex] 
 \hline
\end{tabular} 
\caption{\label{table: NMR} The signal susceptibility for the exotic perturbation $\vec{h}_n$ that interacts only with nuclear spins. The stars $\largestar$ indicate sensitive channels for probing the axion-proton coupling $g_{app}$.  The magnetic system is excited when $| \omega_\mathrm{DM} - \omega_{\tilde{n},+}| \lesssim 1/T_{2n}$ or $| \omega_\mathrm{DM} - \omega_{\tilde{e},+}| \lesssim 1/T_{2e}$ corresponding to the first or second column, respectively.}
\end{center}

\begin{center}
\begin{tabular}{|| c | c c | c ||} 
 \hline
$ \omega_\mathrm{DM} $& $\chi_e^{zy}(\omega_\mathrm{DM})$ & $\chi_e^{zz}(\omega_\mathrm{DM})$  \\ [0.5ex] 
 \hline\hline
 \multirow{2}{*}{$\omega_\mathrm{DM} \sim \omega_{\tilde{n},+}$} & $ \eta^2 \left(\frac{ \omega_n}{\gamma_e (H_0+H_D)} \right) m_0  T_{2n}   \gamma_n$ &  \multirow{2}{*}{$ \largepentagram  \ \eta^2 \frac{\omega_{n}}{\omega_{\tilde{n},+}} m_0 T_{2n}  \gamma_n $} \\ 
 &  $ \times \left( 1-\frac{\omega_{e,+}^2}{2H_E \gamma_e \omega_n} \right)$ &\\
 \hline
 $\omega_\mathrm{DM} \sim \omega_{\tilde{e},+}$   &   $ \largepentagram \ (\sin \psi)  M_0  T_{2e}  \gamma_e$  &  $ \largepentagram \ \left(  \frac{\gamma_e (H_0+H_D)}{\omega_{e,+}} \sin \psi \right) M_0 T_{2e} \gamma_e$ \\[0.5ex] 
 \hline
\end{tabular}
\caption{\label{table: FMR} The signal susceptibility for the exotic perturbation $\vec{h}_e$ that interacts only with electron spins. The stars $\largepentagram$ indicate sensitive channels for probing the axion-electron coupling $g_{aee}$ and kinetic mixing parameter $\epsilon$. The magnetic system is excited when $| \omega_\mathrm{DM} - \omega_{\tilde{n},+}| \lesssim 1/T_{2n}$ or $| \omega_\mathrm{DM} - \omega_{\tilde{e},+}| \lesssim 1/T_{2e}$ corresponding to the first or second column, respectively.}
\end{center}
\end{table}
 %%%%chitable
 
When the DM mass is close to the precession frequency, in particular $| \omega_\mathrm{DM} - \omega_{\tilde{n},+}| \lesssim 1/T_{2n}$ or $| \omega_\mathrm{DM} - \omega_{\tilde{e},+}| \lesssim 1/T_{2e}$, the collective spins would be excited, where
$T_{2n}$ and $T_{2e}$ are the relaxation times for the nuclear-dominated mode and the electron-dominated mode, respectively.
In other words, the excitation bandwidth is given by $\Delta\omega_{n,e}=2/T_{2n,2e}$.
We assume that magnetic relaxation times are much smaller than the coherence time $\tau_\mathrm{DM}$ of the DM field ($T_{2e,2n} \ll \tau_\mathrm{DM}$), which is actually the case for \ce{MnCO3}.
Within the DM coherence time $t \lesssim \tau_\mathrm{DM}$, the magnetic field from DM is a coherent driving field, so the magnetization signal rotates coherently with frequency $\omega_\mathrm{DM}$ whereas its amplitude is determined by the relaxation time $T_{2e,2n}$.
The spread $\Delta \omega_\mathrm{signal}$ in the frequency space of the signal is determined by that of the DM spectrum: 
\begin{equation}
\Delta \omega_\mathrm{signal}= \Delta \omega_\mathrm{DM}=\frac{2\pi}{\tau_\mathrm{DM}}.
\label{SignalBW}
\end{equation}
Numerically, we assume the values of the relaxation time at $T=\SI{0.1}{\K}$ as $T_{2n}=\SI{1}{\us}$ and $T_{2e}=\SI{1}{\ns}$ \cite{bunkovecho,borovik1984spin,PhysRevLett.114.226402,Shiomi:2019aa,doi:10.1143/JPSJ.15.2251}, corresponding to the bandwidth $\Delta\omega_n / 2\pi \sim \SI{0.3}{\MHz} $ and $\Delta\omega_e / 2\pi \sim  \SI{0.3}{\GHz}$, respectively. For the DM mass of interest, $\SI{e-6}{eV}\lesssim m_a\lesssim \SI{e-4}{eV}$, the value of $\tau_\mathrm{DM}$ is $\sim O(0.01)-O(1) \, \si{ms}$ (see Eq.~\eqref{taucohDM}), corresponding to the spread $\Delta \omega_\mathrm{DM}/2\pi \sim O(0.1)-O(10)\, \si{kHz}$.

The susceptibility on resonance is shown in Tables \ref{table: NMR} and \ref{table: FMR}. The star symbols represent the ``sensitive" channels whose responses $\chi$ are large compared with the other channels that can probe the same coupling of DM. We introduce a parameter $\eta$ defined by
\begin{equation}
\eta \equiv H_n  \frac{(H_0+H_D)\gamma_e^2}{\omega_{e,+}^2},
\end{equation}
which is an enhancement factor of signal magnetization compared to the nuclear magnetization, in addition to the effect that nuclear spins are highly polarized by the hyperfine interaction with electron spins.
Note that $\eta\to 0$ for $\sin\psi\to 0$ (see Eq.~(\ref{sinpsi})).
At $T=\SI{0.1}{\K}$, we obtain $\eta \approx 50$.
For example, at $H_0= \SI{1}{\tesla}$ corresponding to $\omega_{\tilde{n},+}/2\pi\approx \SI{500}{\MHz}$ and $\omega_{\tilde{e},+}/2\pi \approx \SI{40}{\GHz}$, we obtain numerically for sensitive channels
\begin{align}
&\chi^{zy}_n(\omega_\mathrm{DM} \approx \omega_{\tilde{n},+}) = 0.2 \times \left( \frac{T_{2n}}{\SI{1}{\us}}\right), &
&\chi^{zy}_n(\omega_\mathrm{DM} \approx \omega_{\tilde{e},+}) = 0.2 \times 10^{-3}  \times \left( \frac{T_{2e}}{\SI{1}{\ns}}\right),\\
&\chi^{zz}_e(\omega_\mathrm{DM} \approx \omega_{\tilde{n},+}) = 10 \times  \left( \frac{T_{2n}}{\SI{1}{\us}}\right),&
&\chi^{zy,zz}_e(\omega_\mathrm{DM} \approx \omega_{\tilde{e},+})  
= 2 \times  \left( \frac{T_{2e}}{\SI{1}{\ns}}\right),
\end{align}
where we assume \ce{MnCO3} of density $\SI{3.7}{g/cm^3}$ and hence the spin density $\rho_s \sim \SI{2e22}{cm^{-3}}$.
The order of magnitude of response does not change around the frequency range of interest (corresponding to $H_0 \sim 0.5$--$\SI{2}{\tesla}$) where the mixing between precessions of electron spins and nuclear spins remains large.
(See, e.g., Fig.~\ref{fig: omega0} for the relation between $H_0$ and the eigenfrequency of the system.)

Through the DM-nuclear interaction, the system is sensitive mostly to the DM-induced field polarized in the $y$ direction. 
On the other hand, through the DM-electron interaction, the system is sensitive mostly to the DM-induced field polarized in the $z$ direction or both $y$ and $z$ direction depending on the mode considered.
We define the angle between the sensitive direction and the polarization direction of the DM-induced magnetic field by a parameter $\theta$. 
To account for the unknown polarization (equivalently the unknown direction of the velocity of DM), we perform a substitution
$
\cos^2 \theta \rightarrow 1/3
$
if the system is sensitive to a single polarization direction, or by
$
\cos^2 \theta \rightarrow 2/3
$
if it is sensitive to any polarization direction in a plane~\cite{PhysRevD.104.095029}.

\subsubsection{Power absorbed to the magnetized sample} \label{subsubsec: power absorbed}
The absorbed power into the material can be estimated by the relation
\begin{equation}
P_\mathrm{absorb}=V  \cdot \sum_{\alpha=y,z} \left( \frac{dm^{\alpha} }{dt}h^{\alpha}_{n} +  \frac{dM^{\alpha} }{dt}h^{\alpha}_{e}\right), \label{Pabsorddef}
\end{equation}
where $V$ is the sample volume.
In Tables \ref{table: NMRpower} and \ref{table: FMRpower}, we show the time-averaged power absorbed under the resonance condition for each channel.
For example, at $H_0=\SI{1}{\tesla}$ corresponding to $\omega_{\tilde{n},+}/2\pi\approx \SI{500}{\MHz}$ and $\omega_{\tilde{e},+}/2\pi \approx \SI{40}{\GHz}$, numerically, we obtain absorbed power for the (most sensitive) $y$ direction of the DM-nucleon interaction ($h_n^y\neq0$) as
\begin{align}
&P_\mathrm{absorb} (\omega_\mathrm{DM} \sim \omega_\mathrm{\tilde{n},+})=\SI{1.5e-21}{\uW} \times \left( \frac{T_{2n}}{\SI{1}{\us}}\right) \left(\frac{h^y_n}{\SI{e-18}{\tesla}}\right)^2 \left( \frac{W_{\ce{MnCO_3}}}{\SI{1}{\kg}} \right),\\
&P_\mathrm{absorb} (\omega_\mathrm{DM} \sim \omega_\mathrm{\tilde{e},+})=\SI{5.4e-25}{\uW} \times\left( \frac{T_{2e}}{\SI{1}{\ns}}\right)\left(\frac{h^y_n}{\SI{e-18}{\tesla}}\right)^2 \left( \frac{W_{\ce{MnCO_3}}}{\SI{1}{\kg}} \right),
\end{align}
where $W_{\ce{MnCO_3}}$ is the total mass of the \ce{MnCO3} sample.
Also, at $H_0=\SI{1}{\tesla}$, we obtain absorbed power for the $y$ and $z$ directions of the DM-electron interaction $(h_e^{y,z}\neq0)$ as:
\begin{align}
&P_\mathrm{absorb} (\omega_\mathrm{DM} \sim \omega_\mathrm{\tilde{n},+})=\SI{4.1e-18}{\uW} \times
\left( \frac{T_{2n}}{\SI{1}{\us}}\right) 
\left(\frac{h^y_e}{\SI{e-18}{\tesla}}\right)^2 \left( \frac{W_{\ce{MnCO_3}}}{\SI{1}{\kg}} \right),\\
&P_\mathrm{absorb} (\omega_\mathrm{DM} \sim \omega_\mathrm{\tilde{e},+})=\SI{6.0e-17}{\uW}
\times\left( \frac{T_{2e}}{\SI{1}{\ns}}\right)
\left(\frac{h^{y,z}_e}{\SI{e-18}{\tesla}}\right)^2 \left( \frac{W_{\ce{MnCO_3}}}{\SI{1}{\kg}} \right).
\end{align}
Within the frequency range of interest (corresponding to $H_0 \sim 0.5$--$\SI{2}{\tesla}$), magnitude of absorption power only slightly changes.

In Sec.~\ref{sec: sensitivity}, we suggest experimental ways to detect the signal through a magnetic system with the hyperfine interaction. We estimate the sensitivity of the method based on the magnetization signal and power absorbed into the system, which we compare with the power of relevant noises.

\begin{table}[t]
\begin{center}
\begin{tabular}{||c| c c ||} 
\hline
 $\omega_\mathrm{DM}$ & $P_\mathrm{absorb}/V(h^y_n)^2$ due to $h^y_n$  &  $P_\mathrm{absorb}/V(h^z_n)^2$ due to $h^z_n$  \\ [0.5ex] 
 \hline\hline
 $\omega_{\tilde{n},+}$ &  $\largestar \  {m_0 T_{2n}  \gamma_n \omega_{\tilde{n},+}^2  }/{2 \omega_n}$ & $ \sin^2 (\psi) m_0 T_{2n} \gamma_n \omega_n /2 $  \\ 
 \hline
 $  \omega_{\tilde{e},+}$ & $\largestar \ m_0 T_{2e} \gamma_n \left( 1- \frac{\omega_{\tilde{n},+}^2}{\omega_n^2} \right) \omega_n /2 $ &  $\left(   \frac{\omega_{n}}{\omega_{e,+}} \sin \psi  (1- \frac{\omega_{e,+}^2}{2H_E \gamma_e \omega_n}) \right)^2 \left( 1- \frac{\omega_{\tilde{n},+}^2}{\omega_n^2}\right)   \omega_n m_0   T_{2e}  \gamma_n /2$ \\[0.5ex] 
 \hline
\end{tabular} 
\caption{\label{table: NMRpower} Power absorbed in the magnetized sample on resonance in the presence of the DM-induced magnetic field interacting only with nuclear spins. The stars $\largestar$ indicate sensitive channels for probing the axion-proton coupling $g_{app}$. The magnetic system is excited when $| \omega_\mathrm{DM} - \omega_{\tilde{n},+}| \lesssim 1/T_{2n}$ or $| \omega_\mathrm{DM} - \omega_{\tilde{e},+}| \lesssim 1/T_{2e}$ corresponding to the first or second column, respectively.}
\end{center}
\begin{center}
\begin{tabular}{||c| c c||} 
 \hline
$\omega_\mathrm{DM} $ & $P_\mathrm{absorb}/V (h^y_e)^2$ due to $h^y_e$ & $P_\mathrm{absorb}/V(h^z_e)^2$ due to $h^z_e$  \\ [0.5ex] 
 \hline\hline
 $\omega_{\tilde{n},+}$ &  $\eta^2 \left( \frac{\omega_n \omega^2_{\tilde{n},+}}{\gamma^2_e (H_0+H_D)^2} \right) m_0  T_{2n}   \gamma_n  \left( 1-\frac{\omega_{e,+}^2}{2H_E \gamma_e \omega_n} \right)^2 /2 $   & $\largepentagram \ \eta^2 \omega_{n} m_0 T_{2n}  \gamma_n /2$ \\ 
 \hline
 $\omega_{\tilde{e},+}$   &   $\largepentagram \ \ M_0  T_{2e}  \gamma_e \frac{\omega_{e,+}^2}{ 2 H_E \gamma_e} /2$  &   $ \largepentagram \ \left(  \gamma_e (H_0+H_D) \sin \psi \right) M_0 T_{2e} \gamma_e /2$\\[0.5ex] 
 \hline
\end{tabular}
\caption{\label{table: FMRpower} Power absorbed in the magnetized sample on resonance in the presence of the DM-induced magnetic field interacting only with electron spins. The stars $\largepentagram$ indicate sensitive channels for probing the axion-electron coupling $g_{aee}$ and the kinetic mixing parameter $\epsilon$. The magnetic system is excited when $| \omega_\mathrm{DM} - \omega_{\tilde{n},+}| \lesssim 1/T_{2n}$ or $| \omega_\mathrm{DM} - \omega_{\tilde{e},+}| \lesssim 1/T_{2e}$ corresponding to the first or second column, respectively.}
\end{center}
\end{table}

%%%%%%%%%%%%%%%%%%%%%%%%%%%%%%%%%%%%%%%%%%%%%%%%%%%%%%%%%%%%%%%%%%%%%%%%%%
\section{Dark matter detection with nuclear magnetic excitation}
\label{sec: sensitivity}
\setcounter{equation}{0}
%%%%%%%%%%%%%%%%%%%%%%%%%%%%%%%%%%%%%%%%%%%%%%%%%%%%%%%%%%%%%%%%%%%%%%%%%%

In this section, we present the sensitivity of DM detection by excitation of hybridized spin precession modes under a low-temperature environment $T=\SI{0.1}{\K}$.
Our basic strategy is as follows. We can set an eigenfrequency of the magnetic system to a desired value by tuning the applied magnetic field $H_0$. When the DM mass is close to the eigenfrequency, collective motion of spins can be excited, since the field of DM axions or DM dark photons oscillates with the frequency $\omega_\mathrm{DM} \simeq m_{a,\gamma'}/\hbar$. 
Then, a wide DM mass range can be searched for by sweeping $H_0$.
The probe range is divided into two much different scales corresponding to the two bands of the nuclear- and electron-dominated modes in \ce{MnCO3}. We discuss them in Secs.~\ref{subsec: nuclear-freq mode} and \ref{subsec: electron-freq mode}, respectively.

\subsection{Sensitivity at the nuclear resonance frequency} \label{subsec: nuclear-freq mode}

We consider the \textit{LC} resonant circuit with a pick-up loop similar to Ref~\cite{Aybas:2021nvn}, combined with the microstrip SQUID amplifier\footnote{
In the frequency range of interest, the dc SQUID is not favorable as a magnetometer because there is a severe parasitic capacitance between the dc SQUID washer and the input coil.
Detection with a reactive (dissipationless) ac SQUID and the microwave resonator \cite{doi:10.1063/1.2803852,JABmates} is proposed to detect dark photons in the frequency range from $\SI{10}{\MHz}$ to $\SI{1}{\GHz}$ in \cite{PhysRevD.92.075012}.
Putting the input coil inside the hole in the SQUID washer might cure the parasitic capacitance problem for the dc SQUID as well \cite{139223}.
Using a microstrip-coupled dc SQUID as the amplifier is also a way to deal with this issue; we pursue this possibility in this work.
}~ \cite{doi:10.1063/1.1347384, doi:10.1063/1.121490,2002PhDT.......116T} to detect magnetization induced by DM.
It has flexibility to tune the resonance frequency in the range of interest and the high quality factor can be achieved. 
Before going into the details of the measurement method and setup, let us consider the frequency range of interest and the quality factor of the resonator.

At $T=\SI{0.1}{\K}$, the spectrum of the nuclear-dominant magnetization precession of \ce{MnCO3} covers a high frequency range from $O(10) \, \si{\MHz}$ to $\sim \SI{600}{\MHz}$ (see Fig.~\ref{fig: omega0}).
For a frequency range $\omega / 2 \pi \gtrsim \SI{500}{\MHz}$, the observed intrinsic relaxation time $T_{2n}$ from spin echoes is of order $O(10) \, \si{\us}$ at a low temperature $O(1)\,\si{\K}$, while below $\SI{500}{\MHz}$ the relaxation time $T_{2n}$ drops rapidly and the signal from DM might be suppressed (see, e.g., Refs.~\cite{bunkovecho,borovik1984spin,PhysRevLett.114.226402,Shiomi:2019aa}).
Here, for simplicity we focus on a frequency range $\omega_\mathrm{DM}/2\pi \approx \omega_{\tilde{n},+}/2\pi = 500$--$\SI{600}{\MHz}$, which can be covered by sweeping the static field $H_0$ from $\SI{0.7}{\tesla}$ to $\SI{2}{\tesla}$ (see Fig.~\ref{fig: omega0}), while assuming constant $T_{2n}=\SI{1}{\us}$.

An important quantity for quantifying the signal and noise in the resonant approach is the quality factor $Q_r$ of the circuit, which is related to the circuit bandwidth $\Delta \omega_r$ and the resonant frequency $\omega_0$ as
\begin{equation}
\Delta \omega_r=\frac{\omega_0}{Q_r}. \label{BWr}
\end{equation}
We desire the value of $Q_r$ as high as possible up to $Q_\mathrm{DM}=10^6$ so that the signal is highly resonant while the circuit bandwidth $\Delta \omega_r$ still covers bandwidth $\Delta \omega_\mathrm{DM}$ of DM signal. Here, in numerical calculation we assume $Q_r=10^5$, while a higher value such as $Q_r\sim 10^6$ might also be achieved by the \textit{LC} circuit, which is discussed in Ref.~\cite{Chaudhuri:2014dla}.

The following steps can be used to scan DM masses within the frequency range $\omega / 2 \pi=500$--$\SI{600}{\MHz}$:
\begin{enumerate}
\item Match eigenfrequencies of the magnetic system and the resonant circuit to the frequency $\omega$ of interest at the same time.
The former can be tuned by changing the external static field $H_0$, while the latter by changing the values of resistance and capacitance of the circuit.
\item Measure the magnetization signal from the magnetic material $\ce{MnCO_3}$ for an interrogation time $\Delta t$.
\item Shift the frequency $\omega$ by an interval $\Delta \omega$ determined by Eq.~\eqref{BWr} after each step of the signal measurement until the frequency range of interest is fully covered. 
\end{enumerate}
The interrogation time $\Delta t$ is determined by the total observation time $T_\mathrm{total}$ and the number of scan steps determined as a function of $Q_r$.

%%circuitone
\begin{figure}[t]
  \centering
  \begin{subfloat}[The original probe circuit with a loaded resistance.] {
    \centering
    \begin{tikzpicture}
      \draw (-2,2) 
      to[short] (-1,2)
      to[R=$R$] (-1,0) 
      to[short] (-2,0);
      \draw (-1,2)
      to[short] (1,2)
      to[C=$C$] (1,0)
      to[short] (-1,0);
      \draw (1,2)
      to[C=$C_c$] (3,2)
      to[R=$R_\mathrm{load}$, o-o] (3,0);
      \draw (-1,0) -- (-1,-0.01) node[ground]{};
      \draw (3,0) -- (3,-0.01) node[ground]{};
      \draw  (-2,0)--++(0,0.9)--++(-0.5, 0.0) arc (340 : 20 : 0.8cm and 0.3cm)--++(0.5, 0)--++(0,0.9);
      \draw (-2.5,0.55) node[left] {$L$};
      \draw (-2.4,1.65) node[left] {sample};
       \def\firstrectangle {(6,6) rectangle (4,4.5)};
    \draw[color=black, fill = lightgray] (-3.5,0.85) rectangle (-2.9,1.4);
    \end{tikzpicture} \label{LCcircuitORIG}
    }
\end{subfloat} 
\begin{subfloat}[The original probe circuit emphasizing the voltage source $V_p$ from the pick-up loop due to a magnetization oscillation of the sample.] {
    \centering
     \begin{tikzpicture}
      \draw (-3,0) 
      to[sV, l=$V_p$] (-3,1)
      to[L=$L$] (-3,2)
      to[short] (-1,2)
      to[R=$R$] (-1,0) 
      to[short] (-3,0);
      \draw (-1,2)
      to[short] (1,2)
      to[C=$C$] (1,0)
      to[short] (-1,0);
      \draw (1,2)
      to[C=$C_c$] (3,2)
      to[R=$R_\mathrm{load}$, o-o] (3,0);
      \draw (-1,0) -- (-1,-0.01) node[ground]{};
      \draw (3,0) -- (3,-0.01) node[ground]{};
    \end{tikzpicture} \label{LCcircuitVp}
        }
\end{subfloat} \quad \quad
    \begin{subfloat}[The Thevenin equivalent circuit.]{
\centering
    \begin{tikzpicture}
      \draw (1,2) 
      to[sV, l=$V_s$] (1,0);
      \draw (1,2)
      to[european resistor,l=$Z(\omega)$] (4,2)
      to[R=$R_\mathrm{load}$,o-o] (4,0);
      \draw (1,0) -- (1,-0.01) node[ground]{};
      \draw (4,0) -- (4,-0.01) node[ground]{};
    \end{tikzpicture} \label{LCcircuitThevenin}
    }       
\end{subfloat} 
    \caption{The \textit{LC} circuit capacitively coupled to an amplifier of impedance $R_\mathrm{load}$ for picking up a magnetization signal from the magnetic sample perturbed by cosmic DM.}\label{LCcircuit}
\end{figure}
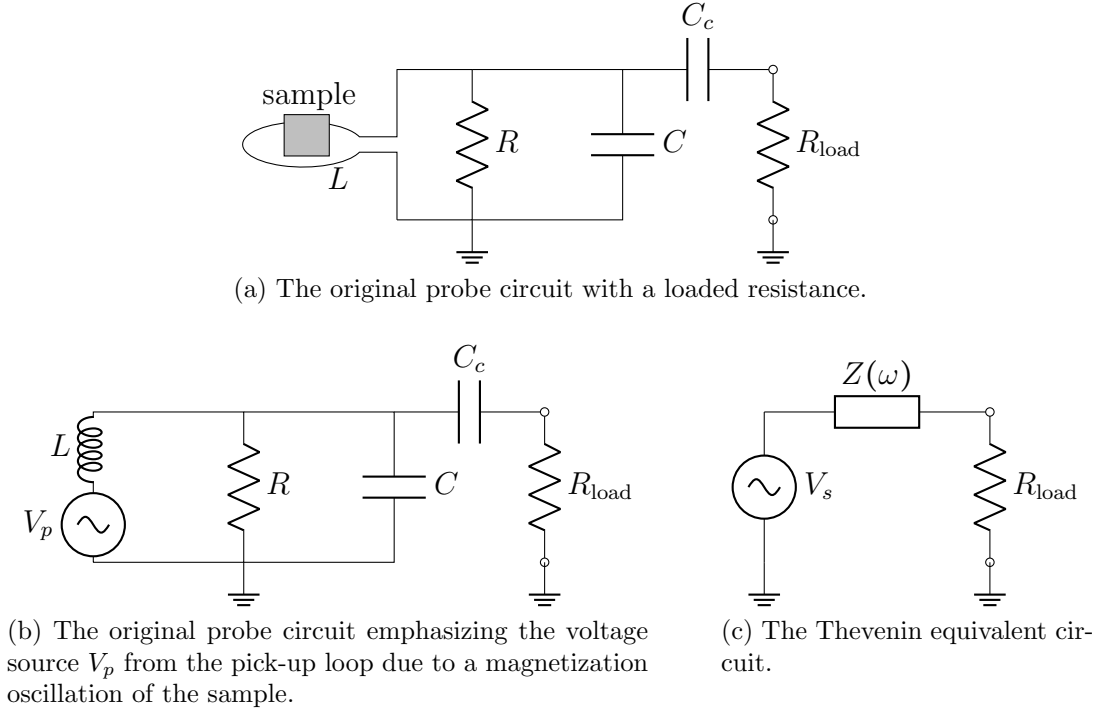
%%circuitone

Now we move to the discussion of experimental setup.
We consider picking up the signal inductively by a pick-up loop $L$ that is in parallel with the resistance $R$ and capacitor $C$, while the resonant circuit is capacitively coupled to the amplifier through the capacitor $C_c$ (Fig.~\ref{LCcircuit}).
When DM induces a magnetic excitation in the magnetized sample, the oscillation of magnetization produces an oscillating flux through the pick-up loop, which generates a voltage $V_p$ that can be detected through the load resistance $R_\mathrm{load}$ associated with the amplifier.

Let us estimate the voltage over the load resistance $R_\mathrm{load}$ (which is the input voltage of the amplifier) so that we can compare the power due to the magnetization signal with that of the thermal noise to derive the signal-to-noise ratio (SNR).
We apply the Thevenin theorem to simplify the task.
For convenience, the Thevenin equivalent circuit is illustrated in Fig.~\ref{LCcircuit} (c). 
The Thevenin impedance $Z(\omega)$ is the impedance of the circuit when we look from the terminal of $R_\mathrm{load}$ with $V_p$ neglected. It is given by
\begin{equation}
Z(\omega)=\frac{1}{(1/ i \omega L)+(1/R)+ i\omega C} +\frac{1}{i\omega C_c}. \label{impedanceThevenin}
\end{equation}
The parameters are to be chosen such that the impedance is matched to the amplifier impedance $R_\mathrm{load}=\SI{50}{\ohm}$ at resonance. 
Requiring $Z=Z_0=\SI{50}{\ohm}$ at an arbitrary resonance frequency $\omega_0$, we can choose \cite{Miller2000InterplayAR,Aybas:2021nvn} 
\begin{gather}
R=Q_0 \omega_0 L,\\
C=\frac{1}{\omega_0^2 L} \left(1- \frac{1}{Q_0}\sqrt{\frac{R-Z_0}{Z_0}} \right),\\
C_c=\frac{1}{\omega_0} \sqrt{\frac{1}{Z_0(R-Z_0)}},
\end{gather}
where $Q_0$ is the unloaded quality factor (or, equivalently, the quality factor of the circuit without $C_c$ and $R_\mathrm{load}$). 
Since the impedance is matched, the loaded quality factor $Q_r$ is determined to be $Q_r=Q_0/2$ \cite{Miller2000InterplayAR}.

On the other hand, according to the Thevenin theorem, the equivalent voltage $V_s$ is equal to the voltage between terminals of $R_\mathrm{load}$ when it is replaced by open terminals (the unloaded probe).
Equivalently, this is the case when we take the limit of $R_\mathrm{load} \rightarrow \infty$.
Therefore, we obtain the Thevenin equivalent voltage $V_s$ in the frequency space as
\begin{align}
\tilde{V}_s(\omega)&= \tilde{V}_p \frac{(1/R+ i\omega C)^{-1}}{ i \omega L+(1/R+ i\omega C)^{-1}} \nonumber\\
&= -i \omega\tilde{\Phi}_{p}  (\omega) \frac{(1/R+ i\omega C)^{-1}}{ i \omega L+(1/R+ i\omega C)^{-1}} \nonumber\\
& \approx -i\omega \tilde{\Phi}_{p}(\omega)  \sqrt{ \frac{Q_0 Z_0}{ \omega_0 L}},
\label{VsignalThevenin}
\end{align}
where $Q_0 \omega_0 L \gg Z_0$ and $\omega\approx\omega_0$ are assumed, and $\Phi_p$ is the flux signal at the pick-up coil induced by the transverse magnetization $M_\mathrm{signal}$ of the sample. The tilde symbol represents the value in the frequency space. One can derive the flux signal $\Phi_p$ from the Faraday induction law \cite{HOULT197671} and the reciprocity theorem:
\begin{equation}
\Phi_p=\int_V dV M_\mathrm{signal} \beta,
\end{equation}
where $\beta$ is the rate of flux induction from one unit current. With the geometry of a one-turn pick-up loop with the sample at the center, we obtain
\begin{equation}
\Phi_p= M_\mathrm{signal} V \frac{\mu_0}{4\pi} \frac{2\pi a^2}{(a^2+d^2)^{3/2}} \approx M_\mathrm{signal} V \frac{\mu_0}{2 a} \ \left( \mathrm{when} \ d\rightarrow 0\right)\;, \label{fluxsig}
\end{equation}
where $a$ is the loop radius, $d$ the distance of the sample from the loop and $V$ the sample volume. Note that the sample volume $V$ is limited by the size of the loop. Assuming a sample of cylindrical shape shrunk by ratio $\nu$ from a cylinder of radius $a$ and height $2a$, the sample volume is given by
\begin{equation}
V= \nu \pi a^2 (2a).
\label{eq:MnCO3_vol}
\end{equation}

We arrive at the Thevenin equivalent circuit with $V_s$ given by Eq.~\eqref{VsignalThevenin} and impedance $Z(\omega)$ given by Eq.~\eqref{impedanceThevenin}, whose value on resonance is set to $R_\mathrm{load}$.
Since the impedance is matched, half of the voltage $V_s$ is applied to the input amplifier.
(See Fig.~\ref{LCcircuit} (c) for the Thevenin equivalent circuit.)
We consider the Johnson-Nyquist thermal noise from the resonator and amplifier with voltage $\tilde{V}_\mathrm{noise}=k_B(T+T_a)Z_0$ at the input amplifier, where $T_a$ is the noise temperature of the amplifier, $T$ the resonator temperature, and $k_B$ the Boltzmann constant.
Combining contributions from the DM signal and noise, we obtain the power density at the input amplifier as
\begin{gather}
\tilde{P}(\omega)=\tilde{P}_s(\omega ) + \tilde{P}_{\mathrm{noise}} (\omega), \\
\tilde{P}_s(\omega ) = \frac{1}{4}\frac{\tilde{V}^2_s (\omega)}{Z_0}, \quad \tilde{P}_{\mathrm{noise}} (\omega)=k_B(T+T_a). \label{powersignoise}
\end{gather}   
At the temperature $T=\SI{0.1}{\K}$, with the microstrip SQUID amplifier tuned for a frequency $\sim  O(100) \,\si{\MHz}$, combined with the heterostructure field-effect transistor amplifier \cite{ADMX:2011hrx}, the noise temperature of the amplifier is less than $T_a=\SI{0.1}{\K}$.

The SNR is estimated from the Dicke radiometer equation \cite{doi:10.1063/1.1770483} with Eqs.~\eqref{fluxsig} and \eqref{powersignoise}.
Assuming $\omega_\mathrm{DM} \approx \omega_0$, we obtain 
\begin{equation}
\mathrm{SNR}= \frac{P_s}{\tilde{P}_\mathrm{noise} (\omega)  \Delta f_r/ \sqrt{\Delta f_r \Delta t }} =\frac{1}{2}
\left( \mu_0 M_\mathrm{signal}  {V}/{2a} \right)^2  \frac{ \omega_{\mathrm{DM}} Q_0}{  L}
\frac{1}{ 4k_B(T+T_a) }  \sqrt{\frac{\Delta t}{ \Delta f_r}},
\end{equation}
where $\Delta f_r$ is the circuit bandwidth, $\Delta t$ is the interrogation time at each frequency, and the overall coefficient $1/2$ is added to account for the time average.
Note that $M_\mathrm{signal}V$ depends on the total number of spin sites and hence on the total mass $W_{\ce{MnCO3}}$ of \ce{MnCO3}.
By requiring $\mathrm{SNR}\geq 1$, we obtain the expected sensitivity for the coupling constant between DM and nucleons or electrons.

\begin{figure}[t!]
  \centering
  \includegraphics[width=0.7 \linewidth]{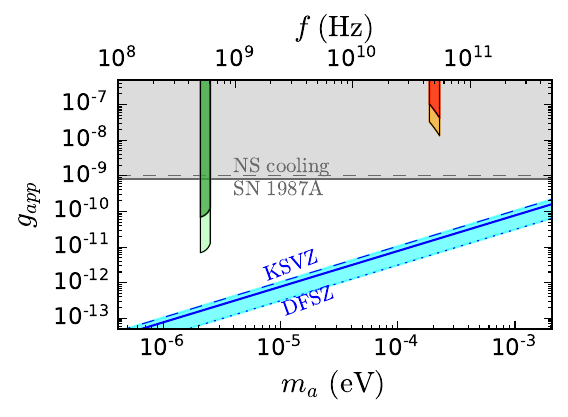}
  \caption{The sensitivity to the axion-proton coupling $g_{app}$ of an experimental setup using a $\SI{1}{\kg}$ canted antiferromagnet \ce{MnCO3}. Two hybridized oscillation modes of nuclear and electron spins are considered with each total observation time $T_\mathrm{total}=1 \,\mathrm{year}$. Green (pale green) projection (derived in Sec.~\ref{subsec: nuclear-freq mode}): the nuclear-dominated mode $\chi^{zy}_n(\omega_{\tilde{n},+})$ is used with $\omega_{\tilde{n},+} / 2 \pi= 500$--$\SI{600}{\MHz}$, $T_{2n}=1 (10) \,\si{\us}$. The macroscopic magnetization signal is assumed to be detected by the inductive pick-up loop associated with the \textit{LC} resonant circuit, of which the quality factor $Q_r= 10^5$ is expected. Each scan step at a fixed frequency takes $\Delta t \approx \SI{1.7e3}{\s}$. Red (orange) projection  (derived in Sec.~\ref{singlephotoncounter}): electron-dominated mode resonance $\chi^{zy}_n(\omega_{\tilde{e},+})$ is used with $\omega_{\tilde{e},+} / 2 \pi=45$--$\SI{55}{\GHz}$, $T_{2e}=1 (10) \,\si{\ns}$ and $\Delta t = 5 \times 10^5 (5 \times 10^4) \,\si{\s}$. In this case, the cavity photons excited in cavity by spin systems are the expected signal. For both projections, the angular average $\cos^2 \theta \rightarrow 1/3$ is assumed to account for the unknown polarization of DM-induced magnetic field. The gray area represents constraints from astrophysical consideration: SN1987A \cite{Carenza:2019pxu,Carenza:2020cis,Fischer:2021jfm}, neutron star cooling \cite{Hamaguchi:2018oqw}. The blue band shows the theoretical prediction from the DFSZ axion model, with $0.28 < \tan \beta < 140$ \cite{Chen:2013kt}.}
  \label{fig:gapp reach}
\end{figure}

\begin{figure}[t!]
  \centering
  \includegraphics[width=0.7 \linewidth]{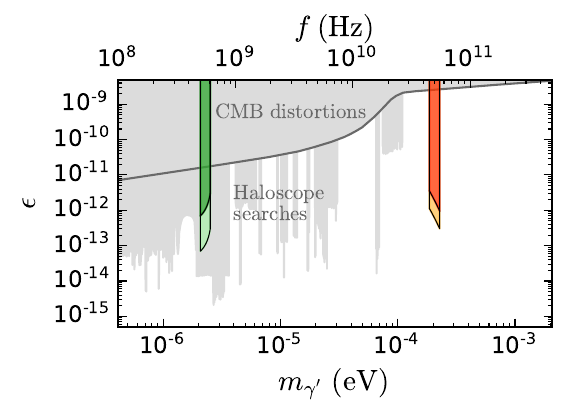}
  \caption{The sensitivity to the dark photon kinetic mixing parameter $\epsilon$ of an experimental setup using a $\SI{1}{\kg}$ canted antiferromagnet \ce{MnCO3}. Two hybridized oscillation modes of nuclear and electron spins are considered with each total observation time $T_\mathrm{total}=1 \,\mathrm{year}$.
  Green (pale green) projection (derived in Sec.~\ref{subsec: nuclear-freq mode}): the nuclear-dominated mode $\chi^{zz}_e (\omega_{\tilde{n},+})$ is used with $\omega_{\tilde{n},+} / 2 \pi= 500$--$\SI{600}{\MHz}$, $T_{2n}=1 (10) \,\si{\us}$. The macroscopic magnetization signal is assumed to be detected by the inductive pick-up loop associated with the \textit{LC} resonant circuit, of which the quality factor $Q_r= 10^5$ is expected. Each scan step at a fixed frequency takes time $\Delta t \approx \SI{1.7e3}{\s}$.
  Red (orange) projection (derived in Sec.~\ref{singlephotoncounter}): electron-dominated mode resonance $\chi^{zz,zy}_e (\omega_{\tilde{e},+})$ is used with $\omega_{\tilde{e},+} / 2 \pi=45$--$\SI{55}{\GHz}$, $T_{2e}=1 (10) \,\si{\ns}$ and $\Delta t = 5 \times 10^5 (5 \times 10^4) \,\si{\s}$. In this case, the cavity photons excited by spin systems are the expected signal. 
  The angular average $\sin^2 \varphi \rightarrow 1/2$ is assumed for both projections, while $\cos^2 \theta \rightarrow 1/3$ and $\cos^2 \theta \rightarrow 2/3$ are assumed for the former and latter, respectively.
  The gray area covered by the bold line represents constraints from cosmological considerations \cite{Arias:2012az}, i.e., distortion of cosmic microwave background and effective neutrino species.
  The other gray areas are constraints from haloscope searches \cite{PhysRevD.104.095029}.}
  \label{fig:epsilon reach}
\end{figure}

Here we assume that the material is placed in the proper direction to read out $M_\mathrm{signal }\equiv  M^z_\mathrm{total}$, and is appropriately shielded from unwanted perturbations.
The sensitivities based on sensitive channels of $\ce{MnCO_3}$ discussed in Sec.~\ref{subsec: response}, along with the present constraints of the relevant DM parameters, are shown in Figs.~\ref{fig:gapp reach}, \ref{fig:gaee reach}, and \ref{fig:epsilon reach}.
We take $T=T_a=\SI{0.1}{\K}$, the pick-up coil inductance $L \approx 0.45 \times (a/\SI{5}{\cm}) \ \mathrm{\mu H}$, $a=\SI{5}{\cm}$, and the total observation time $T_\mathrm{total}=1\,\mathrm{year}$ for probing the range $\omega / 2 \pi=500$--$\SI{600}{\MHz}$. We assume the total \ce{MnCO3} mass of $\SI{1}{\kg}$ that can be achieved with $\nu \approx 0.3$ and the \ce{MnCO3} density of $3.7 \ \si{g/cm^3}$. 
The quality factor $Q_r$ is assumed to be $10^5$.
Although the quality factor of circuit might be affected by the relaxation time of the magnet through the coupling between the circuit and magnet, we neglect this effect just for simplicity.
The interrogation time at each scan step is $\Delta t=\SI{1.7e3}{\s}$.
The relaxation time $T_{2n}$ is assumed to be $\SI{1}{\us}$ for a realistic setup and, in addition, $\SI{10}{\us}$ for an optimistic one.
The latter choice corresponds to a situation where the (effective) magnetic relaxation time is longer thanks to the reduced inhomogeneity in material and applied field.

%%%%%%%%%%Plot sensitivity

%%gann
%%gann

%%gaee
\begin{figure}[t!]
  \centering
  \includegraphics[width=0.7 \linewidth]{figs/data2.pdf}
  \caption{The sensitivity to the axion-electron coupling $g_{aee}$ of an experimental setup using a $\SI{1}{\kg}$ canted antiferromagnet \ce{MnCO3}. Two hybridized oscillation modes of nuclear and electron spins are considered with each total observation time $T_\mathrm{total}=1 \,\mathrm{year}$. 
  Green (pale green) projection (derived in Sec.~\ref{subsec: nuclear-freq mode}): the nuclear-dominated mode $\chi^{zz}_e(\omega_{\tilde{n},+})$ is used with $\omega_{\tilde{n},+} / 2 \pi= 500$--$\SI{600}{\MHz}$, $T_{2n}=1 \ (10) \,\si{\us}$. The macroscopic magnetization signal is assumed to be detected by the inductive pick-up loop associated with the \textit{LC} resonant circuit, of which the quality factor $Q_r= 10^5 $ is expected. Each scan step at a fixed frequency takes time $\Delta t \approx \SI{1.7e3}{\s}$.
  Red (orange) projection (derived in Sec.~\ref{singlephotoncounter}): electron-dominated mode resonance $\chi^{zz,zy}_e(\omega_{\tilde{e},+})$ is used with $\omega_{\tilde{e},+} / 2 \pi=45$--$\SI{55}{\GHz}$, $T_{2e}=1 (10) \,\si{\ns}$ and $\Delta t = 5 \times 10^5 (5 \times 10^4) \,\si{\s}$. In this case, the cavity photons excited by spin systems are the expected signal. 
The angular average $\cos^2 \theta \rightarrow 1/3$ and $2/3$ are assumed for the former and latter projections, respectively.
The gray area represents constraints from astrophysical consideration: the tip of red giants \cite{Capozzi:2020cbu}, luminosity function of white dwarfs \cite{Giannotti:2017hny, MillerBertolami:2014rka}, and the constraints from underground detectors searching for solar axions \cite{LUX:2017glr,PandaX:2017ock,XENON100:2014csq}. The blue band shows the theoretical prediction from the DFSZ axion model, with $0.28 < \tan \beta < 140$ \cite{Chen:2013kt}.}
  \label{fig:gaee reach}
\end{figure}
%%gaee

%%epsilon
%%epsilon

%%%%%%%%%%Plot sensitivity
For the dark photon DM detection, the observed flux at the pick-up loop is due to both the dark photon field itself $h^{\gamma'}$ and the field generated by the magnetization $M_\mathrm{signal}$.
Since typically $\chi_e \gtrsim 1$, the pick-up magnetic flux caused by $M_\mathrm{signal}$ is larger than that of $h^{\gamma'}$; thus, the sensitivity to $\epsilon$ is derived taking into account only the contribution of $M_\mathrm{signal}$, neglecting that of $h^{\gamma'}$.

\subsection{Sensitivity at the electron resonance frequency} \label{subsec: electron-freq mode}

The material \ce{MnCO3} is also sensitive to another frequency range of $\omega / 2 \pi \gtrsim \SI{10}{\GHz}$.
It is due to excitation of the electron-dominated mode (see Fig.~\ref{fig: omega0}).
However, because of the hyperfine interaction, it also contains a nuclear-spin component, and hence we also expect a sensitivity to $g_{app}$.
Here we consider the frequency range $\omega_\mathrm{DM}/2\pi \approx \omega_{\tilde{e},+} /2\pi = 45$--$\SI{55}{\GHz}$. 
The relaxation time of this mode is mainly determined by that of the electron spin which is found to be $T_{2e}\approx O(0.1) \,\si{\ns}$ at $T=\SI{4.2}{\K}$ \cite{doi:10.1143/JPSJ.15.2251} and $T_{2e}\approx O(1) \,\si{\ns}$ at $T=\SI{1.5}{\K}$ \cite{Shiomi:2019aa}.
We use $T_{2e}=1 \,\si{\ns}$ for estimating the sensitivity at $T=\SI{0.1}{\K}$ in the following.

Our strategy is to use a microwave cavity strongly coupled to the magnetization of the material \cite{Barbieri:2016vwg, Chigusa:2020gfs}.
When the DM mass is the same as the excitation energy of the homogeneous magnetic precession, the collective motion of spins is resonantly excited.
With the setup of the microwave cavity strongly coupled to the magnetization of the material, half of the energy of the magnetic resonance is transferred to the excitation of cavity photons which can be detected as the signal.
The readout can be done by a linear amplifier or photon counter, which are discussed in Secs. \ref{linearamp} and \ref{singlephotoncounter}, respectively.
However, our aim is to utilize the latter, a photon counter, to avoid the quantum limit that significantly suppresses the sensitivity at the frequency of interest when we use a linear amplifier \cite{PhysRevD.88.035020}.

The microwave cavity frequency should be tuned to the electron precession frequency, realizing a strong coupling between the magnet and the cavity.
The overall relaxation time of a single mode of this coupled system is given by
\begin{equation}
T_\mathrm{coupled}=2 \left( \frac{1}{T_{2e}} + \frac{1}{\tau_\mathrm{cav}}\right)^{-1} \approx 2 T_{2e}=\SI{2}{\ns}, \label{Tcouple}
\end{equation}
where $\tau_\mathrm{cav}$ is the cavity lifetime, which we assume to be $O(1)\,\si{\us}$ as in Ref.~\cite{Barbieri:2016vwg}. 
The strong coupling between the magnet and the cavity affects the signal calculation discussed in Sec.~\ref{subsec: response} in several ways.
Firstly, the relaxation timescale $T_{2e}$ should be replaced by the coupled scale: $T_{2e} \rightarrow T_\mathrm{coupled}$.
Secondly, the interaction strength between a coupled mode and the DM field is reduced by a factor of $\sqrt{2}$ due to the maximal mixing, which results in a $50\%$ reduction of power absorbed to the material when only a single mode is excited.\footnote{However, with $T_\mathrm{coupled}\approx 2 T_{2e}$, the two mentioned contributions cancel with each other, giving the same absorption power $P_\mathrm{absorb}$ as originally derived in Sec.~\ref{subsec: response}.}
Finally, the coupled relaxation time $T_\mathrm{coupled}$ also determines the bandwidth of the magnon-cavity coupled mode:
\begin{equation}
\Delta \omega_\mathrm{coupled}=2/T_\mathrm{coupled} \approx 1/T_\mathrm{2e}.
\end{equation}

Similarly to the case of the nuclear-frequency range, the following steps can be used to scan the DM mass over the frequency range $\omega / 2 \pi=45$--$\SI{55}{\GHz}$:
\begin{enumerate}
\item Match eigenfrequencies of the magnetic system and the cavity photon to the frequency $\omega$ of interest at the same time, realizing the strong coupling between them. The former can be tuned by changing the external static field $H_0$, while the latter by changing the size of the cavity.
\item Measure the power of cavity photons induced by the magnetic material $\ce{MnCO_3}$ for an interrogation time $\Delta t$.
\item Shift the frequency $\omega$ by an interval $\Delta \omega=\Delta \omega_\mathrm{coupled}$ after each step of the signal measurement until the frequency range of interest is fully covered.
\end{enumerate}
The interrogation time $\Delta t$ is determined by the total observation time $T_\mathrm{total}$ and the number of scan steps, which is fixed from $\Delta \omega$ and the whole frequency range of interest.

\subsubsection{Microwave cavity with linear amplifier} \label{linearamp}

Using a linear amplifier for probing $\omega_\mathrm{DM}/2\pi\sim \SI{50}{\GHz}$, the effective temperature of the quantum noise is $T_Q=\hbar \omega_\mathrm{DM}/k_B \simeq \SI{2.3}{\K}$.
Therefore, under an experimental setup with the temperature $T=\SI{0.1}{\K}$, we assume the quantum noise dominates over others.

Let us assume that the resonance condition is satisfied: $\omega_\mathrm{DM} \approx \omega_{\tilde{e},+}$. 
Input power is discussed in Sec.~\ref{subsec: response} with some subtle care from the magnet-cavity mixing mentioned below Eq.~\eqref{Tcouple}.
In the case of the strong coupling where half of the power is transferred to the cavity \cite{Barbieri:2016vwg, Chigusa:2020gfs}, we have $P_\mathrm{out}= P_\mathrm{absorb}/2.$

The following is the output power calculated at $\omega_\mathrm{DM}/2\pi=\SI{48}{\GHz}$.
For detecting the axion-induced field through the axion-proton coupling $g_{app}$ and axion-electron coupling $g_{aee}$, $\chi_n^{zy}(\omega_{\tilde{e},+})$ and $\chi^{zy,zz}_e(\omega_{\tilde{e},+})$ are the most important, respectively; the corresponding output powers are
\begin{equation}
P^{g_{app}}_\mathrm{out}= \SI{2.2e-26}{\uW} \times \left( \frac{W_{\ce{MnCO_3}}}{\SI{1}{\kg}} \right) \left( \frac{T_{2e}}{\SI{1}{\ns}} \right) \left( \frac{g_{app}}{10^{-10}} \right)^2  \left( \frac{v_\mathrm{DM}/c}{10^{-3}}\right)^2
\left( \frac{\cos^2 \theta}{1/3} \right),
\end{equation}
and
\begin{equation}
P^{g_{aee}}_\mathrm{out}= \SI{5.3e-20}{\uW} \times \left( \frac{W_{\ce{MnCO_3}}}{\SI{1}{\kg}} \right) \left( \frac{T_{2e}}{\SI{1}{\ns}} \right) \left( \frac{g_{aee}}{10^{-12}} \right)^2  \left( \frac{v_\mathrm{DM}/c}{10^{-3}}\right)^2
\left( \frac{\cos^2 \theta}{2/3} \right).
\end{equation}
In the case of dark photon DM, the dominant channels are $\chi^{zy,zz}_e(\omega_{\tilde{e},+})$, and we obtain the output power
\begin{equation}
P^{\epsilon}_\mathrm{out}= \SI{2.4e-21}{\uW} \times \left( \frac{W_{\ce{MnCO_3}}}{\SI{1}{\kg}} \right) \left( \frac{T_{2e}}{\SI{1}{\ns}} \right) \left( \frac{\epsilon}{10^{-12}} \right)^2  \left( \frac{v_\mathrm{DM}/c}{10^{-3}}\right)^2
\left( \frac{\cos^2 \theta}{2/3} \right)
\left( \frac{\sin^2 \varphi}{1/2} \right).
\end{equation}

On the other hand, the power of the quantum noise calculated at $\omega_\mathrm{DM}/2\pi=\SI{48}{\GHz}$ is
\begin{equation}
P_\mathrm{noise}=\hbar \omega_\mathrm{DM} \sqrt{\frac{\Delta f}{\Delta t}} \simeq \SI{5.7e-16}{\uW},
\end{equation}
with the condition of using $T_\mathrm{total}= 1 \,\mathrm{year}$ to probe the frequency range $45$--$\SI{55}{\GHz}$ (and hence fixing the value of $\Delta f/\Delta t$).
We can see that this method is not sensitive enough for detecting the DM signal, suffering from the quantum noise from the linear amplifier. 

\subsubsection{Microwave cavity with photon counter} \label{singlephotoncounter}
Instead of using a linear amplifier, we may utilize a single photon counter avoiding the standard quantum limit \cite{PhysRevD.88.035020}.
In this case, the only relevant noise is the shot noise of the cavity photon of frequency $\omega=\omega_\mathrm{DM}$, whose rate is determined by the effective relaxation time of the coupled system $T_\mathrm{coupled}$ and the cavity temperature $T_c$ as \cite{Barbieri:2016vwg}
\begin{equation}
R_\mathrm{th}= \frac{2/T_\mathrm{coupled}}{e^{\hbar \omega /k_BT_c}-1}.
\end{equation}

In this setup, we define the SNR as
\begin{equation}
\mathrm{SNR} \equiv \frac{R_\mathrm{signal}  \Delta t}{\sqrt{R_\mathrm{th} \Delta t}}.
\end{equation}
The constraints on DM couplings are given by requiring $\mathrm{SNR} \gtrsim 1$.
The output power $P_\mathrm{out}$ transferring to the cavity photon system is related to the excitation rate $R_\mathrm{signal}$ (of spin-flip-induced photons) by the relation 
\begin{equation}
R_\mathrm{signal}\equiv P_\mathrm{out}/\hbar \omega_\mathrm{DM}.
\end{equation}

The sensitivities, based on the sensitive channels of $\ce{MnCO_3}$ discussed in Sec.~\ref{subsec: response}, are shown in Figs.~\ref{fig:gapp reach}, \ref{fig:gaee reach}, and \ref{fig:epsilon reach}  together with the relevant constraints.
We take $T_\mathrm{total}=1\,\mathrm{year}$ for probing the range $\omega/2\pi=45$--$\SI{55}{\GHz}$, and assume the relaxation time $T_{2e}=\SI{1}{\ns}$ as a realistic case. In addition, we add the case with $T_{2e}=\SI{10}{\ns}$, illustrating the optimistic situation when inhomogeneity in the material and the applied field could be reduced and hence the effective magnetic relaxation time is increased.
The interrogation time at each scan step is $\Delta t=\SI{5e5}{\s}$ and $\Delta t=\SI{5e4}{\s}$ for the two cases, respectively.
Again, we assume the total \ce{MnCO_3} mass of $\SI{1}{\kg}$.

\section{Conclusion and discussion}
\label{sec: discuss and conclusion}

We have discussed a possibility of detecting axion or dark photon DM using nuclear magnetization in a magnet, which has a strong hyperfine coupling between electron and nuclear spins.
We focus on the canted antiferromagnet $\ce{MnCO_3}$ as a concrete example of such a magnet.
Both axion and dark photon DMs may interact with nuclear and electron spins and exert (effective) oscillating magnetic fields on them.
This can induce magnetic resonance when the DM mass is equal to the excitation energy of the magnet.
Owing to strong magnetic fields generated by electron spins through the hyperfine interaction, nuclear spins are highly polarized and can give a sizable resonance signal under relatively low external magnetic field.
There is also an electron magnetic resonance, which is also sensitive to DM.
Owing to the hyperfine interaction, the nuclear and electron resonance modes are mixed with each other.
Specifically in $\ce{MnCO_3}$, both modes are sensitive to frequencies much higher than typical values achieved by ordinary nuclear spin precession experiments.
Other materials such as \ce{CsMnF3} \cite{PhysRev.132.144, PhysRev.143.361, PhysRev.156.370},
\ce{CoCO3} \cite{borovik1984spin}, \ce{FeBO3} \cite{borovik1984spin}, and
\ce{Nd_2CuO_4} \cite{Chattopadhyay_1995,Chatterji_2013} are expected to share the same properties. 

We investigated the observable magnetization and energy stored in the magnetic material due to the DM-induced oscillating magnetic field. 
The details of the measurement setup and strategy are discussed in Sec.~\ref{sec: sensitivity} for two frequency ranges: the nuclear frequency $\omega_\mathrm{DM}  / 2 \pi \sim 500$--$\SI{600}{\MHz}$ corresponding to $m_\mathrm{DM}\sim \SI{e-6}{eV}$ (Sec.~\ref{subsec: nuclear-freq mode}) and the electron frequency $\omega_\mathrm{DM} / 2 \pi \sim 45$--$\SI{55}{\GHz}$ corresponding to $m_\mathrm{DM}\sim \SI{e-4}{eV}$ (Sec.~\ref{subsec: electron-freq mode}).
As detection schemes, a pick-up loop with \textit{LC} resonant circuit and cavity photons coupled with the spin systems are considered for the former and latter channels, respectively.
The sensitivity for the 1-year measurement with $\SI{1}{\kg}$ of $\ce{MnCO_3}$ under temperature $\SI{0.1}{\K}$ is shown in Figs.~\ref{fig:gapp reach}, \ref{fig:gaee reach}, and \ref{fig:epsilon reach}. We summarize the results in the following.
 
\begin{itemize}
    \item 
This system is sensitive to the axion-proton interaction. The axion of mass $\sim \SI{e-6}{eV}$ with $g_{app}\gtrsim 10^{-11}$ could be probed by our method through the nuclear-dominated mode. So far it is hardly reached by other experiments. 
For example, a nuclear-precession experiment like CASPEr is sensitive to lower mass regions due to the limitation of the magnitude of the external magnetic field.
We have also examined the electron magnetic resonance excited through the mixing with the nuclear spins, which indeed has a sensitivity to the nucleon-DM interaction at the mass scale $\sim \SI{e-4}{eV}$ or $O(10) \,\si{\GHz}$. However, it turns out that this electron-dominated mode is not sensitive enough to constrain the axion parameter space beyond astrophysical constraints, because the signal is suppressed by the short relaxation scale $T_{2e}$ of the material. Other materials with longer relaxation times should be able to achieve better sensitivity, which is left as a future work.

\item The system is also sensitive to the axion-electron interaction assuming an optimistic magnetic relaxation time achieved by small inhomogeneity of the material and the applied magnetic field.
The nuclear-dominated mode can be used to explore the axion parameter region with $g_{aee}\approx10^{-14}$ and $m_a \sim \SI{e-6}{eV}$.
This region still survives the astrophysical constraints from stellar evolution such as white dwarf luminosity function and the tip of the red giant branch.
In addition, by using the electron magnetic resonance mixed with nuclear spins, we can search for the axion mass of $\SI{e-4}{eV}$ with parameter $g_{aee}\sim 10^{-13}$. 

\item For the kinetic mixing parameter $\epsilon$ of the dark photon, there are already haloscope experiments that are searching for the mass range $m_{\gamma'}\sim \SI{e-6}{eV}$ corresponding to the frequency regions to which the nuclear-dominated mode is sensitive.
Even though we can only probe the already scanned regions with our method, a complementary check of such regions is plausible for $\epsilon \sim 10^{-13}$.
In addition, with the electron magnetic resonance mixed with nuclear spins, we might be able to probe the parameter region down to $\epsilon \sim 10^{-13}$ at $m_{\gamma'}\sim \SI{e-4}{eV}$, which is beyond the current experimental limits and cosmological constraints.
\end{itemize}

As shown above, DM detection with a magnet with a strong hyperfine interaction provides possibilities to detect both axion DM and dark photon DM.
The unique feature of DM detection via the proposed system is the allowed spin transfer between electron and nuclear spins, which provides several available channels and supports the readout of the nuclear signal relying on more accessible electron spins.
Nevertheless, further study is still needed to investigate its real potential and improvability. 
A further study of parameters of magnetic materials is encouraged, such as the magnetic relaxation time $T_{2n,2e}$ and the susceptibility of materials with a strong hyperfine coupling other than $\ce{MnCO_3}$. 
Taking into account the statistical behavior of the DM field would also help one to estimate the result more precisely. 
Further study to cure the weak points associated with the limited band of nuclear-dominated modes and the short relaxation time of electron-spin systems is also definitely useful to maximize the competence of this method.

%%%%%%%%%%%%%%%%%%%%%%
\section*{Acknowledgment}
%%%%%%%%%%%%%%%%%%%%%%
We thank Takahiko Makiuchi, Takashi Kikkawa and Eiji Saitoh for the valuable information about nuclear magnon and magnetic excitation in \ce{MnCO_3}.
S.C. is supported by the Director, Office of Science, Office of High Energy Physics of the U.S. Department of Energy under the Contract No. DE-AC02-05CH1123. 
T.M. is supported by JSPS KAKENHI Grant Number 22H01215. T.S. is supported by the JSPS fellowship Grant number 23KJ0678. 
This work is supported by World Premier International Research Center Initiative (WPI), MEXT, Japan.

\newpage

\appendix

%%%%%%%%%%%%%%%%%%%%%%%%%%%%%%%%%%%%%%%%%%%%%%%%%%%%%%%%%%%%%%%%%%%%%%%%%%%%%%%%%%%%%%%%%%%%%%%%%%%%%%%%%
\section{Exotic interaction between nuclear spins of \ce{Mn} and dark matter} \label{exotic interaction}
\setcounter{equation}{0}
%%%%%%%%%%%%%%%%%%%%%%%%%%%%%%%%%%%%%%%%%%%%%%%%%%%%%%%%%%%%%%%%%%%%%%%%%%%%%%%%%%%%%%%%%%%%%%%%%%%%%%%%%

In this Appendix, we consider the interaction between nuclear spin $\vec{I}$ and axions based on the interaction between the nucleon spin and axion given by Eq.~\eqref{spincouplingaNN}. We mainly follow Ref.~\cite{Kimball_2015} in the discussion below. Natural unit is adopted in this Appendix.

Both the spin and orbital angular momentum of protons and neutrons contribute to the nuclear spin $\vec I$ and nuclear magnetic moment $\vec \mu_I$:
\begin{gather}
\vec{I}=\vec{S}_p+\vec{S}_n+\vec{L}_p+\vec{L}_n,\\
\vec{\mu}_I=g_p \mu_\mathrm{N}  \vec{S}_p+g_n \mu_\mathrm{N}\vec{S}_n + g_{lp} \mu_\mathrm{N}  \vec{L}_p+g_{ln}\mu_\mathrm{N}\vec{L}_n,
\end{gather}
where $\vec{S}_{p,n}$ and $\vec{L}_{p,n}$ are the total spin and orbital angular momentum operators for protons or neutrons, respectively. Spin and orbital $g$ factors are given by $g_p=5.586, g_n=-3.826, g_{lp}=1$, and $g_{ln}=0$.
For convenience, using contribution variable $\sigma_\xi \equiv \langle \vec{I} \cdot \vec{\xi}\rangle / \langle \vec{I}\cdot \vec{I} \rangle $ with the expectation value taken with respect to a nuclear state, and defining the nuclear $g$ factor $g_I$ from $\langle \mu_{zI} \rangle=g_I\mu_\mathrm{N} \langle  I_z\rangle$, we obtain the relation
\begin{gather}
1= \sigma_p+\sigma_n+\sigma_{lp}+\sigma_{ln}, \label{Iconserved}\\
g_I=g_p\sigma_p+g_n \sigma_n + g_{lp}\sigma_{lp}+ g_{ln} \sigma_{ln}. \label{nuclear g-factor}
\end{gather}

Now let us discuss the axion interaction with the nuclear spin. One should note that the axion only couples to the nucleon spin, not to its orbital motion.
Starting from the axion interaction of the form $ (g_{app} \vec{S}_p+g_{ann} \vec{S}_n)\cdot \vec \nabla a$, we obtain the interaction between the nuclear spin $\vec{I}$ and the axion of the form
\begin{equation}
\mathcal{L}_\mathrm{int}=\frac{\tilde{g}_{aI}}{m_N} ( \vec{\nabla} a) \cdot  \vec{I} = \frac{1}{g_I \mu_\mathrm{N} } \frac{\tilde{g}_{aI}}{m_N} (\vec{\nabla}a) \cdot \vec{\mu}_I ,
\end{equation}
where $\tilde{g}_{aI}\equiv g_{app}\sigma_p+g_{ann}\sigma_n$.
Matching this to the magnetic interaction of the form $\mathcal{L}_\mathrm{int}=\vec{\mu}_I \cdot\vec{h}$, with $\vec h$ being the magnetic field and $\vec{\mu}_I$ the nuclear magnetic moment, we can define the effective magnetic field caused by the axion and felt by the nucleus as
\begin{equation}
\vec{h}_I^\mathrm{axion}= \frac{1}{g_I \mu_\mathrm{N} }\frac{\tilde{g}_{aI}}{m_p} \sqrt{2\rho_\mathrm{DM}} \vec{v}_\mathrm{DM}   \sin \left(m_a t + \delta \right),
\end{equation}
where $\rho_\mathrm{DM}$ and $\vec{v}_\mathrm{DM}$ are the density and velocity vector of DM, respectively. We have used $\vec{\nabla} a \approx m \vec{v}_\mathrm{DM} a(t)$ and assumed that DM is composed solely of axions, hence $\rho_\mathrm{DM}=m^2_a a_0^2/2$ where $a_0$ is the amplitude of axion coherent oscillation.

The remaining task is to find a spin contribution for the case of the magnetic isotope of interest; here we focus on \ce{^{55}Mn}. 
For example, we can adopt the semiempirical approach following Engel and Vogel \cite{PhysRevD.40.3132}, by assuming that the nucleon species with even number (in this case, neutrons) contributes little to the angular momentum of the system, and taking $g_I$ as the observed value.
With $g_I=1.38$ as the observed nuclear $g$ factor for \ce{^{55}Mn} and $\sigma_n = 0, \sigma_{ln}=0$, we obtain the spin contribution as
$
\sigma_p \approx 0.1, \sigma_{lp}\approx 0.9.
$

However, there are some uncertainties in the above semiempirical estimation. One of them comes from the deformed shell structure of the \ce{^{55}Mn} nucleus \cite{Ye_2023,PhysRevC.103.014303}, which implies the collective rotation of the nucleus in addition to the intrinsic state of nucleons. This can affect the ratio of the spin and orbital contribution to nuclear spin.
In particular, concerning the $z$-direction contribution to the magnetic moment, the nuclear magnetic moment is given by \cite{nilsson1955binding, bohr1998nuclear}
\begin{equation}
\mu=\frac{I}{I+1} \langle \Omega | \mu_z |  \Omega \rangle + \frac{1}{I+1}g_RI.
\end{equation}
The first term is the magnetic moment from the angular momentum of the proton in the state $| \Omega \rangle$, while the second term accounts for the collective rotation of nucleons with the rotational $g$ factor $g_R$. 
Owing to the deformed configuration of the \ce{^{55}Mn} nucleus, particles experience axially symmetric potential (with $z$ being the axial direction), and degenerate spherical shells split.
Almost all nucleons couple in pairwise contributing little to the angular momentum, and the last odd proton occupies the Nilsson state with the asymptotic quantum number $[312]5/2^-$.
(For the Nilsson state $[N n_z \Lambda]\Omega^\pi$ of a nucleon \cite{nilsson1955binding}, $N$ is the total quantum number, $n_z$ the number of oscillator quanta in the $z$ direction, $\Lambda$ the orbital angular momentum in the $z$ direction, $\Omega$ the total angular momentum in the $z$ direction, and $\pi$ the parity.)
This last odd proton and collective rotation of nucleons contribute to the nuclear spin $I=5/2$.
For the ground state of \ce{^{55}Mn}, the state of the last proton can be approximated by $ | \Omega \rangle=|l_z=2,S_z=1/2 \rangle$. Since
\begin{equation}
\langle \Omega | \mu_z |  \Omega \rangle= \frac{1}{2}g_p+2 g_{lp},
\end{equation}
and the rotation $g$ factor $g_R$ of the nucleus should not have a contribution from the spin of the nucleons,
we know that 
$
\sigma_p =0.5/ (I+1) \approx 0.14
$ is the proton-spin contribution realizing the magnetic moment $\mu$.
The value of $\sigma_p$ from this deformed shell model is slightly higher than the one obtained from the simple semiempirical model.

Another uncertainty is the neutron-spin contribution which has been neglected so far. For the axion coupling to the nuclear spin, the above two approximations imply that the \ce{^{55}Mn} nucleus is sensitive to the axion-proton coupling constant $g_{app}$, but not to the axion-neutron coupling $g_{ann}$.
However, in reality there might also be a significant spin contribution from the core neutrons through its spin polarization via the spin-spin interaction \cite{PhysRevC.73.055501}.
It may make the \ce{^{55}Mn} nucleus also sensitive to the axion-neutron coupling. In this paper, we neglect this effect for simplicity.

In this work, we use the value
\begin{equation}
\sigma_p=0.1,\quad \sigma_n=0,
\end{equation}
to numerically estimate the magnitude of the axion-induced magnetic field acting on the \ce{^{55}Mn} nuclear spin.

 %%%%%%%%%%%%%%%%%%%%%%%%%%%%%%%%%%%%%%%%%%%%%%%%%%%%%%%%%%%%%%%%%%%%%%%%%%%%%%%%%%%%%%%%%%%%%%%%%%
\section{Detailed calculation of magnetic dynamics in \ce{MnCO3}} \label{appendix: classic}
\setcounter{equation}{0}
%%%%%%%%%%%%%%%%%%%%%%%%%%%%%%%%%%%%%%%%%%%%%%%%%%%%%%%%%%%%%%%%%%%%%%%%%%%%%%%%%%%%%%%%%%%%%%%%%%%

In this Appendix we calculate the magnetic dynamics of \ce{MnCO3} under the DM background.
In Appendices \ref{appendix: classic} and \ref{appendix: magnonpic}, SI unit is adopted with vacuum permeability factor $\mu_0$ is omitted in the formulas for convenience.

To estimate the eigensystem of the magnetic system in \ce{MnCO3}, we first note that there is the orders-of-magnitude relation of the effective fields and other constants:
\begin{gather}
H_E \rightarrow H_E, \quad H_n \rightarrow H_n, \quad \gamma_e \rightarrow \gamma_e\\
H_0 \rightarrow H_0 \delta, \quad H_D \rightarrow H_D \delta, \quad H_K\rightarrow H_{K}\delta\\
H_a \rightarrow H_a \delta^2, \quad H_{K'} \delta^2, \quad \gamma_n \rightarrow \gamma_n \delta^2
\end{gather}
where $\delta \sim 10^{-2}$ roughly accounts for the hierarchy among these parameters. Below we will consider only the leading order result in $\delta$, while $\delta$ appearance is omitted for notational simplicity. 

Recall that the potential of the system and equation of motions are given by Eqs.~\eqref{potentialpervolume} and \eqref{EOMgeneral}. The dynamics of the magnetic system concerning only in-phase modes following Eqs.~\eqref{M-inphase-EOM} and \eqref{m-inphase-EOM} reads as
\begin{equation}
\frac{d \vec{u} (t) }{dt} = \Omega \vec{u}(t) + \vec{R}(t) +   \vec{D}(t),
\end{equation}
where $\vec{u} \equiv \left( M^x_{+} , M^y_{+} , m^x_{+} , m^y_{+} \right)^T$ represents a collect of the perturbation of magnetization parameters defined by Eq.~\eqref{defMnpm}, and at leading order
\begin{equation}
\Omega=
\begin{pmatrix}
0& -2 \gamma_e H_E& 0 &\gamma_e H_n \\
{\gamma_e^2 \left( H_0(H_0+H_D) +2  H_E (H_a+H_{K'})\right)}/{2 \gamma_e H_E} &0& -\gamma_e H_n&\\
0&-\gamma_n H_a&0& \gamma_n H_n\\
\gamma_n H_a & 0 & -\gamma_n H_n & 0
\end{pmatrix}.
\end{equation}
Term $\vec{R}$ represents relaxation of the magnetizations. $\vec{D}$ is source term due to the oscillating magnetic field induced by DM and is given by
\begin{equation}
\vec{D}(t)=
\begin{pmatrix}
h_e^{y} M_0\\
h_e^{z} M_0 \sin \psi\\
h_n^{y} m_0\\
h_n^{z} m_0 \sin \psi
\end{pmatrix} \left(e^{i\omega_\mathrm{DM} t}+e^{-i\omega_\mathrm{DM} t}\right).
\end{equation}

In the case that $\vec{R}=0$ and $\vec{D}=0$, with the ansatz 
\begin{equation}
\vec{u}(t)=\vec{u}e^{i\omega t},
\end{equation}
the equation of motion reads as $i\omega \vec{u}=\Omega \vec{u}$ of which the solution gives eigenvalues of the precession mode of magnetization as
\begin{equation}
\omega_{1,3}=\pm \omega_{\tilde{n},+}, \quad  \omega_{2,4}=\pm \omega_{\tilde{e},+},
\end{equation}
where $\omega_{\tilde{n},+}$ and $\omega_{\tilde{e},+}$ are given by Eqs.~\eqref{ninMnCO3} and \eqref{einMnCO3}.
The corresponding eigenmodes are
\begin{gather}
\vec{u}_{1}=\vec{u}_3^*= \left( -\frac{2 i H_E \gamma_e^2 \omega_n^2}{\gamma_n \omega_{\tilde{n},+} \omega^2_{e,+}}, - \frac{\omega_n \left( 2H_E\gamma_e \omega_n - \omega^2_{e,+}\right)}{2H_E \gamma_n \omega_{e,+}^2} , -\frac{i \omega_n}{\omega_2},1 \right)^T, \\
\vec{u}_{2}= \vec{u}_4^*= \left(\frac{i \omega_{e,+}}{H_a \gamma_n}, \frac{\omega_{e,+}^2}{2H_a H_E \gamma_e \gamma_n}, -\frac{i (2 H_E \gamma_e \omega_n-\omega_{e,+}^2)}{2H_E \gamma_e \omega_{e,+}},1 \right)^T,
\end{gather}
up to an arbitrary normalization factor.

Now, we take into account the damping effect $\vec{R}$ and source term $\vec{D}$. We adopt here for simplicity the relaxation term of the form
\begin{equation}
\vec{R}=
-u_i/T_{i}, \label{dampingR}
\end{equation}
where $T_{1,3}=T_{2n}$ and $T_{2,4}=T_{2e}$ denoting the relaxation timescale of the nuclear-dominated mode and electron-dominated mode, respectively.
With ansatz
\begin{equation}
\vec{u} (t)=\sum_i c_i \vec{u}_i e^{i\omega_\mathrm{DM} t}+\mathrm{H.c.},  \label{steadysolutionclassic}
\end{equation}
the equation of motion reduces to the problem finding $c_i$'s satisfying $\sum_i \left(i( \omega_\mathrm{DM} - \omega_i ) + 1/T_{i} \right) c_i \vec{u}_i=\vec{D}$.

The signal magnetization $(M_\mathrm{signal}  \equiv M^z_\mathrm{total}  )$ is then given by
\begin{align}
M_\mathrm{signal} (t)&=\sin \psi \sum_i c_i (\vec{u}_i)_1 e^{i\omega_\mathrm{DM} t}+\mathrm{H.c.}
\end{align}
When $\omega_\mathrm{DM} \sim \omega_1$, the $i=1$ term is dominant with corresponding steady-state solution of the form
\begin{equation}
c_1 \propto \frac{T_{2n}}{(\omega_1-\omega_\mathrm{DM})^2 T_{2n}^2+1}.
\end{equation}
It is approximately the solution at time $ t \gtrsim T_{2n} $ after begin excited. On the other hand, when $\omega_\mathrm{DM} \sim \omega_2$ the $i=2$ term is dominant with steady-state solution of the form
\begin{equation}
c_2 \propto \frac{T_{2e}}{(\omega_2-\omega_\mathrm{DM})^2 T_{2e}^2+1},
\end{equation}
which is a good approximation for the solution at the time $t  \gtrsim T_{2e}$ after being excited. That is the resonance occurs when $| \omega_\mathrm{DM} - \omega_{\tilde{n},+}| \lesssim 1/T_{2n}$ or $| \omega_\mathrm{DM} - \omega_{\tilde{e},+}| \lesssim 1/T_{2e}$.
Note that when $t \lesssim \tau_\mathrm{DM}$, the magnetization rotates with a coherent phase with frequency $\omega_\mathrm{DM}$, where $\tau_\mathrm{DM}$ is the coherence time of DM assumed to be larger than $T_{2n,2e}$. Owing to the velocity distribution of DM, the signal spreads with bandwidth $\Delta \omega_\mathrm{DM}$ in the frequency space.
For the power estimation, we take into account the relevant component of magnetization and use Eq.~\eqref{Pabsorddef}. The signal susceptibility and power are listed in Tables~\ref{table: NMR}, \ref{table: FMR},  and \ref{table: NMRpower}, \ref{table: FMRpower}, respectively.

If one uses the relaxation term $R=\left(-1/\tau_{2n},-/\tau_{2n}, -/\tau_{2e}, -\tau_{2e} \right)^T$ of Bloch's kind, the effective relaxation time $T_{2n},T_{2e}$ can be expressed as
\begin{gather}
T_{2n}=\frac{\tau_{2n} \tau_{2e} (H_a \gamma_e \omega_n (2H_E \gamma_e \omega_n-\omega_{e,+}^2)+\omega_{e,+}^4)}{\tau_{2e}\omega_{e,+}^4+H_a \tau_{2n} \gamma_e \omega_n (2H_E\gamma_e \omega_n - \omega_{e,+}^2)},
\\
T_{2e}=\frac{\tau_{2n} \tau_{2e} (H_a \gamma_e \omega_n (2H_E \gamma_e \omega_n-\omega_{e,+}^2)+\omega_{e,+}^4)}{\tau_{2n}\omega_{e,+}^4+H_a \tau_{2e} \gamma_e \omega_n (2H_E\gamma_e \omega_n - \omega_{e,+}^2)} .
\end{gather}

%%%%%%%%%%%%%%%%%%%%%%%%%%%%%%%%%%%%%%%%%%%%%%%%%%%%%%%%%%%%%%%%
\section{Nuclear magnon picture}  \label{appendix: magnonpic}
%%%%%%%%%%%%%%%%%%%%%%%%%%%%%%%%%%%%%%%%%%%%%%%%%%%%%%%%%%%%%%%%

In this Appendix, we discuss the dynamics of the magnetic system of \ce{MnCO3} using the quantum magnon picture. Based on the following derivation, we check that the response derived in the magnon picture is consistent with that derived in the classical picture illustrated in the main text (Sec.~\ref{sec: nuclear MD}).

\setcounter{equation}{0}
\subsection{Hamiltonian in the magnon picture}

The following is the Hamiltonian for the spin of \ce{MnCO3} in the antiferromagnetic phase \cite{doi:10.1143/JPSJ.15.2251, PhysRev.136.A218,turov1973nuclear, PhysRevB.97.024425}:
\begin{align}
\begin{split}
\mathcal{H}=&2J \sum_{ i,j  \neq i} \vec{S}_i \cdot \vec{S}_j + 2 \sum_{  i,j\neq i } \vec{S}_j \cdot(\vec{S}_i \cross \vec{e}_y D)\\
& + \frac{K}{2} \left[  \sum_i (S_i^y)^2 + \sum_j (S_j^y)^2  \right]- \frac{K'}{2} \left[ \sum_i (S_i^z)^2 + \sum_j (S_j^z)^2 \right] \\
& + \gamma_e \hbar \left[ \sum_i \vec{S}_i + \sum_j \vec{S}_j  \right] \cdot (\vec{H}+\vec{h}_e) - \gamma_n \hbar \left[ \sum_i \vec{I}_i  + \sum_j \vec{I}_j  \right] \cdot (\vec{H}+\vec{h}_n) \\
&+ \tilde{A}_\mathrm{hy} [ \sum_i \vec{S}_i \cdot \vec{I}_i + \sum_j \vec{S}_j \cdot \vec{I}_j],
\end{split} \label{hamill}
\end{align}
where $i$ and $j$ represent lattice sites in two sublattices; $J,K,K'$, and $\tilde{A}_\mathrm{hy}$ are positive constants; and $\gamma_n,\gamma_e$ are gyromagnetic ratios for nuclear and electron spin, respectively.
The Hamiltonian includes (1) isotropic exchange interaction between electron spins of nearest sites between $\vec{S}_i$ and $\vec{S}_j$; (2) Dzyaloshinskii--Moriya interactions; (3) easy-plane anisotropy; (4) in-plane uniaxial anisotropy; (5) Zeeman effects for electron spins $\vec{S}$ and nuclear spins $\vec{I}$; and (6) hyperfine interaction between electron spins $\vec{S}$ and nuclear spins $\vec{I}$. Here, we discuss the situation where there are static fields 
\begin{equation}
\vec{H}=H_0 \vec{e}_x
\end{equation}
applied in the $x$ direction and an exotic oscillating field induced by DM of which $\vec{h}_e(t)$ is interacting with electron spin, and $\vec{h}_n(t)$ is interacting with nuclear spin. The coordinate is set to coincide with the situation discussed in the main text using classical theory. The relations between spins and magnetizations are given by Eqs.~\eqref{MSrelation} and \eqref{mIrelation}. 
It reduces to the classical Hamiltonian (\ref{potentialpervolume}) when all the spins in each sublattice are aligned, which gives the relation
\begin{gather}
H_E=\frac{2JzS}{\gamma_e \hbar}, \quad H_D=\frac{2DzS}{\gamma_e \hbar}, \quad
H_{K'}=\frac{K' S}{\gamma_e\hbar},\quad H_K=\frac{KS}{\gamma_e\hbar},\quad A_\mathrm{hy}M_0= \frac{\tilde{A}_\mathrm{hy}S}{\gamma_n\hbar}.
\end{gather}
Note that the direction between magnetization and spin of electron is opposite while we take gyromagnetic ratio parameter $\gamma_e,\gamma_n$ to be positive.

In the ground state, electron spins align in the $xz$ plane pointing (almost in) to $+z$ and $-z$ direction for each sublattice due to easy-plane and in-plane anisotropy. They slightly tilt toward the $-x$ direction due to the applied static field $H_0$ and the Dzyaloshinskii-Moriya interactions. For nuclear spins, they align opposite to the electron spin of the same \ce{Mn} due to hyperfine interaction. The configuration (of magnetizations) is shown in Fig.~\ref{fig: groundstateM}.

For convenience to consider fluctuation around ground state, we rotate our spin parameters for each sublattice:
\begin{subequations}  \label{coordinate trans S}
\begin{gather}
S_i^x=S_i^{x_1} \cos \psi + S_i^{z_1} \sin \psi,\\
S_i^z=-S_i^{x_1} \sin \psi + S_i^{z_1} \cos \psi,\\
S_j^x=-S_j^{x_2} \cos \psi + S_j^{z_2} \sin \psi, \\
S_j^z= -S_j^{x_2} \sin \psi - S_j^{z_2} \cos \psi,
\end{gather}
\end{subequations}
where the ground state expectation value gives $\langle S_i^{z_{1}}\rangle =\langle S_j^{z_{2}}\rangle =-S=-5/2$ and $\langle S_{i,j}^{x_{1,2},y} \rangle=0$. The coordinate transformation is shown in Fig.~\ref{fig: coortransf}.
Owing to hyperfine interaction, at ground state nuclear spins lie in the opposite direction to the electron spins of the same sites and similarly form two sublattices. We perform the same coordinate transformation for nuclear spins $\vec{I}$ (Eq.~\eqref{coordinate trans S} but with $S\leftrightarrow I$).
In this case, the thermal expectation value for nuclear spin temperature $T$ is 
\begin{equation}
\langle I_i^{z_{1}}\rangle =\langle I_j^{z_{2}}\rangle =  \langle I\rangle
\end{equation}
given by Eq.~\eqref{Iavg}, and $\langle I_{i,j}^{x_{1,2},y} \rangle=0$. The function $\langle I \rangle/I$ is shown in Fig.~\ref{fig: BF}.
Now the hyperfine interaction can be written in the form
\begin{equation}
\mathcal{H}_\mathrm{hy}= \mathcal{H}_{\parallel} +\mathcal{H}_{\mathrm{mix}},
\end{equation}
where
\begin{gather}
\mathcal{H}_{\parallel} =\tilde{A}_\mathrm{hy} \sum_i \left(  S^{z1}_i I^{z1}_i  \right)+ \sum_j \left(  S^{z2}_j I^{z2}_j \right), \label{Hparahy}\\
\mathcal{H}_{\mathrm{mix}}=\tilde{A}_\mathrm{hy} \sum_i \left(  S^{x1}_i I^{x1}_i + S^{y}_i I^{y}_i \right)+ \sum_j \left(  S^{x2}_j I^{x2}_j + S^{y}_j I^{y}_j \right). \label{Hmixhy}
\end{gather}
The term $\mathcal{H}_{\parallel}$ represents the hyperfine interaction in the direction of spin alignment in the ground state, which makes both the eigenfrequency of nuclear- and electron-spin precession higher. On the other hand, the term $\mathcal{H}_{\mathrm{mix}}$ causes the mixing between the nuclear- and electron-spin precession modes. 
Then we perform the Holstein-Primakoff transformation for expressing perturbations in terms of magnons:
\begin{subequations} \label{Holstein}
\begin{align} 
(-S_{i}^{z_{1}})&=S- a_i^\dagger a_i, & I_{i}^{z_{1}}&= \langle I \rangle - c_i^\dagger c_i \\
S_{i}^{+}= (-S_{i}^{x_{1}})&+i S_{i}^{y}  =\sqrt{2S} a_i, &   I_{i}^{+}= I_{i}^{x_{1}}&+i I_{i}^{y}  =  \sqrt{2\langle I \rangle} c_i,\\
S_{i}^{-}= (-S_{i}^{x_{1}})&-i S_{i}^{y}=\sqrt{2S} a^\dagger_i, &  I_{i}^{-}= I_{i}^{x_{1}}&-i I_{i}^{y}  =  \sqrt{2\langle I \rangle} c_i^\dagger
\end{align}
\end{subequations}
for the first sublattice,
where $a,a^\dagger$ and $c,c^\dagger$ are magnon annihilation, creation operators satisfying bosonic commutation relation.
Here, we consider only a small perturbation and hence the higher order of creation/annihilation in the Holstein-Primakoff is neglected.
We perform the same transformation for the other sublattice by using $b_j$ and $d_j$ representing the electron and nuclear magnon operators respectively. 
We also perform the Fourier transformation:
\begin{gather} \label{fouriermagnon}
a_{i}=\sqrt{\frac{1}{N}}\sum_{\vec{k}} e^{i \vec{k} \vec{r}_{i}} a_k, \quad b_{j}=\sqrt{\frac{1}{N}}\sum_{\vec{k}} e^{i \vec{k} \vec{r}_{j}} b_k, \quad
c_{i}=\sqrt{\frac{1}{N}}\sum_{\vec{k}} e^{i \vec{k} \vec{r}_{i}} c_k, \quad d_{j}=\sqrt{\frac{1}{N}}\sum_{\vec{k}} e^{i \vec{k} \vec{r}_{j}} d_k,
\end{gather}
with $N$ denoting the number of spin sites in each sublattice. However, we focus only on the $\vec{k}=0$ mode since it is dominantly excited by DM which is assumed to be almost spatially uniform. As long as only the $k=0$ mode is considered, we simply omit the subscript $k$ for the annihilation and creation operators.

\subsection{Diagonalization}

The quadratic terms in the Hamiltonian can be divided into several terms as
\begin{equation}
\mathcal{H}=\mathcal{H}_e+\mathcal{H}_n +\mathcal{H}_\mathrm{mix}+\mathcal{H}_\mathrm{DM},
\end{equation}
where $\mathcal{H}_e$ is the electron-magnon term including the $\mathcal{H}_\parallel$ part of the hyperfine interaction, $\mathcal{H}_n$ is the nuclear-magnon term, $\mathcal{H}_\mathrm{mix}$ is the mixing term given by Eq.~\eqref{Hmixhy}, and $\mathcal{H}_\mathrm{DM}$ is the interaction term between the spin and DM.
The electron-magnon Hamiltonian $\mathcal{H}_e$ reads as
\begin{equation}
\mathcal{H}_e/\hbar= A_e (a^\dagger a +b^\dagger b) + B_e (a b+ a^\dagger b^\dagger)+ \frac{1}{2} C_e(a a+ b b + \mathrm{h.c.}) + D_e (a b^\dagger + a^\dagger b),
\end{equation}
with
\begin{gather}
A_e =\gamma_e \left[  H_E - \frac{H_0^2 - H_D^2}{2 H_E} + H_{K'}  + H_0  \sin{\psi} + H_a + \frac{H_K}{2} \right], \nonumber \\
B_e=\gamma_e \left( -H_E + \frac{1}{2} \frac{H_0^2 - H_D^2}{2 H_E} \right), \quad C_e= -\gamma_e \frac{H_K}{2},  \quad
D_e=\gamma_e \frac{1}{2}  \frac{H_0^2 - H_D^2}{2 H_E}. \nonumber
\end{gather}
The nuclear-magnon Hamiltonian $\mathcal{H}_n$ reads as
\begin{equation}
\mathcal{H}_n /\hbar=\omega_n(c^\dagger c+d^\dagger d),
\end{equation}
where
\begin{equation}
\omega_n= \tilde{A}_\mathrm{hy} S/\hbar.
\end{equation}
The interaction Hamiltonian between the DM and spin is 
\begin{equation}
\mathcal{H}_\mathrm{DM}= \gamma_e \hbar \left[ \sum_i \vec{S}_i + \sum_j \vec{S}_j  \right] \cdot \vec{h}_e (t)- \gamma_n \hbar \left[ \sum_i \vec{I}_i  + \sum_j \vec{I}_j  \right] \cdot \vec{h}_n(t)
\end{equation}
with $\vec{h}_n=\vec{h}_n^\mathrm{axion}$ and $\vec{h}_e^\mathrm{axion}$ given by Eqs.~\eqref{axionfieldI} and \eqref{axionfielde} for the case of axion DM, and $\vec{h}_n=\vec{h}_e=\vec{h}^{\gamma'}$ given by Eq.~ \eqref{Beff_DP} for the case of dark photon DM.

We can diagonalize $\mathcal{H}_e$ with the following generalized Bogoliubov transformation \cite{Rezende1,PhysRev.139.A450}
\begin{equation}
X=QZ,
\end{equation}
where 
$
X\equiv
\left(
a  \ b \ a^{\dagger}\  b^{\dagger}
\right)^T
$,$
Z
\equiv
\left(
\alpha \ \beta \  \alpha^{\dagger} \ \beta^{\dagger}
\right)^T
$,
and
\begin{align}
Q
=
\begin{pmatrix}
Q_1& Q_2  \\
Q_2^* & Q_1^* 
\end{pmatrix}
,\quad
 Q_1=
\begin{pmatrix}
Q_{11}& Q_{12}  \\
-Q_{11} & Q_{12} 
\end{pmatrix}
,\quad
Q_2=
\begin{pmatrix}
Q_{13} & Q_{14}  \\
-Q_{13}  & Q_{14} 
\end{pmatrix},
\end{align}
with the following elements:
\begin{align}
Q_{11}&=- \left[ \frac{(A_e-D_e) + \omega_{e,-}}{4 \omega_{e,-}} \right]^{1/2}, & Q_{12}&=\left[ \frac{(A_e+ D_e) + \omega_{e,+}}{4 \omega_{e,+ }} \right]^{1/2}, \nonumber\\
Q_{13}&=\left[ \frac{(A_e-D_e) - \omega_{e,-}}{4 \omega_{e,-}} \right]^{1/2}, & Q_{14}&=\left[ \frac{(A_e+ D_e) - \omega_{e,+}}{4 \omega_{e,+}} \right]^{1/2}.
\end{align}
Note that the matrix $Q$ ensures the canonical quantization properties of $\alpha$ and $\beta$ through the relation
\begin{equation}
Q
\begin{pmatrix}
I_2 & 0\\
0 & -I_2
\end{pmatrix}
Q^{\dagger}=
\begin{pmatrix}
I_2 & 0\\
0 & -I_2
\end{pmatrix}.
\end{equation}
The diagonalized Hamiltonian is written as 
\begin{equation}
\mathcal{H}_e/\hbar=\omega_{e,-} \alpha^\dagger \alpha+\omega_{e,+} \beta^\dagger \beta,
\end{equation}
 where $\omega_{e,-}$ and $\omega_{e,+}$ are equal to those given by Eqs.~\eqref{e-freq} and \eqref{e+freq}, respectively. Magnons $\alpha$ and $\beta$ correspond to the out-phase and in-phase mode of electron-spin precession, respectively (see Sec.~\ref{subsec: MDMnCO3}). Because the eigenenergy of the out-phase mode is far from that of the nuclear magnon, its mixing with the nuclear magnon is expected to be small. We then focus only on the in-phase mode $\beta$.

Nuclear magnons and electron magnons mix with each other by the precession component of hyperfine interaction:
\begin{equation}
\mathcal{H}_{\rm mix}=-\tilde{A}_{\rm hy} \sqrt{S \langle I\rangle} (a c+a^{\dagger} c^{\dagger}  +b d+  b^{\dagger} d^{\dagger} ).
\end{equation}
When we focus only on the in-phase mode $\beta$ of the electron magnon, we obtain
\begin{equation}
\mathcal{H}_{\rm mix}=-\tilde{A}_{\rm hy} \sqrt{S \langle I \rangle}\left[ ( Q_{12} \beta + Q_{14} \beta^{\dagger} ) \eta +  ( Q_{12} \beta^{\dagger} + Q_{14} \beta ) \eta^{\dagger}  \right],
\end{equation}
where 
\begin{equation}
\eta \equiv (c+d)/\sqrt{2}. \label{nuclearinphase}
\end{equation}

Let us focus on the in-phase magnons. First neglecting the DM part, the Hamiltonian of interest is given by
\begin{align}
\mathcal{H}_0/\hbar&=\omega_{e,+}\beta^\dagger \beta+\omega_{n} \eta^\dagger \eta + \mathcal{H}_\mathrm{mix}/\hbar \nonumber\\
&=A \beta^\dagger \beta +A' \eta^\dagger \eta + B (\beta \eta + \beta^\dagger \eta^\dagger) +  D(\beta \eta^\dagger + \beta^\dagger \eta)
\end{align}
with
\begin{gather}
A=\omega_{e,+}, \quad A'=\omega_n \\
B=-\tilde{A}_{\mathrm{hy}}\sqrt{2\langle I\rangle S} Q_{12} / \hbar, \quad D=-\tilde{A}_{\mathrm{hy}}\sqrt{2 \langle I\rangle S} Q_{14} / \hbar.
\end{gather}
We can  diagonalize it by the transformation
\begin{equation}
Y=RW,
\end{equation}
 where 
$
Y \equiv
\left(
\beta  \ \eta \ \beta^{\dagger}\  \eta^{\dagger}
\right)^T
$ and $
W
\equiv
\left(
\tilde{\beta}  \ \tilde{\eta} \ \tilde{\beta}^{\dagger}\  \tilde{\eta}^{\dagger}
\right)^T$. We find that matrix $R$ that preserves canonical quantization of magnon operators is in the form
\begin{equation}
R=
\begin{pmatrix}
U & V \\
V^* & U^*
\end{pmatrix}, \label{Rmatrix}
\end{equation}
with the elements given by
\begin{gather}
U=
\left(u_1  \quad u_2\right)
, \quad
u_i= k_{i}
\begin{bmatrix}
\textcircled{+}_i\\
2AD
\end{bmatrix},
\quad
\textcircled{+}_i=\omega_i^2+( A-A') \omega_i -(A A' + D^2 -B^2), \\
V=
\left(v_1 \quad v_2 \right)
, \quad
v_i= C_i k_{i}
\begin{bmatrix}
\textcircled{-}_i\\
2AD
\end{bmatrix},
\quad
\textcircled{-}_i=\omega_i^2-(A-A') \omega_i -(A A' + D^2 -B^2),
\end{gather}
where 
\begin{gather}
C_i=-\frac{D}{B} \left[ \frac{(\omega_i-A)(\omega_i-A')-(D^2-B^2)}{(\omega_i-A)(\omega_i+A')-(D^2-B^2)} \right],\\
k_{1}=\frac{1}{2A D} \left( \frac{C_2 \omega_2}{-C_1 \omega_1+C_1C_2^2 \omega_1 + C_2 \omega_2 -C_1^2C_2 \omega_2 } \right)^{1/2},\\
k_{2}=\frac{1}{2A D} \left( \frac{-C_1 \omega_1}{-C_1 \omega_1+C_1C_2^2 \omega_1 + C_2 \omega_2 -C_1^2C_2 \omega_2 } \right)^{1/2},
\end{gather}
and 
\begin{equation}
 \omega^2_{1,2} = D^2-B^2 + \frac{A^2+A'^2}{2 } \pm \frac{1}{2}\sqrt{(A-A')^2[(A+A')^2 + 4 (D^2-B^2)] + 16 A A' D^2}. \label{eigenmix}
\end{equation}
Note that $\omega_1=\omega_{\tilde{e},+}$ and $\omega_2=\omega_{\tilde{n},+}$, which are equal to those given by Eqs.~\eqref{einMnCO3} and \eqref{ninMnCO3}, respectively. At last, a diagonalized full Hamiltonian of electron and nuclear magnon is given in the form
\begin{equation}
\mathcal{H}_0/\hbar=\omega_{\tilde{e},+} \tilde{\beta}^\dagger \tilde{\beta}+ \omega_{\tilde{n},+}\tilde{\eta}^\dagger
\tilde{\eta}.
\end{equation}

For the convenience of later discussion, we define the following mixing angle matrix $\phi^{\pm}$:
\begin{equation}
\begin{pmatrix}
\beta^\dagger\pm \beta\\
\eta^\dagger \pm \eta
\end{pmatrix}
=
\begin{pmatrix}
\phi^\pm_{\beta \tilde{\beta}} & \phi^\pm_{\beta \tilde{\eta}}\\
\phi^\pm_{\eta \tilde{\beta}} & \phi^\pm_{\eta \tilde{\eta}}
\end{pmatrix}
\begin{pmatrix}
\tilde{\beta}^\dagger\pm \tilde{\beta}\\
\tilde{\eta}^\dagger \pm \tilde{\eta},
\end{pmatrix}
\end{equation}
which can be obtained from matrix $R$ defined in Eq.~\eqref{Rmatrix}.

\subsection{Response}
Next, we discuss the magnetization response due to the DM-induced magnetic field.
As defined in Eq.~\eqref{MsigMztotal}, the magnetization signal $M_\mathrm{signal}$ is defined by the total magnetization in the $z$ direction $M^z_\mathrm{total}$ which is the oscillation component in this setup. We can write it in terms of the magnon operator $\tilde{\beta},\tilde{\eta}$ as
\begin{align}
M_{\mathrm{signal}}V&=\langle M_\mathrm{total}^z\rangle  V = -\gamma_e \hbar  \langle  \sum_i S^{z}_i + \sum_j S^{z}_j \rangle \nonumber \\
&=- \gamma_e \hbar \sin \psi \sqrt{N_{\mathrm{total}}S} (Q_{12}+Q_{14})  \left( \phi^+_{\beta \tilde{\eta}}\langle \tilde{\eta} + \tilde{\eta}^{\dagger} \rangle + \phi^+_{\beta \tilde{\beta}} \langle \tilde{\beta} + \tilde{\beta}^{\dagger} \rangle  \right), \label{signalmacroderi1}
\end{align}
where $N_\mathrm{total}= 2N$ is the number of total spin sites. On the other hand, the Hamiltonian interaction between DM-induced field and magnon $\tilde{\beta},\tilde{\eta}$ reads as
\begin{align}
\begin{split}
\mathcal{H}_\mathrm{DM}= &  -i\gamma_n \hbar h_n^y (t)   \sqrt{N_\mathrm{total} \langle I\rangle/2}   \left( \phi^-_{ \eta \tilde{\eta} } ( \tilde{\eta}^{\dagger} - \tilde{\eta} ) + \phi^-_{ \eta \tilde{\beta}} (\tilde{\eta}^\dagger- \tilde{\eta}) \right)   \\
&+ \gamma_n \hbar  h_n^z (t) \sin \psi \sqrt{N_\mathrm{total} \langle I \rangle/2}  \left( \phi^+_{ \eta \tilde{\eta} } ( \tilde{\eta}^{\dagger} +\tilde{\eta} ) + \phi^+_{ \eta \tilde{\beta}} (\tilde{\beta}^\dagger+ \tilde{\beta}) \right)  \\
&+\gamma_e \hbar h^z_e(t) \sin \psi \sqrt{N_{\mathrm{total}}S} (Q_{12}+Q_{14})  \left( \phi^+_{\beta \tilde{\eta}} ( \tilde{\eta} + \tilde{\eta}^{\dagger}) + \phi^+_{\beta \tilde{\beta}}  (\tilde{\beta}+ \tilde{\beta}^{\dagger}  ) \right)  \\
&+i \gamma_e \hbar  h_e^y(t) \sqrt{ N_\mathrm{total} S} (Q_{12}-Q_{14}) \left( \phi^-_{\beta \tilde{\eta}} (\tilde{\eta}^\dagger-\tilde{\eta})+  \phi^-_{\beta \tilde{\beta}} (\tilde{\beta}^\dagger-\tilde{\beta}) \right).
\end{split}
\end{align}

One can then apply the linear excitation theory to find the expectation value of operators in Eq.~\eqref{signalmacroderi1}, and hence the magnetization signal. Focusing on one particular mode $\zeta=\tilde{\beta},\tilde{\eta}$, the Hamiltonian is
\begin{gather}
H=H_0+H_\mathrm{DM};\\
H_0= \hbar\omega_\zeta \zeta^\dagger \zeta, \quad H_\mathrm{DM}=[ g_\zeta \zeta e^{i\omega_\mathrm{DM} t}+ \mathrm{H.c.} ],
\end{gather}
where $g_\zeta$ is the coupling constant depending on the amplitude of the DM-induced field $\vec{h}_e, \vec{h}_n$ (see Eqs.~\eqref{axionfieldI},\eqref{axionfielde}, and \eqref{Beff_DP}), while $\omega_\mathrm{DM}$ is equal to the axion mass $m_a$ or dark photon mass $m_\mathrm{\gamma'}$. 
The Heisenberg equation of motion reads as
\begin{equation}
\frac{d \zeta}{dt}=-i\omega_\zeta \zeta-\frac{1}{T_{2\zeta}}\zeta - \frac{i}{\hbar} g^*_\zeta e^{-i \omega_\mathrm{DM} t},
\end{equation}
with $T_{2\zeta}$ representing the relaxation time of the precession mode $\zeta$ which is twice a value of the magnon lifetime. The stationary steady-state solution for this system is given by
\begin{equation}
\langle \zeta \rangle= \frac{g^*_\zeta/\hbar}{(\omega_\mathrm{DM}-\omega_\zeta) + i(1/T_{2\zeta})} e^{-i\omega_\mathrm{DM} t}, \label{solution1zeta}
\end{equation}
which is a good approximation for the solution at $t \gtrsim T_{2\zeta}$. Substituting the expectation value to the expression of magnetization signal (Eq.~\eqref{signalmacroderi1}), one can derive the response of the system to DM.
Owing to velocity distribution of DM, the magnetization signal spreads with width $\Delta \omega_\mathrm{DM}$ determined by that of the DM field.
We find that the response of the system derived from the quantum magnon picture is consistent with that of classical theory given in Tables \ref{table: NMR} and \ref{table: FMR}.
On the other hand, the power of the system can be derived from the relation $P= \hbar \omega_\zeta \langle n_\zeta \rangle {2}/{T_{2\zeta}}$ where $n_\zeta=\zeta^\dagger\zeta$ is the number operator.

\newpage

%%%%%%%%%%%%%%%%%%%%%%%%%%%%%%%%%%%%%%%

\bibliographystyle{jhep}
\bibliography{ref.bib}

\providecommand{\href}[2]{#2}\begingroup\raggedright\begin{thebibliography}{100}

\bibitem{Workman:2022ynf}
{\scshape Particle Data Group} collaboration, \emph{{Review of Particle Physics}}, \href{https://doi.org/10.1093/ptep/ptac097}{\emph{PTEP} {\bfseries 2022} (2022) 083C01}.

\bibitem{Peccei:1977hh}
R.D.~Peccei and H.R.~Quinn, \emph{{CP Conservation in the Presence of Instantons}}, \href{https://doi.org/10.1103/PhysRevLett.38.1440}{\emph{Phys. Rev. Lett.} {\bfseries 38} (1977) 1440}.

\bibitem{Peccei:1977ur}
R.D.~Peccei and H.R.~Quinn, \emph{{Constraints Imposed by CP Conservation in the Presence of Instantons}}, \href{https://doi.org/10.1103/PhysRevD.16.1791}{\emph{Phys. Rev. D} {\bfseries 16} (1977) 1791}.

\bibitem{Weinberg:1977ma}
S.~Weinberg, \emph{{A New Light Boson?}}, \href{https://doi.org/10.1103/PhysRevLett.40.223}{\emph{Phys. Rev. Lett.} {\bfseries 40} (1978) 223}.

\bibitem{Wilczek:1977pj}
F.~Wilczek, \emph{{Problem of Strong $P$ and $T$ Invariance in the Presence of Instantons}}, \href{https://doi.org/10.1103/PhysRevLett.40.279}{\emph{Phys. Rev. Lett.} {\bfseries 40} (1978) 279}.

\bibitem{Svrcek:2006yi}
P.~Svrcek and E.~Witten, \emph{{Axions In String Theory}}, \href{https://doi.org/10.1088/1126-6708/2006/06/051}{\emph{JHEP} {\bfseries 06} (2006) 051} [\href{https://arxiv.org/abs/hep-th/0605206}{{\ttfamily hep-th/0605206}}].

\bibitem{Arvanitaki:2009fg}
A.~Arvanitaki, S.~Dimopoulos, S.~Dubovsky, N.~Kaloper and J.~March-Russell, \emph{{String Axiverse}}, \href{https://doi.org/10.1103/PhysRevD.81.123530}{\emph{Phys. Rev. D} {\bfseries 81} (2010) 123530} [\href{https://arxiv.org/abs/0905.4720}{{\ttfamily 0905.4720}}].

\bibitem{Cicoli:2012sz}
M.~Cicoli, M.~Goodsell and A.~Ringwald, \emph{{The type IIB string axiverse and its low-energy phenomenology}}, \href{https://doi.org/10.1007/JHEP10(2012)146}{\emph{JHEP} {\bfseries 10} (2012) 146} [\href{https://arxiv.org/abs/1206.0819}{{\ttfamily 1206.0819}}].

\bibitem{Jaeckel:2010ni}
J.~Jaeckel and A.~Ringwald, \emph{{The Low-Energy Frontier of Particle Physics}}, \href{https://doi.org/10.1146/annurev.nucl.012809.104433}{\emph{Ann. Rev. Nucl. Part. Sci.} {\bfseries 60} (2010) 405} [\href{https://arxiv.org/abs/1002.0329}{{\ttfamily 1002.0329}}].

\bibitem{Jaeckel:2012mjv}
J.~Jaeckel, \emph{{A force beyond the Standard Model - Status of the quest for hidden photons}}, {\emph{Frascati Phys. Ser.} {\bfseries 56} (2012) 172} [\href{https://arxiv.org/abs/1303.1821}{{\ttfamily 1303.1821}}].

\bibitem{Arias:2012az}
P.~Arias, D.~Cadamuro, M.~Goodsell, J.~Jaeckel, J.~Redondo and A.~Ringwald, \emph{{WISPy Cold Dark Matter}}, \href{https://doi.org/10.1088/1475-7516/2012/06/013}{\emph{JCAP} {\bfseries 06} (2012) 013} [\href{https://arxiv.org/abs/1201.5902}{{\ttfamily 1201.5902}}].

\bibitem{Fabbrichesi:2020wbt}
M.~Fabbrichesi, E.~Gabrielli and G.~Lanfranchi, \emph{{The Dark Photon}},  \href{https://arxiv.org/abs/2005.01515}{{\ttfamily 2005.01515}}.

\bibitem{CDMS-II:2009ktb}
{\scshape CDMS-II} collaboration, \emph{{Dark Matter Search Results from the CDMS II Experiment}}, \href{https://doi.org/10.1126/science.1186112}{\emph{Science} {\bfseries 327} (2010) 1619} [\href{https://arxiv.org/abs/0912.3592}{{\ttfamily 0912.3592}}].

\bibitem{CDMS:2009fba}
{\scshape CDMS} collaboration, \emph{{Search for Axions with the CDMS Experiment}}, \href{https://doi.org/10.1103/PhysRevLett.103.141802}{\emph{Phys. Rev. Lett.} {\bfseries 103} (2009) 141802} [\href{https://arxiv.org/abs/0902.4693}{{\ttfamily 0902.4693}}].

\bibitem{SuperCDMS:2018mne}
{\scshape SuperCDMS} collaboration, \emph{{First Dark Matter Constraints from a SuperCDMS Single-Charge Sensitive Detector}}, \href{https://doi.org/10.1103/PhysRevLett.121.051301}{\emph{Phys. Rev. Lett.} {\bfseries 121} (2018) 051301} [\href{https://arxiv.org/abs/1804.10697}{{\ttfamily 1804.10697}}].

\bibitem{XENON:2007uwm}
{\scshape XENON} collaboration, \emph{{First Results from the XENON10 Dark Matter Experiment at the Gran Sasso National Laboratory}}, \href{https://doi.org/10.1103/PhysRevLett.100.021303}{\emph{Phys. Rev. Lett.} {\bfseries 100} (2008) 021303} [\href{https://arxiv.org/abs/0706.0039}{{\ttfamily 0706.0039}}].

\bibitem{XENON:2020kmp}
{\scshape XENON} collaboration, \emph{{Projected WIMP sensitivity of the XENONnT dark matter experiment}}, \href{https://doi.org/10.1088/1475-7516/2020/11/031}{\emph{JCAP} {\bfseries 11} (2020) 031} [\href{https://arxiv.org/abs/2007.08796}{{\ttfamily 2007.08796}}].

\bibitem{PandaX-II:2017hlx}
{\scshape PandaX-II} collaboration, \emph{{Dark Matter Results From 54-Ton-Day Exposure of PandaX-II Experiment}}, \href{https://doi.org/10.1103/PhysRevLett.119.181302}{\emph{Phys. Rev. Lett.} {\bfseries 119} (2017) 181302} [\href{https://arxiv.org/abs/1708.06917}{{\ttfamily 1708.06917}}].

\bibitem{PandaX-II:2018xpz}
{\scshape PandaX-II} collaboration, \emph{{Constraining Dark Matter Models with a Light Mediator at the PandaX-II Experiment}}, \href{https://doi.org/10.1103/PhysRevLett.121.021304}{\emph{Phys. Rev. Lett.} {\bfseries 121} (2018) 021304} [\href{https://arxiv.org/abs/1802.06912}{{\ttfamily 1802.06912}}].

\bibitem{SENSEI:2020dpa}
{\scshape SENSEI} collaboration, \emph{{SENSEI: Direct-Detection Results on sub-GeV Dark Matter from a New Skipper-CCD}}, \href{https://doi.org/10.1103/PhysRevLett.125.171802}{\emph{Phys. Rev. Lett.} {\bfseries 125} (2020) 171802} [\href{https://arxiv.org/abs/2004.11378}{{\ttfamily 2004.11378}}].

\bibitem{Barbieri:2016vwg}
R.~Barbieri, C.~Braggio, G.~Carugno, C.S.~Gallo, A.~Lombardi, A.~Ortolan et~al., \emph{{Searching for galactic axions through magnetized media: the QUAX proposal}}, \href{https://doi.org/10.1016/j.dark.2017.01.003}{\emph{Phys. Dark Univ.} {\bfseries 15} (2017) 135} [\href{https://arxiv.org/abs/1606.02201}{{\ttfamily 1606.02201}}].

\bibitem{Budker:2013hfa}
D.~Budker, P.W.~Graham, M.~Ledbetter, S.~Rajendran and A.~Sushkov, \emph{{Proposal for a Cosmic Axion Spin Precession Experiment (CASPEr)}}, \href{https://doi.org/10.1103/PhysRevX.4.021030}{\emph{Phys. Rev. X} {\bfseries 4} (2014) 021030} [\href{https://arxiv.org/abs/1306.6089}{{\ttfamily 1306.6089}}].

\bibitem{JacksonKimball:2017elr}
D.F.~Jackson~Kimball et~al., \emph{{Overview of the Cosmic Axion Spin Precession Experiment (CASPEr)}}, \href{https://doi.org/10.1007/978-3-030-43761-9_13}{\emph{Springer Proc. Phys.} {\bfseries 245} (2020) 105} [\href{https://arxiv.org/abs/1711.08999}{{\ttfamily 1711.08999}}].

\bibitem{PhysRevLett.126.141802}
D.~Aybas, J.~Adam, E.~Blumenthal, A.V.~Gramolin, D.~Johnson, A.~Kleyheeg et~al., \emph{Search for axionlike dark matter using solid-state nuclear magnetic resonance}, \href{https://doi.org/10.1103/PhysRevLett.126.141802}{\emph{Phys. Rev. Lett.} {\bfseries 126} (2021) 141802}.

\bibitem{Hochberg:2015pha}
Y.~Hochberg, Y.~Zhao and K.M.~Zurek, \emph{{Superconducting Detectors for Superlight Dark Matter}}, \href{https://doi.org/10.1103/PhysRevLett.116.011301}{\emph{Phys. Rev. Lett.} {\bfseries 116} (2016) 011301} [\href{https://arxiv.org/abs/1504.07237}{{\ttfamily 1504.07237}}].

\bibitem{Hochberg:2015fth}
Y.~Hochberg, M.~Pyle, Y.~Zhao and K.M.~Zurek, \emph{{Detecting Superlight Dark Matter with Fermi-Degenerate Materials}}, \href{https://doi.org/10.1007/JHEP08(2016)057}{\emph{JHEP} {\bfseries 08} (2016) 057} [\href{https://arxiv.org/abs/1512.04533}{{\ttfamily 1512.04533}}].

\bibitem{Hochberg:2016sqx}
Y.~Hochberg, T.~Lin and K.M.~Zurek, \emph{{Absorption of light dark matter in semiconductors}}, \href{https://doi.org/10.1103/PhysRevD.95.023013}{\emph{Phys. Rev. D} {\bfseries 95} (2017) 023013} [\href{https://arxiv.org/abs/1608.01994}{{\ttfamily 1608.01994}}].

\bibitem{Bloch:2016sjj}
I.M.~Bloch, R.~Essig, K.~Tobioka, T.~Volansky and T.-T.~Yu, \emph{{Searching for Dark Absorption with Direct Detection Experiments}}, \href{https://doi.org/10.1007/JHEP06(2017)087}{\emph{JHEP} {\bfseries 06} (2017) 087} [\href{https://arxiv.org/abs/1608.02123}{{\ttfamily 1608.02123}}].

\bibitem{Hochberg:2017wce}
Y.~Hochberg, Y.~Kahn, M.~Lisanti, K.M.~Zurek, A.G.~Grushin, R.~Ilan et~al., \emph{{Detection of sub-MeV Dark Matter with Three-Dimensional Dirac Materials}}, \href{https://doi.org/10.1103/PhysRevD.97.015004}{\emph{Phys. Rev. D} {\bfseries 97} (2018) 015004} [\href{https://arxiv.org/abs/1708.08929}{{\ttfamily 1708.08929}}].

\bibitem{Coskuner:2019odd}
A.~Coskuner, A.~Mitridate, A.~Olivares and K.M.~Zurek, \emph{{Directional Dark Matter Detection in Anisotropic Dirac Materials}}, \href{https://doi.org/10.1103/PhysRevD.103.016006}{\emph{Phys. Rev. D} {\bfseries 103} (2021) 016006} [\href{https://arxiv.org/abs/1909.09170}{{\ttfamily 1909.09170}}].

\bibitem{Trickle:2019nya}
T.~Trickle, Z.~Zhang, K.M.~Zurek, K.~Inzani and S.M.~Griffin, \emph{{Multi-Channel Direct Detection of Light Dark Matter: Theoretical Framework}}, \href{https://doi.org/10.1007/JHEP03(2020)036}{\emph{JHEP} {\bfseries 03} (2020) 036} [\href{https://arxiv.org/abs/1910.08092}{{\ttfamily 1910.08092}}].

\bibitem{Griffin:2019mvc}
S.M.~Griffin, K.~Inzani, T.~Trickle, Z.~Zhang and K.M.~Zurek, \emph{{Multichannel direct detection of light dark matter: Target comparison}}, \href{https://doi.org/10.1103/PhysRevD.101.055004}{\emph{Phys. Rev. D} {\bfseries 101} (2020) 055004} [\href{https://arxiv.org/abs/1910.10716}{{\ttfamily 1910.10716}}].

\bibitem{Geilhufe:2019ndy}
R.M.~Geilhufe, F.~Kahlhoefer and M.W.~Winkler, \emph{{Dirac Materials for Sub-MeV Dark Matter Detection: New Targets and Improved Formalism}}, \href{https://doi.org/10.1103/PhysRevD.101.055005}{\emph{Phys. Rev. D} {\bfseries 101} (2020) 055005} [\href{https://arxiv.org/abs/1910.02091}{{\ttfamily 1910.02091}}].

\bibitem{Trickle:2020oki}
T.~Trickle, Z.~Zhang and K.M.~Zurek, \emph{{Effective field theory of dark matter direct detection with collective excitations}}, \href{https://doi.org/10.1103/PhysRevD.105.015001}{\emph{Phys. Rev. D} {\bfseries 105} (2022) 015001} [\href{https://arxiv.org/abs/2009.13534}{{\ttfamily 2009.13534}}].

\bibitem{Hochberg:2021pkt}
Y.~Hochberg, Y.~Kahn, N.~Kurinsky, B.V.~Lehmann, T.C.~Yu and K.K.~Berggren, \emph{{Determining Dark-Matter\textendash{}Electron Scattering Rates from the Dielectric Function}}, \href{https://doi.org/10.1103/PhysRevLett.127.151802}{\emph{Phys. Rev. Lett.} {\bfseries 127} (2021) 151802} [\href{https://arxiv.org/abs/2101.08263}{{\ttfamily 2101.08263}}].

\bibitem{Knapen:2021run}
S.~Knapen, J.~Kozaczuk and T.~Lin, \emph{{Dark matter-electron scattering in dielectrics}}, \href{https://doi.org/10.1103/PhysRevD.104.015031}{\emph{Phys. Rev. D} {\bfseries 104} (2021) 015031} [\href{https://arxiv.org/abs/2101.08275}{{\ttfamily 2101.08275}}].

\bibitem{Knapen:2021bwg}
S.~Knapen, J.~Kozaczuk and T.~Lin, \emph{{python package for dark matter scattering in dielectric targets}}, \href{https://doi.org/10.1103/PhysRevD.105.015014}{\emph{Phys. Rev. D} {\bfseries 105} (2022) 015014} [\href{https://arxiv.org/abs/2104.12786}{{\ttfamily 2104.12786}}].

\bibitem{Mitridate:2021ctr}
A.~Mitridate, T.~Trickle, Z.~Zhang and K.M.~Zurek, \emph{{Dark matter absorption via electronic excitations}}, \href{https://doi.org/10.1007/JHEP09(2021)123}{\emph{JHEP} {\bfseries 09} (2021) 123} [\href{https://arxiv.org/abs/2106.12586}{{\ttfamily 2106.12586}}].

\bibitem{Chen:2022pyd}
H.-Y.~Chen, A.~Mitridate, T.~Trickle, Z.~Zhang, M.~Bernardi and K.M.~Zurek, \emph{{Dark matter direct detection in materials with spin-orbit coupling}}, \href{https://doi.org/10.1103/PhysRevD.106.015024}{\emph{Phys. Rev. D} {\bfseries 106} (2022) 015024} [\href{https://arxiv.org/abs/2202.11716}{{\ttfamily 2202.11716}}].

\bibitem{Mitridate:2022tnv}
A.~Mitridate, T.~Trickle, Z.~Zhang and K.M.~Zurek, \emph{{Snowmass White Paper: Light Dark Matter Direct Detection at the Interface With Condensed Matter Physics}},  in \emph{{Snowmass 2021}}, 3, 2022 [\href{https://arxiv.org/abs/2203.07492}{{\ttfamily 2203.07492}}].

\bibitem{Trickle:2022fwt}
T.~Trickle, \emph{{Extended calculation of electronic excitations for direct detection of dark matter}}, \href{https://doi.org/10.1103/PhysRevD.107.035035}{\emph{Phys. Rev. D} {\bfseries 107} (2023) 035035} [\href{https://arxiv.org/abs/2210.14917}{{\ttfamily 2210.14917}}].

\bibitem{Knapen:2017ekk}
S.~Knapen, T.~Lin, M.~Pyle and K.M.~Zurek, \emph{{Detection of Light Dark Matter With Optical Phonons in Polar Materials}}, \href{https://doi.org/10.1016/j.physletb.2018.08.064}{\emph{Phys. Lett. B} {\bfseries 785} (2018) 386} [\href{https://arxiv.org/abs/1712.06598}{{\ttfamily 1712.06598}}].

\bibitem{Griffin:2018bjn}
S.~Griffin, S.~Knapen, T.~Lin and K.M.~Zurek, \emph{{Directional Detection of Light Dark Matter with Polar Materials}}, \href{https://doi.org/10.1103/PhysRevD.98.115034}{\emph{Phys. Rev. D} {\bfseries 98} (2018) 115034} [\href{https://arxiv.org/abs/1807.10291}{{\ttfamily 1807.10291}}].

\bibitem{Cox:2019cod}
P.~Cox, T.~Melia and S.~Rajendran, \emph{{Dark matter phonon coupling}}, \href{https://doi.org/10.1103/PhysRevD.100.055011}{\emph{Phys. Rev. D} {\bfseries 100} (2019) 055011} [\href{https://arxiv.org/abs/1905.05575}{{\ttfamily 1905.05575}}].

\bibitem{Mitridate:2020kly}
A.~Mitridate, T.~Trickle, Z.~Zhang and K.M.~Zurek, \emph{{Detectability of Axion Dark Matter with Phonon Polaritons and Magnons}}, \href{https://doi.org/10.1103/PhysRevD.102.095005}{\emph{Phys. Rev. D} {\bfseries 102} (2020) 095005} [\href{https://arxiv.org/abs/2005.10256}{{\ttfamily 2005.10256}}].

\bibitem{Marsh:2022fmo}
D.J.E.~Marsh, J.I.~McDonald, A.J.~Millar and J.~Sch\"utte-Engel, \emph{{Axion detection with phonon-polaritons revisited}}, \href{https://doi.org/10.1103/PhysRevD.107.035036}{\emph{Phys. Rev. D} {\bfseries 107} (2023) 035036} [\href{https://arxiv.org/abs/2209.12909}{{\ttfamily 2209.12909}}].

\bibitem{Trickle:2019ovy}
T.~Trickle, Z.~Zhang and K.M.~Zurek, \emph{{Detecting Light Dark Matter with Magnons}}, \href{https://doi.org/10.1103/PhysRevLett.124.201801}{\emph{Phys. Rev. Lett.} {\bfseries 124} (2020) 201801} [\href{https://arxiv.org/abs/1905.13744}{{\ttfamily 1905.13744}}].

\bibitem{Chigusa:2020gfs}
S.~Chigusa, T.~Moroi and K.~Nakayama, \emph{{Detecting light boson dark matter through conversion into a magnon}}, \href{https://doi.org/10.1103/PhysRevD.101.096013}{\emph{Phys. Rev. D} {\bfseries 101} (2020) 096013} [\href{https://arxiv.org/abs/2001.10666}{{\ttfamily 2001.10666}}].

\bibitem{Esposito:2022bnu}
A.~Esposito and S.~Pavaskar, \emph{{Optimal anti-ferromagnets for light dark matter detection}},  \href{https://arxiv.org/abs/2210.13516}{{\ttfamily 2210.13516}}.

\bibitem{Chigusa:2023hms}
S.~Chigusa, M.~Hazumi, E.D.~Herbschleb, N.~Mizuochi and K.~Nakayama, \emph{{Light Dark Matter Search with Nitrogen-Vacancy Centers in Diamonds}},  \href{https://arxiv.org/abs/2302.12756}{{\ttfamily 2302.12756}}.

\bibitem{Marsh:2018dlj}
D.J.E.~Marsh, K.-C.~Fong, E.W.~Lentz, L.~Smejkal and M.N.~Ali, \emph{{Proposal to Detect Dark Matter using Axionic Topological Antiferromagnets}}, \href{https://doi.org/10.1103/PhysRevLett.123.121601}{\emph{Phys. Rev. Lett.} {\bfseries 123} (2019) 121601} [\href{https://arxiv.org/abs/1807.08810}{{\ttfamily 1807.08810}}].

\bibitem{Schutte-Engel:2021bqm}
J.~Sch\"utte-Engel, D.J.E.~Marsh, A.J.~Millar, A.~Sekine, F.~Chadha-Day, S.~Hoof et~al., \emph{{Axion quasiparticles for axion dark matter detection}}, \href{https://doi.org/10.1088/1475-7516/2021/08/066}{\emph{JCAP} {\bfseries 08} (2021) 066} [\href{https://arxiv.org/abs/2102.05366}{{\ttfamily 2102.05366}}].

\bibitem{Chigusa:2021mci}
S.~Chigusa, T.~Moroi and K.~Nakayama, \emph{{Axion/hidden-photon dark matter conversion into condensed matter axion}}, \href{https://doi.org/10.1007/JHEP08(2021)074}{\emph{JHEP} {\bfseries 08} (2021) 074} [\href{https://arxiv.org/abs/2102.06179}{{\ttfamily 2102.06179}}].

\bibitem{Dixit:2020ymh}
A.V.~Dixit, S.~Chakram, K.~He, A.~Agrawal, R.K.~Naik, D.I.~Schuster et~al., \emph{{Searching for Dark Matter with a Superconducting Qubit}}, \href{https://doi.org/10.1103/PhysRevLett.126.141302}{\emph{Phys. Rev. Lett.} {\bfseries 126} (2021) 141302} [\href{https://arxiv.org/abs/2008.12231}{{\ttfamily 2008.12231}}].

\bibitem{Chen:2022quj}
S.~Chen, H.~Fukuda, T.~Inada, T.~Moroi, T.~Nitta and T.~Sichanugrist, \emph{{Detection of hidden photon dark matter using the direct excitation of transmon qubits}},  \href{https://arxiv.org/abs/2212.03884}{{\ttfamily 2212.03884}}.

\bibitem{PhysRev.129.1105}
P.G.~de~Gennes, P.A.~Pincus, F.~Hartmann-Boutron and J.M.~Winter, \emph{Nuclear magnetic resonance modes in magnetic material. i. theory}, \href{https://doi.org/10.1103/PhysRev.129.1105}{\emph{Phys. Rev.} {\bfseries 129} (1963) 1105}.

\bibitem{borovik1984spin}
A.~Borovik-Romanov, Y.M.~Bun'kov, B.S.~Dumesh, M.I.~Kurkin, M.P.~Petrov and V.~Chekmarev, \emph{The spin echo in systems with a coupled electron-nuclear precession}, {\emph{Soviet Physics Uspekhi} {\bfseries 27} (1984) 235}.

\bibitem{Shiomi:2019aa}
Y.~Shiomi, J.~Lustikova, S.~Watanabe, D.~Hirobe, S.~Takahashi and E.~Saitoh, \emph{Spin pumping from nuclear spin waves}, \href{https://doi.org/10.1038/s41567-018-0310-x}{\emph{Nature Physics} {\bfseries 15} (2019) 22}.

\bibitem{Kikkawa:2021aa}
T.~Kikkawa, D.~Reitz, H.~Ito, T.~Makiuchi, T.~Sugimoto, K.~Tsunekawa et~al., \emph{Observation of nuclear-spin seebeck effect}, \href{https://doi.org/10.1038/s41467-021-24623-6}{\emph{Nature Communications} {\bfseries 12} (2021) 4356}.

\bibitem{doi:10.1063/5.0107157}
S.M.~Rezende, \emph{Introduction to nuclear spin waves in ferro- and antiferromagnets}, \href{https://doi.org/10.1063/5.0107157}{\emph{Journal of Applied Physics} {\bfseries 132} (2022) 091101}.

\bibitem{PhysRev.109.606}
H.~Suhl, \emph{Effective nuclear spin interactions in ferromagnets}, \href{https://doi.org/10.1103/PhysRev.109.606}{\emph{Phys. Rev.} {\bfseries 109} (1958) 606}.

\bibitem{10.1143/PTP.20.542}
T.~Nakamura, \emph{{Indirect Coupling of Nuclear Spins in Antiferromagnet with Particular Reference to MnF2 at Very Low Temperatures}}, \href{https://doi.org/10.1143/PTP.20.542}{\emph{Progress of Theoretical Physics} {\bfseries 20} (1958) 542}.

\bibitem{doi:10.1143/JPSJ.15.2251}
M.~Date, \emph{Magnetic resonance in mnco3}, \href{https://doi.org/10.1143/JPSJ.15.2251}{\emph{Journal of the Physical Society of Japan} {\bfseries 15} (1960) 2251}.

\bibitem{doi:10.1063/1.1713505}
D.~Shaltiel and H.J.~Fink, \emph{Nuclear antiferromagnetic double resonance in mnco3}, \href{https://doi.org/10.1063/1.1713505}{\emph{Journal of Applied Physics} {\bfseries 35} (1964) 848}.

\bibitem{PhysRev.132.144}
K.~Lee, A.M.~Portis and G.L.~Witt, \emph{Magnetic properties of the hexagonal antiferromagnet csmn${\mathrm{f}}_{3}$}, \href{https://doi.org/10.1103/PhysRev.132.144}{\emph{Phys. Rev.} {\bfseries 132} (1963) 144}.

\bibitem{PhysRev.143.361}
V.~Minkiewicz and A.~Nakamura, \emph{Direct observation of ${\mathrm{mn}}^{55}$ nmr in antiferromagnetic csmn${\mathrm{f}}_{3}$}, \href{https://doi.org/10.1103/PhysRev.143.361}{\emph{Phys. Rev.} {\bfseries 143} (1966) 361}.

\bibitem{PhysRev.156.370}
L.B.~Welsh, \emph{Properties of the ${\mathrm{mn}}^{55}$ nuclear-magnetic-resonance modes in csmn${\mathrm{f}}_{3}$}, \href{https://doi.org/10.1103/PhysRev.156.370}{\emph{Phys. Rev.} {\bfseries 156} (1967) 370}.

\bibitem{Gorghetto:2018ocs}
M.~Gorghetto and G.~Villadoro, \emph{{Topological Susceptibility and QCD Axion Mass: QED and NNLO corrections}}, \href{https://doi.org/10.1007/JHEP03(2019)033}{\emph{JHEP} {\bfseries 03} (2019) 033} [\href{https://arxiv.org/abs/1812.01008}{{\ttfamily 1812.01008}}].

\bibitem{PhysRevLett.43.103}
J.E.~Kim, \emph{Weak-interaction singlet and strong $\mathrm{CP}$ invariance}, \href{https://doi.org/10.1103/PhysRevLett.43.103}{\emph{Phys. Rev. Lett.} {\bfseries 43} (1979) 103}.

\bibitem{SHIFMAN1980493}
M.~Shifman, A.~Vainshtein and V.~Zakharov, \emph{Can confinement ensure natural cp invariance of strong interactions?}, \href{https://doi.org/https://doi.org/10.1016/0550-3213(80)90209-6}{\emph{Nuclear Physics B} {\bfseries 166} (1980) 493}.

\bibitem{DINE1981199}
M.~Dine, W.~Fischler and M.~Srednicki, \emph{A simple solution to the strong cp problem with a harmless axion}, \href{https://doi.org/https://doi.org/10.1016/0370-2693(81)90590-6}{\emph{Physics Letters B} {\bfseries 104} (1981) 199}.

\bibitem{Zhitnitsky:1980tq}
A.R.~Zhitnitsky, \emph{{On Possible Suppression of the Axion Hadron Interactions. (In Russian)}}, {\emph{Sov. J. Nucl. Phys.} {\bfseries 31} (1980) 260}.

\bibitem{Chen:2013kt}
C.-Y.~Chen and S.~Dawson, \emph{{Exploring Two Higgs Doublet Models Through Higgs Production}}, \href{https://doi.org/10.1103/PhysRevD.87.055016}{\emph{Phys. Rev. D} {\bfseries 87} (2013) 055016} [\href{https://arxiv.org/abs/1301.0309}{{\ttfamily 1301.0309}}].

\bibitem{Ema:2016ops}
Y.~Ema, K.~Hamaguchi, T.~Moroi and K.~Nakayama, \emph{{Flaxion: a minimal extension to solve puzzles in the standard model}}, \href{https://doi.org/10.1007/JHEP01(2017)096}{\emph{JHEP} {\bfseries 01} (2017) 096} [\href{https://arxiv.org/abs/1612.05492}{{\ttfamily 1612.05492}}].

\bibitem{Calibbi:2016hwq}
L.~Calibbi, F.~Goertz, D.~Redigolo, R.~Ziegler and J.~Zupan, \emph{{Minimal axion model from flavor}}, \href{https://doi.org/10.1103/PhysRevD.95.095009}{\emph{Phys. Rev. D} {\bfseries 95} (2017) 095009} [\href{https://arxiv.org/abs/1612.08040}{{\ttfamily 1612.08040}}].

\bibitem{Graham:2015rva}
P.W.~Graham, J.~Mardon and S.~Rajendran, \emph{{Vector Dark Matter from Inflationary Fluctuations}}, \href{https://doi.org/10.1103/PhysRevD.93.103520}{\emph{Phys. Rev. D} {\bfseries 93} (2016) 103520} [\href{https://arxiv.org/abs/1504.02102}{{\ttfamily 1504.02102}}].

\bibitem{Ema:2019yrd}
Y.~Ema, K.~Nakayama and Y.~Tang, \emph{{Production of purely gravitational dark matter: the case of fermion and vector boson}}, \href{https://doi.org/10.1007/JHEP07(2019)060}{\emph{JHEP} {\bfseries 07} (2019) 060} [\href{https://arxiv.org/abs/1903.10973}{{\ttfamily 1903.10973}}].

\bibitem{Agrawal:2018vin}
P.~Agrawal, N.~Kitajima, M.~Reece, T.~Sekiguchi and F.~Takahashi, \emph{{Relic Abundance of Dark Photon Dark Matter}}, \href{https://doi.org/10.1016/j.physletb.2019.135136}{\emph{Phys. Lett. B} {\bfseries 801} (2020) 135136} [\href{https://arxiv.org/abs/1810.07188}{{\ttfamily 1810.07188}}].

\bibitem{Co:2018lka}
R.T.~Co, A.~Pierce, Z.~Zhang and Y.~Zhao, \emph{{Dark Photon Dark Matter Produced by Axion Oscillations}}, \href{https://doi.org/10.1103/PhysRevD.99.075002}{\emph{Phys. Rev. D} {\bfseries 99} (2019) 075002} [\href{https://arxiv.org/abs/1810.07196}{{\ttfamily 1810.07196}}].

\bibitem{Bastero-Gil:2018uel}
M.~Bastero-Gil, J.~Santiago, L.~Ubaldi and R.~Vega-Morales, \emph{{Vector dark matter production at the end of inflation}}, \href{https://doi.org/10.1088/1475-7516/2019/04/015}{\emph{JCAP} {\bfseries 04} (2019) 015} [\href{https://arxiv.org/abs/1810.07208}{{\ttfamily 1810.07208}}].

\bibitem{Kitajima:2023pby}
N.~Kitajima and F.~Takahashi, \emph{{Resonant production of dark photons from axion without a large coupling}},  \href{https://arxiv.org/abs/2303.05492}{{\ttfamily 2303.05492}}.

\bibitem{Dror:2018pdh}
J.A.~Dror, K.~Harigaya and V.~Narayan, \emph{{Parametric Resonance Production of Ultralight Vector Dark Matter}}, \href{https://doi.org/10.1103/PhysRevD.99.035036}{\emph{Phys. Rev. D} {\bfseries 99} (2019) 035036} [\href{https://arxiv.org/abs/1810.07195}{{\ttfamily 1810.07195}}].

\bibitem{Nakayama:2021avl}
K.~Nakayama and W.~Yin, \emph{{Hidden photon and axion dark matter from symmetry breaking}}, \href{https://doi.org/10.1007/JHEP10(2021)026}{\emph{JHEP} {\bfseries 10} (2021) 026} [\href{https://arxiv.org/abs/2105.14549}{{\ttfamily 2105.14549}}].

\bibitem{Long:2019lwl}
A.J.~Long and L.-T.~Wang, \emph{{Dark Photon Dark Matter from a Network of Cosmic Strings}}, \href{https://doi.org/10.1103/PhysRevD.99.063529}{\emph{Phys. Rev. D} {\bfseries 99} (2019) 063529} [\href{https://arxiv.org/abs/1901.03312}{{\ttfamily 1901.03312}}].

\bibitem{Kitajima:2022lre}
N.~Kitajima and K.~Nakayama, \emph{{Dark Photon Dark Matter from Cosmic Strings and Gravitational Wave Background}},  \href{https://arxiv.org/abs/2212.13573}{{\ttfamily 2212.13573}}.

\bibitem{Kitajima:2023fun}
N.~Kitajima and K.~Nakayama, \emph{{Viable Vector Coherent Oscillation Dark Matter}},  \href{https://arxiv.org/abs/2303.04287}{{\ttfamily 2303.04287}}.

\bibitem{PhysRevD.104.095029}
A.~Caputo, A.J.~Millar, C.A.J.~O'Hare and E.~Vitagliano, \emph{Dark photon limits: A handbook}, \href{https://doi.org/10.1103/PhysRevD.104.095029}{\emph{Phys. Rev. D} {\bfseries 104} (2021) 095029}.

\bibitem{Chaudhuri:2014dla}
S.~Chaudhuri, P.W.~Graham, K.~Irwin, J.~Mardon, S.~Rajendran and Y.~Zhao, \emph{{Radio for hidden-photon dark matter detection}}, \href{https://doi.org/10.1103/PhysRevD.92.075012}{\emph{Phys. Rev. D} {\bfseries 92} (2015) 075012} [\href{https://arxiv.org/abs/1411.7382}{{\ttfamily 1411.7382}}].

\bibitem{PhysRevB.86.224407}
J.B.~Lee, W.G.~Hong, H.J.~Kim, Z.~Jagli\ifmmode \check{c}\else \v{c}\fi{}i\ifmmode~\acute{c}\else \'{c}\fi{}, S.~Jazbec, M.~Wencka et~al., \emph{Canted antiferromagnetism on a nanodimensional spherical surface geometry: The case of mnco${}_{3}$ small hollow nanospheres}, \href{https://doi.org/10.1103/PhysRevB.86.224407}{\emph{Phys. Rev. B} {\bfseries 86} (2012) 224407}.

\bibitem{Liang:2020aa}
W.~Liang, L.~Li, R.~Li, Y.~Yin, Z.~Li, X.~Liu et~al., \emph{Crystal structure of impurity-free rhodochrosite (mnco3) and thermal expansion properties}, \href{https://doi.org/10.1007/s00269-019-01078-2}{\emph{Physics and Chemistry of Minerals} {\bfseries 47} (2020) 9}.

\bibitem{DZYALOSHINSKY1958241}
I.~Dzyaloshinsky, \emph{A thermodynamic theory of ``weak'' ferromagnetism of antiferromagnetics}, \href{https://doi.org/https://doi.org/10.1016/0022-3697(58)90076-3}{\emph{Journal of Physics and Chemistry of Solids} {\bfseries 4} (1958) 241}.

\bibitem{PhysRev.120.91}
T.~Moriya, \emph{Anisotropic superexchange interaction and weak ferromagnetism}, \href{https://doi.org/10.1103/PhysRev.120.91}{\emph{Phys. Rev.} {\bfseries 120} (1960) 91}.

\bibitem{PhysRev.117.635}
T.~Moriya, \emph{Theory of magnetism of ni${\mathrm{f}}_{2}$}, \href{https://doi.org/10.1103/PhysRev.117.635}{\emph{Phys. Rev.} {\bfseries 117} (1960) 635}.

\bibitem{jiseikei}
K.~Yoshida, \emph{Magnetism (translated from Japanese)}, Iwanami Shoten (2015).

\bibitem{PhysRevLett.114.226402}
L.V.~Abdurakhimov, Y.M.~Bunkov and D.~Konstantinov, \emph{Normal-mode splitting in the coupled system of hybridized nuclear magnons and microwave photons}, \href{https://doi.org/10.1103/PhysRevLett.114.226402}{\emph{Phys. Rev. Lett.} {\bfseries 114} (2015) 226402}.

\bibitem{PhysRevB.97.024425}
L.V.~Abdurakhimov, M.A.~Borich, Y.M.~Bunkov, R.R.~Gazizulin, D.~Konstantinov, M.I.~Kurkin et~al., \emph{Nonlinear nmr and magnon bec in antiferromagnetic materials with coupled electron and nuclear spin precession}, \href{https://doi.org/10.1103/PhysRevB.97.024425}{\emph{Phys. Rev. B} {\bfseries 97} (2018) 024425}.

\bibitem{PhysRev.184.574}
W.J.~Ince, \emph{Coupled antiferromagnetic-nuclear-magnetic resonance in ${\mathrm{rbmnf}}_{3}$}, \href{https://doi.org/10.1103/PhysRev.184.574}{\emph{Phys. Rev.} {\bfseries 184} (1969) 574}.

\bibitem{PhysRevB.4.1572}
J.B.~Merry and D.I.~Bolef, \emph{Nuclear acoustic resonance of ${\mathrm{mn}}^{55}$ in antiferromagnetic rbmn${\mathrm{f}}_{3}$}, \href{https://doi.org/10.1103/PhysRevB.4.1572}{\emph{Phys. Rev. B} {\bfseries 4} (1971) 1572}.

\bibitem{doi:10.1063/1.1729366}
A.M.~Portis, G.L.~Witt and A.J.~Heeger, \emph{Excitation of nuclear magnetic resonance modes in antiferromagnetic kmnf3}, \href{https://doi.org/10.1063/1.1729366}{\emph{Journal of Applied Physics} {\bfseries 34} (1963) 1052}.

\bibitem{PhysRevLett.37.533}
A.R.~King, V.~Jaccarino and S.M.~Rezende, \emph{Nuclear magnons and nuclear magnetostatic modes in mn${\mathrm{f}}_{2}$}, \href{https://doi.org/10.1103/PhysRevLett.37.533}{\emph{Phys. Rev. Lett.} {\bfseries 37} (1976) 533}.

\bibitem{PhysRev.135.A661}
A.J.~Heeger and T.W.~Houston, \emph{Nuclear magnetic resonance in ferrimagnetic mn${\mathrm{fe}}_{2}$${\mathrm{o}}_{4}$}, \href{https://doi.org/10.1103/PhysRev.135.A661}{\emph{Phys. Rev.} {\bfseries 135} (1964) A661}.

\bibitem{PhysRev.136.A218}
H.~Fink and D.~Shaltiel, \emph{Nuclear frequency pulling in a dzialoshinskii-moriya-type weak ferromagnet: Mnc${\mathrm{o}}_{3}$}, \href{https://doi.org/10.1103/PhysRev.136.A218}{\emph{Phys. Rev.} {\bfseries 136} (1964) A218}.

\bibitem{turov1973nuclear}
E.~Turov, I.~Kurkin and V.~Nikolaev, \emph{Nuclear spin motion with allowance for suhl-nakamura interaction}, {\emph{Soviet Physics JETP} {\bfseries 37} (1973) 147}.

\bibitem{PhysRev.82.651}
R.E.~Sheriff and D.~Williams, \emph{Nuclear gyromagnetic ratios. iii}, \href{https://doi.org/10.1103/PhysRev.82.651}{\emph{Phys. Rev.} {\bfseries 82} (1951) 651}.

\bibitem{STONE200575}
N.~Stone, \emph{Table of nuclear magnetic dipole and electric quadrupole moments}, \href{https://doi.org/https://doi.org/10.1016/j.adt.2005.04.001}{\emph{Atomic Data and Nuclear Data Tables} {\bfseries 90} (2005) 75}.

\bibitem{PhysRev.142.300}
D.~Shaltiel, \emph{Nuclear magnetic resonance of mnc${\mathrm{o}}_{3}$ in the canted spin state}, \href{https://doi.org/10.1103/PhysRev.142.300}{\emph{Phys. Rev.} {\bfseries 142} (1966) 300}.

\bibitem{RevModPhys.24.63}
P.F.A.~Klinkenberg, \emph{Tables of nuclear shell structure}, \href{https://doi.org/10.1103/RevModPhys.24.63}{\emph{Rev. Mod. Phys.} {\bfseries 24} (1952) 63}.

\bibitem{Ye_2023}
H.~Ye, J.~Li, , D.~Yang, H.~Jin and X.~Huang, \emph{Low-lying state investigations of odd-a mn isotopes around n = 28}, \href{https://doi.org/10.1088/1572-9494/aca07f}{\emph{Communications in Theoretical Physics} {\bfseries 75} (2023) 025302}.

\bibitem{bunkovecho}
Y.M.~Bun'kov and B.S.~Dumesh, \emph{Dynamic effects in pulsed nmr in easy plane antiferromagnets with large dynamic frequency shifts}, {\emph{Zh. Eksp. Teor. Fiz.} {\bfseries 68} (1975) }.

\bibitem{Aybas:2021nvn}
D.~Aybas et~al., \emph{{Search for Axionlike Dark Matter Using Solid-State Nuclear Magnetic Resonance}}, \href{https://doi.org/10.1103/PhysRevLett.126.141802}{\emph{Phys. Rev. Lett.} {\bfseries 126} (2021) 141802} [\href{https://arxiv.org/abs/2101.01241}{{\ttfamily 2101.01241}}].

\bibitem{doi:10.1063/1.2803852}
J.A.B.~Mates, G.C.~Hilton, K.D.~Irwin, L.R.~Vale and K.W.~Lehnert, \emph{Demonstration of a multiplexer of dissipationless superconducting quantum interference devices}, \href{https://doi.org/10.1063/1.2803852}{\emph{Applied Physics Letters} {\bfseries 92} (2008) 023514}.

\bibitem{JABmates}
J.A.B.~Mates, \emph{The Microwave Squid Multiplexer}, Ph.D. thesis, University of Colorado, 2011.

\bibitem{PhysRevD.92.075012}
S.~Chaudhuri, P.W.~Graham, K.~Irwin, J.~Mardon, S.~Rajendran and Y.~Zhao, \emph{Radio for hidden-photon dark matter detection}, \href{https://doi.org/10.1103/PhysRevD.92.075012}{\emph{Phys. Rev. D} {\bfseries 92} (2015) 075012}.

\bibitem{139223}
M.~Tarasov, V.~Belitsky and G.~Prokopenko, \emph{Dc squid rf amplifiers}, \href{https://doi.org/10.1109/77.139223}{\emph{IEEE Transactions on Applied Superconductivity} {\bfseries 2} (1992) 79}.

\bibitem{doi:10.1063/1.1347384}
M.~M{\"u}ck, J.B.~Kycia and J.~Clarke, \emph{Superconducting quantum interference device as a near-quantum-limited amplifier at 0.5 ghz}, \href{https://doi.org/10.1063/1.1347384}{\emph{Applied Physics Letters} {\bfseries 78} (2001) 967}.

\bibitem{doi:10.1063/1.121490}
M.~M{\"u}ck, M.-O.~Andr{\'e}, J.~Clarke, J.~Gail and C.~Heiden, \emph{Radio-frequency amplifier based on a niobium dc superconducting quantum interference device with microstrip input coupling}, \href{https://doi.org/10.1063/1.121490}{\emph{Applied Physics Letters} {\bfseries 72} (1998) 2885}.

\bibitem{2002PhDT.......116T}
R.~{Therrien}, \emph{{The microstrip SQUID amplifier}}, Ph.D. thesis, University of California, Berkeley, Jan., 2002.

\bibitem{Miller2000InterplayAR}
J.B.~Miller, B.H.~Suits, A.N.~Garroway and M.A.~Hepp, \emph{Interplay among recovery time, signal, and noise: series- and parallel-tuned circuits are not always the same}, {\emph{Concepts in Magnetic Resonance} {\bfseries 12} (2000) 125}.

\bibitem{HOULT197671}
D.~Hoult and R.~Richards, \emph{The signal-to-noise ratio of the nuclear magnetic resonance experiment}, \href{https://doi.org/https://doi.org/10.1016/0022-2364(76)90233-X}{\emph{Journal of Magnetic Resonance (1969)} {\bfseries 24} (1976) 71}.

\bibitem{ADMX:2011hrx}
{\scshape ADMX} collaboration, \emph{{Design and performance of the ADMX SQUID-based microwave receiver}}, \href{https://doi.org/10.1016/j.nima.2011.07.019}{\emph{Nucl. Instrum. Meth. A} {\bfseries 656} (2011) 39} [\href{https://arxiv.org/abs/1105.4203}{{\ttfamily 1105.4203}}].

\bibitem{doi:10.1063/1.1770483}
R.H.~Dicke, \emph{The measurement of thermal radiation at microwave frequencies}, \href{https://doi.org/10.1063/1.1770483}{\emph{Review of Scientific Instruments} {\bfseries 17} (1946) 268}.

\bibitem{Carenza:2019pxu}
P.~Carenza, T.~Fischer, M.~Giannotti, G.~Guo, G.~Mart\'\i{}nez-Pinedo and A.~Mirizzi, \emph{{Improved axion emissivity from a supernova via nucleon-nucleon bremsstrahlung}}, \href{https://doi.org/10.1088/1475-7516/2019/10/016}{\emph{JCAP} {\bfseries 10} (2019) 016} [\href{https://arxiv.org/abs/1906.11844}{{\ttfamily 1906.11844}}].

\bibitem{Carenza:2020cis}
P.~Carenza, B.~Fore, M.~Giannotti, A.~Mirizzi and S.~Reddy, \emph{{Enhanced Supernova Axion Emission and its Implications}}, \href{https://doi.org/10.1103/PhysRevLett.126.071102}{\emph{Phys. Rev. Lett.} {\bfseries 126} (2021) 071102} [\href{https://arxiv.org/abs/2010.02943}{{\ttfamily 2010.02943}}].

\bibitem{Fischer:2021jfm}
T.~Fischer, P.~Carenza, B.~Fore, M.~Giannotti, A.~Mirizzi and S.~Reddy, \emph{{Observable signatures of enhanced axion emission from protoneutron stars}}, \href{https://doi.org/10.1103/PhysRevD.104.103012}{\emph{Phys. Rev. D} {\bfseries 104} (2021) 103012} [\href{https://arxiv.org/abs/2108.13726}{{\ttfamily 2108.13726}}].

\bibitem{Hamaguchi:2018oqw}
K.~Hamaguchi, N.~Nagata, K.~Yanagi and J.~Zheng, \emph{{Limit on the Axion Decay Constant from the Cooling Neutron Star in Cassiopeia A}}, \href{https://doi.org/10.1103/PhysRevD.98.103015}{\emph{Phys. Rev. D} {\bfseries 98} (2018) 103015} [\href{https://arxiv.org/abs/1806.07151}{{\ttfamily 1806.07151}}].

\bibitem{Capozzi:2020cbu}
F.~Capozzi and G.~Raffelt, \emph{{Axion and neutrino bounds improved with new calibrations of the tip of the red-giant branch using geometric distance determinations}}, \href{https://doi.org/10.1103/PhysRevD.102.083007}{\emph{Phys. Rev. D} {\bfseries 102} (2020) 083007} [\href{https://arxiv.org/abs/2007.03694}{{\ttfamily 2007.03694}}].

\bibitem{Giannotti:2017hny}
M.~Giannotti, I.G.~Irastorza, J.~Redondo, A.~Ringwald and K.~Saikawa, \emph{{Stellar Recipes for Axion Hunters}}, \href{https://doi.org/10.1088/1475-7516/2017/10/010}{\emph{JCAP} {\bfseries 10} (2017) 010} [\href{https://arxiv.org/abs/1708.02111}{{\ttfamily 1708.02111}}].

\bibitem{MillerBertolami:2014rka}
M.M.~Miller~Bertolami, B.E.~Melendez, L.G.~Althaus and J.~Isern, \emph{{Revisiting the axion bounds from the Galactic white dwarf luminosity function}}, \href{https://doi.org/10.1088/1475-7516/2014/10/069}{\emph{JCAP} {\bfseries 10} (2014) 069} [\href{https://arxiv.org/abs/1406.7712}{{\ttfamily 1406.7712}}].

\bibitem{LUX:2017glr}
{\scshape LUX} collaboration, \emph{{First Searches for Axions and Axionlike Particles with the LUX Experiment}}, \href{https://doi.org/10.1103/PhysRevLett.118.261301}{\emph{Phys. Rev. Lett.} {\bfseries 118} (2017) 261301} [\href{https://arxiv.org/abs/1704.02297}{{\ttfamily 1704.02297}}].

\bibitem{PandaX:2017ock}
{\scshape PandaX} collaboration, \emph{{Limits on Axion Couplings from the First 80 Days of Data of the PandaX-II Experiment}}, \href{https://doi.org/10.1103/PhysRevLett.119.181806}{\emph{Phys. Rev. Lett.} {\bfseries 119} (2017) 181806} [\href{https://arxiv.org/abs/1707.07921}{{\ttfamily 1707.07921}}].

\bibitem{XENON100:2014csq}
{\scshape XENON100} collaboration, \emph{{First Axion Results from the XENON100 Experiment}}, \href{https://doi.org/10.1103/PhysRevD.90.062009}{\emph{Phys. Rev. D} {\bfseries 90} (2014) 062009} [\href{https://arxiv.org/abs/1404.1455}{{\ttfamily 1404.1455}}].

\bibitem{PhysRevD.88.035020}
S.K.~Lamoreaux, K.A.~van Bibber, K.W.~Lehnert and G.~Carosi, \emph{Analysis of single-photon and linear amplifier detectors for microwave cavity dark matter axion searches}, \href{https://doi.org/10.1103/PhysRevD.88.035020}{\emph{Phys. Rev. D} {\bfseries 88} (2013) 035020}.

\bibitem{Chattopadhyay_1995}
T.~Chattopadhyay and K.~Siemensmeyer, \emph{Hyperfine induced nuclear polarization in nd2cuo4}, \href{https://doi.org/10.1209/0295-5075/29/7/012}{\emph{Europhysics Letters} {\bfseries 29} (1995) 579}.

\bibitem{Chatterji_2013}
T.~Chatterji, O.~Holderer and H.~Schneider, \emph{Direct evidence for nuclear spin waves in nd2cuo4 by high-resolution neutron-spin-echo spectroscopy}, \href{https://doi.org/10.1088/0953-8984/25/47/476002}{\emph{Journal of Physics: Condensed Matter} {\bfseries 25} (2013) 476002}.

\bibitem{Kimball_2015}
D.F.J.~Kimball, \emph{Nuclear spin content and constraints on exotic spin-dependent couplings}, \href{https://doi.org/10.1088/1367-2630/17/7/073008}{\emph{New Journal of Physics} {\bfseries 17} (2015) 073008}.

\bibitem{PhysRevD.40.3132}
J.~Engel and P.~Vogel, \emph{Spin-dependent cross sections of weakly interacting massive particles on nuclei}, \href{https://doi.org/10.1103/PhysRevD.40.3132}{\emph{Phys. Rev. D} {\bfseries 40} (1989) 3132}.

\bibitem{PhysRevC.103.014303}
B.~Hern\'andez, P.~Sarriguren, O.~Moreno, E.~Moya~de Guerra, D.N.~Kadrev and A.N.~Antonov, \emph{Nuclear shape transitions and elastic magnetic electron scattering}, \href{https://doi.org/10.1103/PhysRevC.103.014303}{\emph{Phys. Rev. C} {\bfseries 103} (2021) 014303}.

\bibitem{nilsson1955binding}
S.G.~Nilsson, \emph{Binding states of individual nucleons in strongly deformed nuclei}, {\emph{Dan. Mat. Fys. Medd.} {\bfseries 29} (1955) 1}.

\bibitem{bohr1998nuclear}
A.~Bohr and B.~Mottelson, \emph{Nuclear Structure}, no.~v. 2 in Nuclear Structure, World Scientific (1998).

\bibitem{PhysRevC.73.055501}
V.V.~Flambaum and A.F.~Tedesco, \emph{Dependence of nuclear magnetic moments on quark masses and limits on temporal variation of fundamental constants from atomic clock experiments}, \href{https://doi.org/10.1103/PhysRevC.73.055501}{\emph{Phys. Rev. C} {\bfseries 73} (2006) 055501}.

\bibitem{Rezende1}
S.M.~Rezende, \emph{Fundamentals of Magnonics}, Lecture Notes in Physics, 969, Springer International Publishing, Cham, 1st ed. 2020.~ed. (2020).

\bibitem{PhysRev.139.A450}
R.M.~White, M.~Sparks and I.~Ortenburger, \emph{Diagonalization of the antiferromagnetic magnon-phonon interaction}, \href{https://doi.org/10.1103/PhysRev.139.A450}{\emph{Phys. Rev.} {\bfseries 139} (1965) A450}.

\end{thebibliography}\endgroup

%%%%%%%%%%%%%%%%%%%%%%%%%%%%%%%%%%%%%%%

\end{document}